\documentclass[a4paper,11pt]{article}
\usepackage{epsfig,amssymb,amstext}

\newenvironment{Filesave}[1]{}{} \newcommand{\OPBIL}{} \newcommand{\CLOBIL}{ }

\renewcommand{\arraystretch}{1.3}

   \def\bege{\begin{equation}}
   \def\ende{\end{equation}}
   \def\bega{\begin{eqnarray}}
   \def\enda{\end{eqnarray}}
   \def\began{\begin{eqnarray*}}
   \def\endan{\end{eqnarray*}}

   \def\ds{\displaystyle}

\newcommand{\nNewt}{}
\newcommand{\tabstart}{}
\newcommand{\tabend}{}

\newcommand{\bm}[1]{\mbox{\boldmath$#1$\unboldmath}}
\newcommand{\N}{{\mathbb N}} 

\newcommand{\R}{{\mathbb R}}
\newcommand{\Z}{{\mathbb Z}}

\newcommand{\rem}[1]{}

\newcommand{\todo}[1]{}
\newcommand{\dof}{f}
\newcommand{\Eell}{{\cal E}}   
\newcommand{\Fell}{{\cal F}}
\newcommand{\Kell}{{\cal K}}

\newcommand{\rad}{\rho}  
\newcommand{\ang}{\phi}     
\newcommand{\prad}{p_{\rad}}
\newcommand{\pradt}{p_{\rad}^{\ 2}}
\newcommand{\pang}{p_{\ang}}
\newcommand{\pangt}{p_{\ang}^{\ 2}}
\newcommand{\action}{S}
\newcommand{\BId}{{\bm {\tilde I}}}
\newcommand{\Idrad}{\tilde{I}_{\rad}}
\newcommand{\Idang}{\tilde{I}_{\ang}}
\newcommand{\omdrad}{\tilde{\omega}_{\rad}}
\newcommand{\omdang}{\tilde{\omega}_{\ang}}
\newcommand{\wrd}{\tilde{w}}
\newcommand{\BIf}{{\bm I}}  
\newcommand{\Ifrad}{I_{\rad}}
\newcommand{\Ifang}{I_{\ang}}
\newcommand{\Bomf}{{\bm \omega}}
\newcommand{\omfrad}{\omega_{\rad}}
\newcommand{\omfang}{\omega_{\ang}}
\newcommand{\wrf}{w}
\newcommand{\BJf}{{\bm J}}  
\newcommand{\Jfrad}{J_{\rad}}
\newcommand{\Jfang}{J_{\ang}}

\newcommand{\Bn}{{\bm {\tilde n}}} 
\newcommand{\wrfn}{w_{\text{n}}}
\newcommand{\sds}{n_{\text sc}(E)}
\newcommand{\sdsg}{n_{\text sc}(E;\gamma)}
\newcommand{\Bmaslovf}{{\bm \alpha}}

\newcommand{\Bmaslovd}{{\bm {\tilde \alpha}}}

\newcommand{\Brtopo}{{\bm \mu}}
\newcommand{\rtopo}{\mu}
\newcommand{\psiwkbrad}{F}
\newcommand{\psiwkbang}{G}
\newcommand{\nrad}{r}
\newcommand{\nang}{l}
\newcommand{\tO}{O}
\newcommand{\tR}{R}
\newcommand{\tTO}{TO}
\newcommand{\tTR}{TR}
\newcommand{\tOc}{O_c}

\newcommand{\tTOc}{TO_c}
\newcommand{\os}{o_s}
\newcommand{\ou}{o_u}
\newcommand{\ob}{o_b}
\newcommand{\lengths}{L_s}
\newcommand{\lengthu}{L_u}
\newcommand{\lengthb}{L_b}
\newcommand{\Veff}{V_{\text{eff}}}
\newcommand{\const}{\text{const}}
\newcommand{\curv}{C}
\newcommand{\trace}{\text{tr}\,}

\newcommand{\capsty}{\footnotesize}

\newcommand{\hono}[1]{}
\newcommand{\nono}[1]{}

\newcommand{\nootnote}[1]{}
\newcommand{\EBK}{{\em EBK}}
\newcommand{\WKB}{{\em WKB}}
\newcommand{\Sec}{sec.}
\newcommand{\equ}{Eq.}
\newcommand{\fig}{Fig.}
\newcommand{\tab}{Tab.}
\newcommand{\EBKBT}{{\em EBKBT}}
\newcommand{\WKBBT}{{\em semiuniform WKBBT}}
\newcommand{\uniWKBBT}{{\em uniform WKBBT}}

\newcommand{\NEU}{}

\newcommand{\Fang}{G}
\newcommand{\Frad}{F}
\newcommand{\Fxi}{f}
\newcommand{\Feta}{g}

\newcommand{\psieta}{\Fang}
\newcommand{\alteta}{\ang}
\newcommand{\psixi}{\Frad}

\newcommand{\sFxi}{h_m} 
\newcommand{\sxi}{x}  
\newcommand{\D}[1]{\partial_{#1}}
\newcommand{\DD}[1]{\partial^2_{#1}}
\newcommand{\SeKo}{\lambda}
\newcommand{\ec}{c^2}

\newcommand{\fcos}{f_{\text{cos}}}
\newcommand{\fsin}{f_{\text{sin}}}

\newtheorem{platz}{{\bf Fig.}} 
\rem{
\newcommand{\Capts}[1]{#1}
\newcommand{\FIGo}[3]{\vspace*{-4ex}%
\marginpar{ \begin{platz} \label{#1} ~ \end{platz} }\vspace*{4ex}}
\newcommand{\printfigcap}[2]{\rule{0cm}{1cm}{\bf Figure~\ref{#1}:} #2 \\ }
\newcommand{\showfig}[1]{\begin{figure} #1 \caption{Elliptic Quantum Billiard} \end{figure} } 
} 
\newcommand{\Capts}[1]{}
\newcommand{\FIGo}[3]{\begin{figure}%
#3%
\caption[]{\capsty #2}%
\label{#1}%
\end{figure}}
\rem{
\includeonly{
  introdu
 ,classic
 ,qm
 ,miller
 ,bt
 ,invprob
 ,conclus 
}
} 

\begin{document}

\OPBIL

\vspace*{1cm}
\begin{center}
{\huge Elliptic Quantum Billiard} \\[8ex]
{\large Holger Waalkens, Jan Wiersig, Holger R.~Dullin}\\[5ex]
{\large Institut f\"ur Theoretische Physik and}\\[1ex]
{\large Institut f\"ur Dynamische Systeme}\\[1ex]
{\large University of Bremen}\\[1ex]
{\large Postfach 330~440}\\[1ex]
{\large 28334 Bremen, Germany}\\[5ex]
E-mail: jwiersig@physik.uni-bremen.de \\[5ex]
\today
\end{center}
\vspace*{1cm}

\section*{Abstract}
The exact and semiclassical quantum mechanics of the elliptic 
billiard is investigated.
The classical system is integrable and exhibits a separatrix, dividing the phase
space into regions of oscillatory and rotational motion. 
The classical separability carries over to quantum mechanics, and
the Schr\"odinger equation is shown to be equivalent to the
spheroidal wave equation. The quantum eigenvalues show a clear
pattern when transformed into the classical action space.
The implication of the separatrix on the wave functions is illustrated.
A uniform \WKB\ quantization taking into account complex orbits is shown
to be adequate for the semiclassical quantization in the presence
of a separatrix. The pattern of states in classical action space 
is nicely explained by this quantization procedure.
We extract an effective Maslov phase varying smoothly on the
energy surface, which is used to modify the Berry-Tabor trace formula, 
resulting in a summation over non-periodic orbits. 
This modified trace formula produces the correct number of 
states, even close to the separatrix.
The Fourier transform of the density of states is explained in
terms of classical orbits, and the amplitude and form of
the different kinds of peaks is analytically calculated.

\clearpage



\section{Introduction}
\label{sec:introduction}

The semiclassical quantization of a Hamiltonian system is deeply connected to 
the structure of its phase space. 
The generic Hamiltonian system contains a complicated mixture of 
near-integrable and chaotic motion, and a consistent semiclassical quantization
scheme does not exist for the generic case. It does exist, however,
in the non-generic limiting cases, the integrable and the ergodic systems.

During the last two decades much progress has been made for
ergodic systems. The Gutzwiller trace formula gives the density of 
states as a sum over periodic orbits \cite{Gutz67,Gutz70,Gutz71}. 
It works well if all periodic orbits are unstable and isolated, i.e.\ for
hyperbolic systems, and is the starting point for most
calculations in this field. 

The investigation of integrable systems reaches back to
the beginning of quantum mechanics.
Bohr and  Sommerfeld, among others, succeeded in the quantization of actions.
The hydrogen atom is the most famous example. The deficiency of
the old quantum mechanics to calculate, e.g., the spectrum of the
helium atom was set into clear light by Einstein \cite{Einstein17}. He
formulated the Bohr-Sommerfeld quantization conditions in terms of invariant
tori which foliate the phase space of integrable systems.
Moreover he noted that this foliation is absent in generic systems,
such that this quantization scheme fails.
The same year when Schr{\"o}dinger introduced  his famous equation
Brillouin  explained that the quantization of tori is a consequence of the
single-valuedness of the quantum mechanical wave functions. 
Keller showed how the quantization conditions have to be modified because 
of caustics in the classical motion \cite{Keller58}. 
For an integrable system with $f$ degrees of freedom
described by its phase space variables 
$({\bf q,p}) = (q_1,...,q_f,p_1,...,p_f)$ 
the  Einstein-Brillouin-Keller (\EBK)
quantization conditions for the actions $I_i$ read
\begin{eqnarray}\label{eq:EBK}
 I_i=\frac{1}{2\pi}\oint_{\gamma_i} {\bf p}\,d{\bf q}=\hbar(n_i+\frac{\alpha_i}{4})\,,
\end{eqnarray}
where the integration is to be taken along $f$ topologically independent paths
$\gamma_i$ around the $f$-torus, 
and $\alpha_i$ are the Maslov indices due to the classical caustics.

This paper deals  with a nontrivial example of an integrable system: 
the planar elliptic billiard. Classically
``billiard'' refers to a system in which a point particle moves freely 
inside a domain and is elastically reflected at the domain boundary.
The corresponding quantum mechanical problem is to solve the Laplace
equation inside the domain with Dirichlet boundary condition.
Billiards have become popular because for them on the one hand the classical
calculations are easier than for systems with smooth potential, 
and on the other hand they allow for experimental measurements, 
e.g.\  
\cite{SS90,SS92,AGHHLRRS94}.
%
Jacobi \cite{Jacobi1866} showed that elliptic coordinates separate 
the geodesic flow on the ellipsoid, which contains our billiard as
a limiting case.
The same coordinate system leads to the separation of 
Schr\"odinger's equation, including the Dirichlet boundary condition,
into two Mathieu equations.

Integrability allows for a semiclassical quantization \`a la \EBK. 
In our case, however, there are problems due to the presence of a separatrix.
For all energies it divides phase space into regions of rotational and 
oscillatory motion \cite{RW95}. 
The Maslov indices are different for the two classical regions,
resulting in a discontinuity in the \EBK\ quantization condition which 
in turn leads to ambiguities for states close to the separatrix.
To overcome this problem it is necessary to introduce a uniform
quantization condition. This can be done by
investigating the asymptotics of the solutions of Schr\"odinger's equation
close to the separatrix, which was carried out for the
elliptic billiard in \cite{AyantArvieu86}. 
We will follow a different approach based on connection matrices 
between the amplitudes of \WKB\ wave functions, see \cite{Miller68,Child74},
the review of Berry and Mount \cite{BM72}, and the references therein.
This approach can be interpreted in terms of orbits with complex classical 
action.
A transformation of the quantum mechanical eigenvalues to classical action
space allows for a unique mapping of the eigenvalues to the quantum numbers.

The discontinuity in the \EBK\ quantization condition carries over to the
Berry-Tabor trace formula  \cite{BerryTabor76} - the analogue of the 
Gutzwiller trace formula for integrable systems.
The resonant tori, foliated by families of periodic orbits, 
take over the role of the isolated periodic orbits 
as the objects to be summed over.
The uniformization employed in \cite{BerryTabor76} incorporates the
influence of orbits with negative classical action in the vicinity
of isolated stable periodic orbits.
We will introduce a modification of the Berry-Tabor trace formula which takes
care of the separatrix in terms of an effective Maslov phase varying smoothly
across. It will turn out that in order to correctly produce eigenstates close 
to the separatrix the summation has to be taken over non-periodic orbits,
i.e.\ over classical tori with in general irrational winding number.
This trace formula incorporates all kinds of classically non-real orbits.
\nNewt

\NEU
The Berry-Tabor trace formula was further investigated by 
Richens~\cite{Richens82}, who showed that it contains 
contributions of the stable isolated periodic orbits,
whose contributions are equal to the
corresponding terms in the Gutzwiller's trace
formula~\cite{Gutz71}. We will use his results in the study of the
length spectrum, extending the ``inverse quantum chaology'', see e.g.~\cite{Wintgen87,BSS95}, to integrable systems.

After completion of this work 
there independently appeared a preprint by 
M.~Sieber \cite{Sieber96}, whose first part contains considerations
similar to parts of our paper.
Nevertheless, the key point of his paper being the semiclassical consequences
of the deformation of an elliptic billiard to an oval, while we 
concentrate on the separatrix and on complex orbits,
the overlap is only mild.

The organization of our paper is as follows. We start with a short
summary of the classical facts in \Sec~\ref{sec:classic}
and perform the exact quantum mechanical calculations in 
\Sec~\ref{sec:qmechanic}. In \Sec~\ref{sec:semiclassic} we introduce a
uniform \WKB\ quantization condition. This method gives an
effective Maslov phase which is used to modify the Berry-Tabor trace 
formula for systems with a separatrix in \Sec~\ref{sec:bt}, resulting
in a sum over non-periodic orbits.
The appearance of resonant tori in the length spectrum, i.e. in
the Fourier transform of the density of states,
is investigated in \Sec~\ref{sec:invproblem}.
The conclusion and an outlook are given in \Sec~\ref{sec:conclusions}.

\section{Classical mechanics}
\label{sec:classic}

The classical dynamics of the planar elliptic billiard has been investigated
by many authors, e.g.~\cite{Berry81,RW95,KT91,Chang88} and the references 
therein. We just give a summary of the facts important for our purpose.
Scaling the longer semimajor axis to one, the boundary of the billiard 
is described by
\bege
 x^2+\frac{y^2}{1-a^2} = 1, \qquad 0 \leq a < 1 \ ,
\ende
with foci at $(x,y) = (\pm a,0)$. 
The boundary of the ellipse is a $\ang$-coordinate line of 
the elliptic coordinates $(\rad,\ang)$ given by
\bege
\label{eq:radang}
 \begin{array}{r}
  (x,y) = \left(\ds a
  \cos{\ang}\cosh{\rad},a\sin{\ang}\sinh{\rad} \right) \ , 
 \end{array} 
\ende
where the coordinate ranges are  
\bege \label{eqn:rhomax}
  0 \leq \rad \leq \rad_{\text{max}} = {\text{arccosh}}(1/a), \quad 
  0 \leq \ang \leq 2\pi\ .
\ende
The lines $\rad = \const$ are confocal ellipses and the lines $\ang
= \const$ are confocal hyperbolas.
Introducing the conjugate momenta $(\prad,\pang)$, a reflection at the
boundary $\rad = \rad_{\text{max}}$ is simply described by 
$ 
  (\rad,\ang,\prad,\pang) \rightarrow 
  (\rad,\ang,-\prad,\pang)
$. 
The Hamiltonian of a freely moving particle of unit mass 
reads in these coordinates
\bege \label{eq:HamEll}
 H = \frac{\pradt+\pangt}{2a^2(\cosh^2 \rad-\cos^2 \ang)}\ .
\ende
Multiplying \equ~(\ref{eq:HamEll}) with the denominator of the
right hand side yields the
separation constant $K$:
\bega 
\label{eq:separation2}
        \pradt + (K - 2Ea^2\cosh^2 \rad) & = & 0, \\
\label{eq:separation1}
        \pangt - (K - 2Ea^2\cos^2 \ang) & = & 0 \,.
\enda 

\begin{Filesave}{bilder}
\def\figeffpot{Effective potentials
    $\Veff (\rad)$ and $\Veff (\ang)$ and the tori of the
    elliptic billiard. Real tori are marked by $\tR$ (rotations)
    or $\tO$ (oscillations). 
     The classically forbidden motion are 
    $\tTR$ (tunneling from region 1 to 2 in the radial direction),
    $\tTO$ (tunneling from region 1 to 2 in the angular direction),
    and $\tOc$ and $\tTOc$, the complex continuations of $\tO$ 
    and $\tTO$ below the minimum of $\Veff (\ang)$.
    Orbits with complex position are indicated by a circled cross.
    The reason for taking negative values of $\rad$ into account
    will become clear in \Sec~\ref{sec:semiclassic}.}
\def\FIGeffpot{\centerline{\psfig{figure=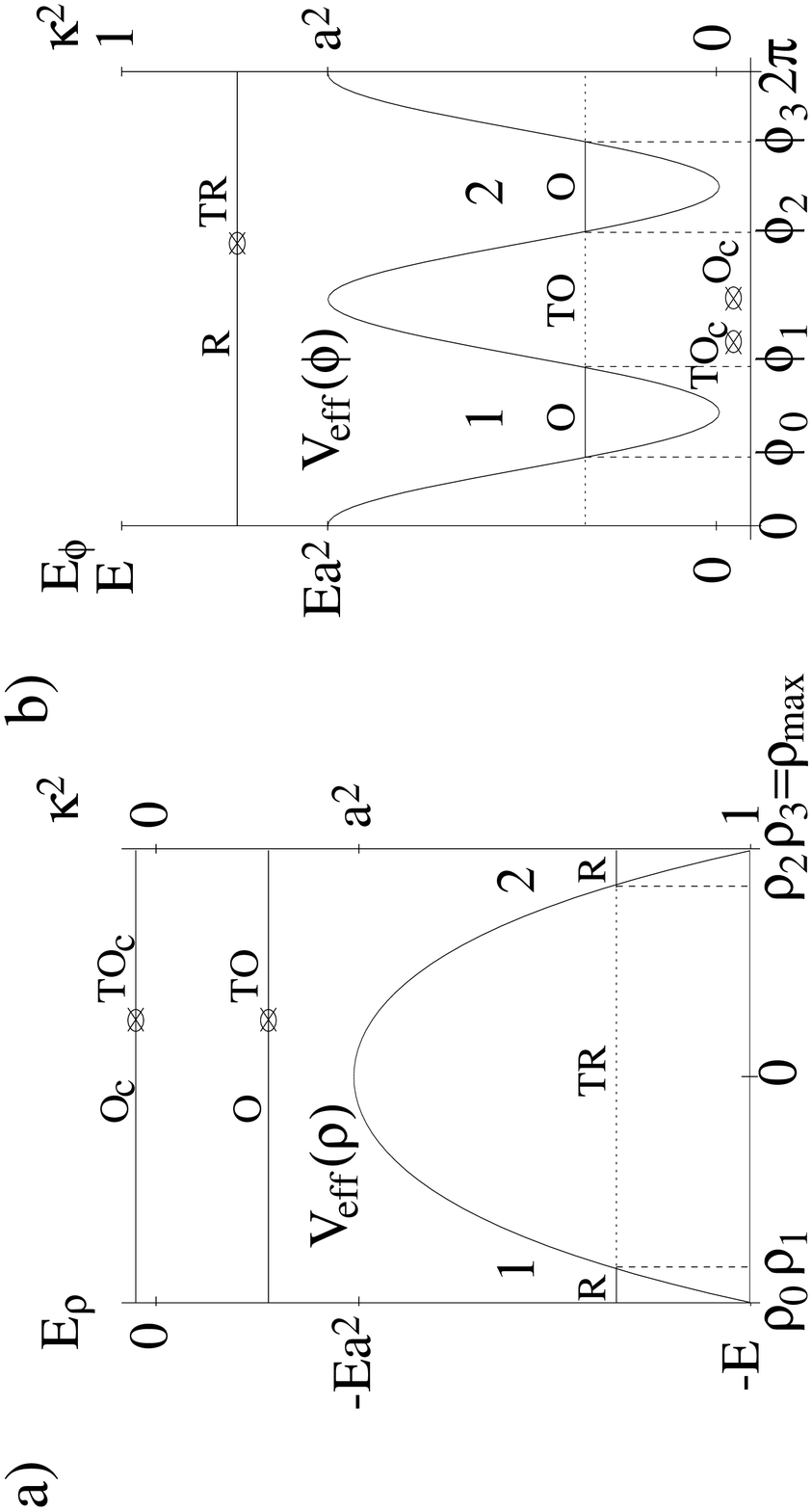,angle=-90,width=13.5cm}}}
\FIGo{fig:effpot}{\figeffpot}{\FIGeffpot}
\end{Filesave}

Here $E > 0$ is the energy, and $\kappa^2 = K/(2E)$ is the second
constant of the motion. Since $H$ and $K$ are in involution,
the system is integrable and therefore the energy
surface is foliated by invariant Liouville 2-tori. Each of the 
equations~(\ref{eq:separation2}) and (\ref{eq:separation1}) can be
interpreted as a Hamiltonian system with one degree of freedom,
with effective energy 
$E_\rad = -E\kappa^2$ and $E_\ang = E\kappa^2$ respectively,
and a sum of kinetic term and effective potential 
$\Veff (\rad) = -Ea^2\cosh^2(\rad)$ or $\Veff (\ang) = Ea^2\cos^2(\ang)$. 
The topologically different types of motions can
be discussed in terms of the effective potentials shown in
\fig~\ref{fig:effpot}. It is obvious that there are only two types
of classically allowed motion. For $\kappa^2 > a^2$ the 
trajectories avoid the interior of the ellipse $\cosh(\rad) = \kappa/a$, 
touching its boundary between every two consecutive reflections at the 
billiard boundary $\rad = \rad_{\text{max}}$.
This ``type $\tR$'' motion is similar to the rotational motion in a 
planar circular billiard. For $\kappa^2 < a^2$ the
trajectories always cross the $x$-axis between the foci; they are
confined to the domain enclosed by the hyperbolas 
$\cos \ang = \pm \kappa/a$.
This ``type $\tO$'' motion involves an oscillation in $\ang$.
The special values $\kappa^2 = 0$ and $\kappa^2 = 1$ represent the
stable oscillation along the $y$-axis ($\os$) and the sliding motion along the
boundary ($\ob$), respectively. $\kappa^2 = a^2$ characterizes the separatrix
motion and the unstable isolated periodic orbit.
An orbit on the separatrix alternately passes through one or the other focus
between reflections.
The unstable periodic orbit in the center of the separatrix performs 
an oscillation along the $x$-axis ($\ou$). 
It is the only orbit which passes through both foci between consecutive 
reflections.
The length of this orbit is  $\lengthu = 4$, the stable orbit has
$\lengths = 4\sqrt{1-a^2}$ and the length of the sliding orbit is just
the circumference $\lengthb = 4\Eell(a)$ of the billiard. $\Eell(k)$ is
the complete elliptic integral of the second kind in the notation of
\cite{GradRyzh65,Byrd71}.
The action $\action$ and the period $T$ of these orbits can easily be
obtained from $\action = \sqrt{2E}L$ and 
$ T = 
        \partial S/ \partial E = 
        \action/(2E) =  
        L/\sqrt{2E}$.

The calculation of the action variables $\BIf = (\Ifrad,\Ifang) =
(\frac{1}{2\pi}\oint \prad d\rad,\frac{1}{2\pi}\oint \pang d\ang)$, the
frequencies $\Bomf = \partial H/\partial \BIf$, and the
winding number $\wrf = \omfang / \omfrad$ of the elliptic billiard can
be found  in
\cite{RW95}. Since the system is invariant under reflections about the
$x$- and $y$-axis, we will also consider the desymmetrized
elliptic billiard, that is a quarter of the full ellipse. 
For type $\tR$ $(\kappa^2 > a^2)$ the actions $\BId$ and the winding number
$\wrd = \omdang / \omdrad$ of the desymmetrized billiard are
\bege
\begin{array}{lllrl}
 \label{eq:IWI}
 \Idrad & = & \frac{\sqrt{2E}}{\pi}\;\left( \sin \chi - 
             \kappa\; \Eell(\chi,\frac{a}{\kappa}) \right) 
             & = & \Ifrad \ , \\
 \Idang & = & \frac{\sqrt{2E}}{\pi}\kappa\;\Eell(\frac{a}{\kappa}) & =
 & \pm \frac{1}{2}\Ifang \ , \\
 \wrd & = & \Fell(\chi,\frac{a}{\kappa})/\Kell(\frac{a}{\kappa}) & = & \pm 2
 \wrf \ ,  \nonumber
\end{array}
\ende
with $\sin^2 \chi = \ds (1-\kappa^2)/(1-a^2)$.
For type $\tO$ $(\kappa^2 < a^2)$ they are given by
\bege
\begin{array}{lllrl}
 \label{eq:IWII}
 \Idrad & = & \frac{\sqrt{2E}}{\pi}\;\left( \sin \psi  
         + \frac{a^2-\kappa^2}{a}\;\Fell(\psi,\frac{\kappa}{a}) 
         - a\; \Eell(\psi,\frac{\kappa}{a}) \right) & = 
         & \frac{1}{2}\Ifrad \ , \nonumber \\
 \Idang & = & \frac{\sqrt{2E}}{\pi}\; \left( a\;\Eell(\frac{\kappa}{a})
           - \frac{a^2-\kappa^2}{a}\;\Kell(\frac{\kappa}{a}) \right)
           & = & \frac{1}{2}\Ifang \ , \\
 \wrd & = & \Fell(\psi,\frac{\kappa}{a})/\Kell(\frac{\kappa}{a}) & = & \wrf\ ,\nonumber 
\end{array}
\ende
with $\sin^2 \psi = \ds (1-a^2)/(1-\kappa^2)$. $\Kell(k)$ is the
complete elliptic integral of the first kind, $\Fell(\mu,k)$ and 
$\Eell(\mu,k)$ are incomplete elliptic integrals of first and second kind. 

\begin{Filesave}{bilder}
\def\figesurface{%
    Energy surfaces (a)
    for the values $E=0.5,1,1.5,2,2.5,3$ and the winding number (b) of the 
    desymmetrized elliptic billiard with parameter $a=0.7$. The
    line $\kappa^2 = a^2$ ($\Idrad = \frac{1-a}{a}\Idang$) represents 
    the separatrix motion and the
    unstable oscillation along the $x$-axis. The line
    $\kappa^2 = 0$ ($\Idang = 0$) marks the
    stable oscillation along the $y$-axis, and $\kappa^2 = 1$ ($\Idrad = 0$) 
    corresponds to the sliding motion along the
    billiard boundary. The dotted lines represent the real complex tori of
    type $\tOc$.} 
\def\FIGesurface{\centerline{\psfig{figure=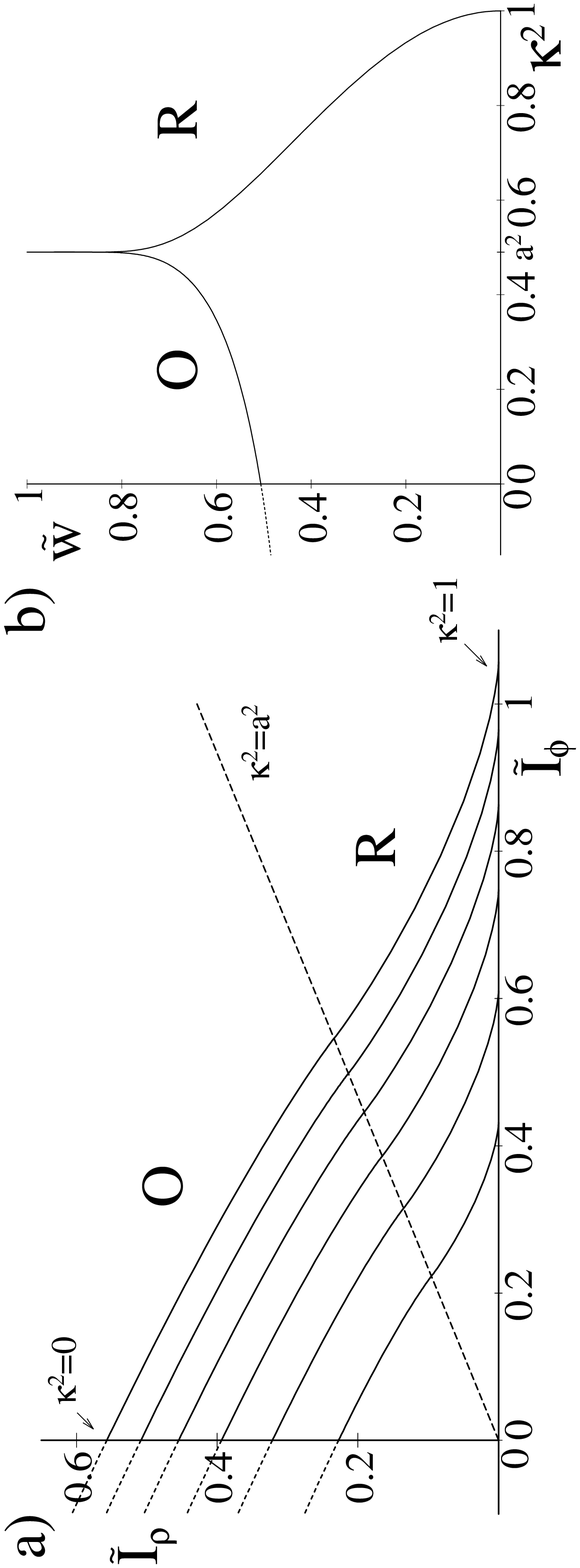,angle=-90,width=14.0cm}}}
\FIGo{fig:esurface}{\figesurface}{\FIGesurface}
\end{Filesave}
Figure~\ref{fig:esurface}a shows the energy surfaces $H(\BId) = E$ in
action space. All these lines have the same shape,
because the actions scale with $\sqrt{E}$. We denote the curvature of
the energy surfaces by $\curv$ and its sign by $\beta$. 
This means $\beta = +1$ for the patch being concave away from the origin
(type $\tR$) and $\beta = -1$ for the convex patch (types $\tO$).
In the limit $\kappa^2 \to -\infty$ the curvature tends to zero. 
The energy independent winding number $\wrd$ is shown in
\fig~\ref{fig:esurface}b. 
The cusp at $\kappa^2 = a^2$ has the limiting value $\wrd = 1$. 
The winding number of the stable isolated orbit $\os$ with 
$\kappa^2 = 0$ is ${2}\arccos{(a)}/\pi$.

Beside the three special periodic orbits $\os$, $\ou$ and $\ob$ there are 
families of periodic orbits on
resonant tori. A $\Brtopo$-resonant torus is determined by its
rational winding number $\wrf(\kappa) =
\rtopo_{\ang}/\rtopo_{\rad}$ or equivalently by the frequency vector $\Bomf$
being
proportional to $\Brtopo = (\rtopo_{\rad},\rtopo_{\ang})$, where
$\rtopo_{\rad}$ and $\rtopo_{\ang}$ are relatively prime integers.
We denote the action vector of a $\Brtopo$-resonant torus by $\BIf^\rtopo$. 
The action of a prime non-isolated orbit is given by
$\action(\BIf^\rtopo) = 2\pi \Brtopo\BIf^\rtopo$.
Figure~\ref{fig:elliorb} shows prime
orbits of $(\rtopo_{\rad},\rtopo_{\ang})$-resonant tori for type $\tR$
and $\tO$ for the full elliptic billiard. According to 
(\ref{eq:IWI}), the winding number for type $\tR$ is reduced
by a factor $2$ if compared with \fig~\ref{fig:esurface}b.
The orbits are chosen in such a way that they are always symmetric to the
$y$-axis and, if possible, also to the $x$-axis.
For type $\tR$ the integer $\rtopo_\rad$ counts the number of
reflections at the boundary and $\rtopo_\ang$ the rotations about the
origin. 
For type $\tO$ the integer $\rtopo_\rad$
gives half the number of reflections, and $\rtopo_\ang$ is half the
number of times that the orbit touches the caustics $\cos \ang = \pm \kappa/a$.
\begin{Filesave}{bilder}
\def\figelliorb{%
    Periodic orbit representatives
    of low $(\rtopo_{\rad},\rtopo_{\ang})$-resonant tori for $a=0.7$, 
    shown together with the corresponding caustic. For type $\tR$ we have
    $\rtopo_\ang / \rtopo_\rad = w = \tilde w/2 < 1/2$, while
    $w\pi \ge 2 \arccos a$ for type $\tO$.}
\def\FIGelliorb{%
 \centerline{\raisebox{5cm}{R}\psfig{figure=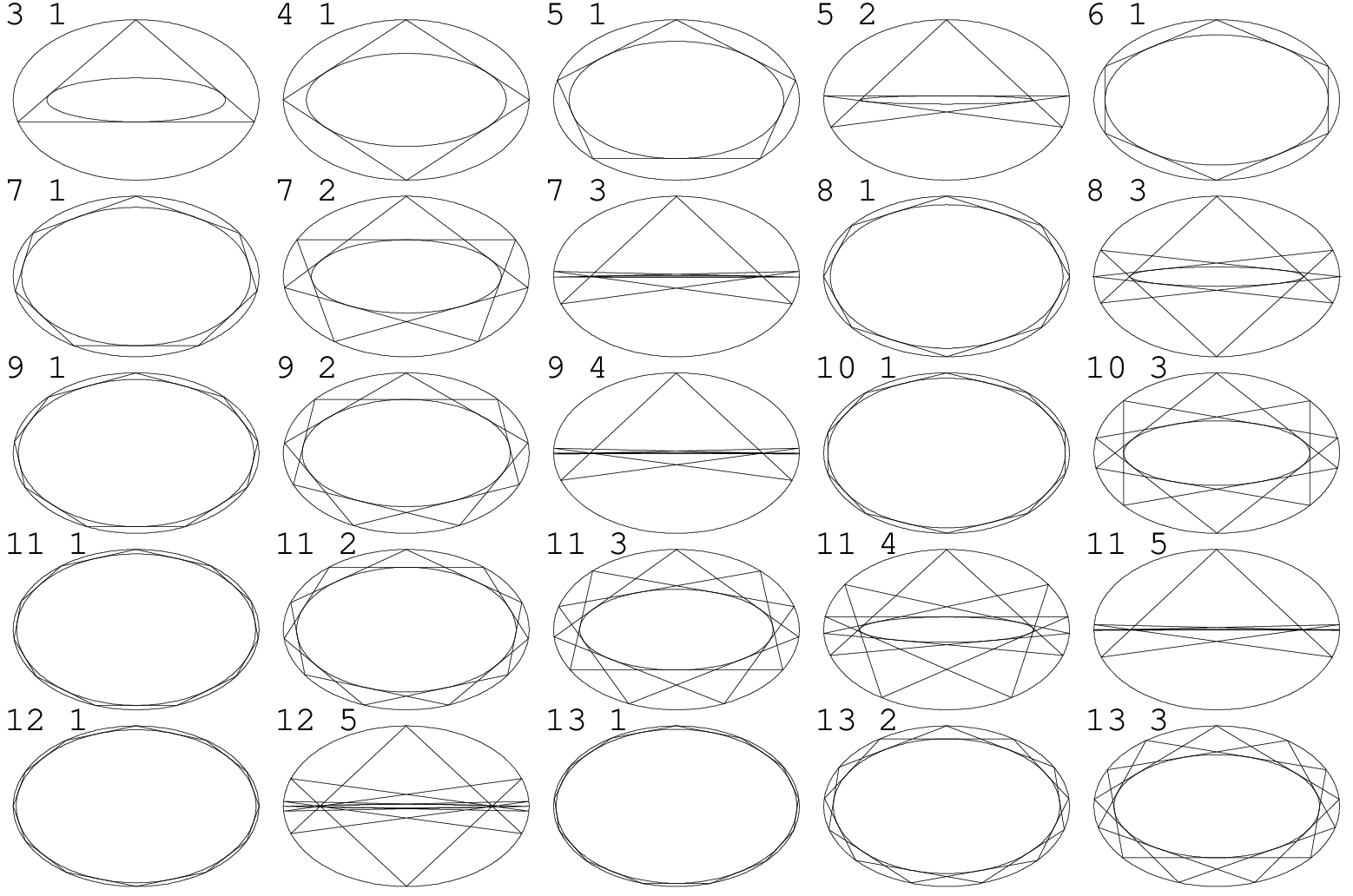,width=7.5cm,angle=0} 
        \raisebox{5cm}{O}\psfig{figure=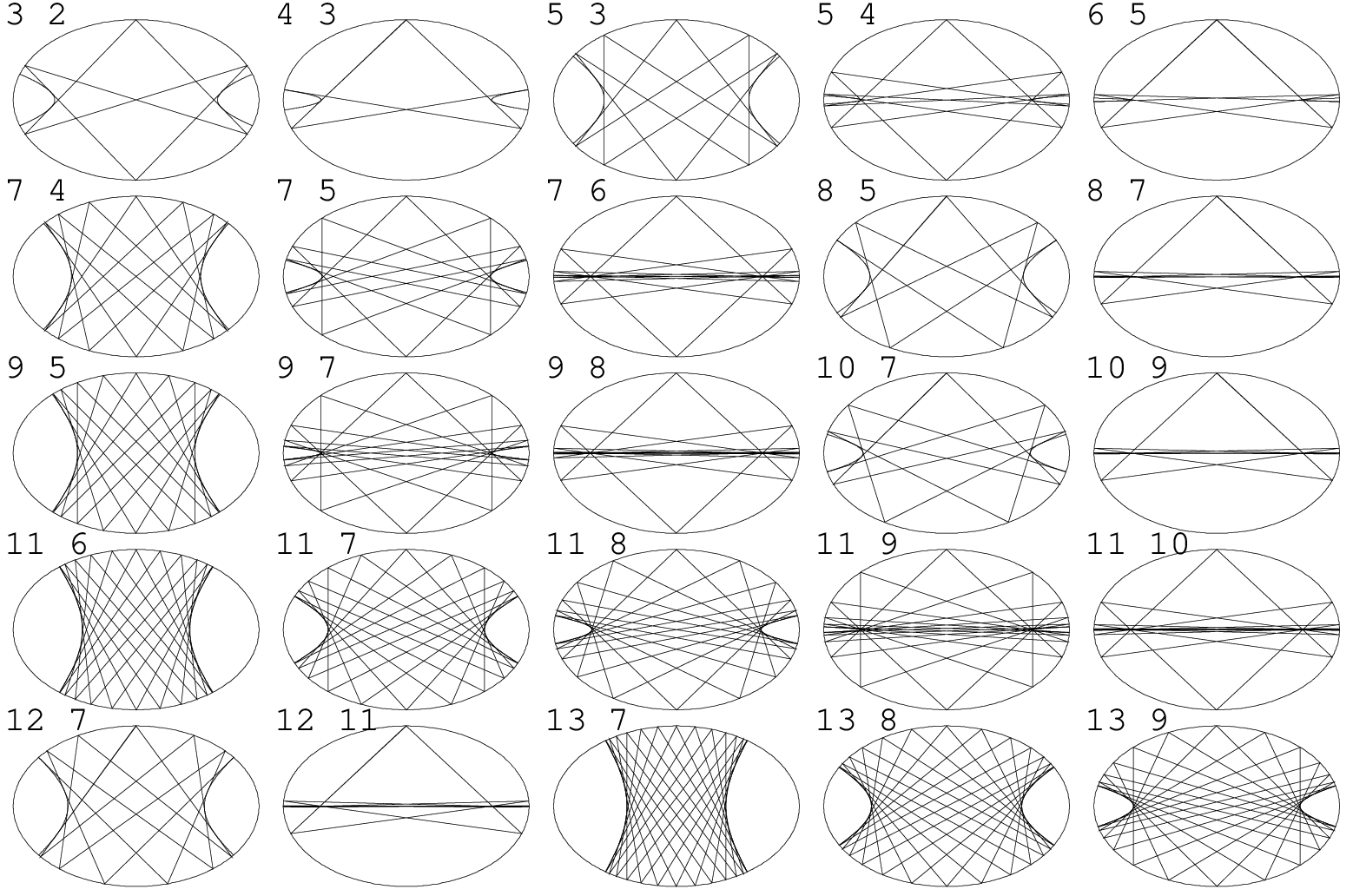,width=7.5cm,angle=0}} }
\FIGo{fig:elliorb}{\figelliorb}{\FIGelliorb}
\end{Filesave}

In the course of this paper it will be important to distinguish 
different kinds of non-real orbits which have to be taken into account
in the uniformization. Since we are dealing with a separable system, 
it is sufficient to classify non-real orbits for one degree of freedom
systems. We define and distinguish non-real orbits via the way a 
(generalized) action integral can be assigned to them.
Allowing complex values for the momentum $p$ and the position $q$,
we are dealing with $2\pi I = \oint p(q)\,dq$, where $p(q)^2=2(E-V(q))$, 
and the integral is taken between (possibly complex) turning points, 
i.e.\ the (possibly complex) zeroes of $E-V(q)$.
A real orbit has an action integral connecting real turning points
on a path (in the complex $q$-plane) for which $p$ is real, i.e.\ just
along the real $q$-axis. Of course the value of this integral is the
same for any path in the complex plane encircling the two turning 
points. The special integration path chosen can be thought of as
the orbit of the particle.
A ``tunneling orbit'' also connects real turning points, however, the energy
$E$ is smaller than $V(q)$, such that $p$ is purely imaginary along
the real $q$-axis, hence giving a purely imaginary action.
There are two more non-real orbits connecting complex turning points.
The ``scattering orbit'' is related to quantum mechanical 
scattering resonances forming above a potential maximum. 
Its path of integration 
is taken to be an anti-Stokes line. 
It is the line of $q$ in the complex plane for which 
the integral $\int_{q_1}^{q} p(q')\,dq'$ is purely imaginary, 
where $q_1$ is a turning point.
In the case of an even potential with maximum at 0, as it is the case 
for the elliptic billiard, the anti-Stokes line is just the imaginary $q$-axis,
and the momentum is real.
By construction the action of a scattering orbit is purely imaginary.
In the following, tunneling and scattering orbits will be treated similarly,
we refer to both of them as ``imaginary complex orbits''.
They are related to the ``ghost orbits'' described in \cite{KHD93}
and the barrier penetration integral of Miller \cite{Miller68} gives 
$i$ times their action ($i=\sqrt{-1}$).
\NEU
The final type of non-real orbit is related to orbits below a
potential minimum. They are obtained by integrating along
a Stokes line (on which the integral is always real) 
between complex turning points, hence their action will
be real.
In the case of a symmetric potential with minimum at 0 the Stokes
line is just the imaginary $q$-axis, and the momentum is purely imaginary.
We call this kind of non-real orbit ``real complex orbit'', because
its action is real. These orbits are the ``complex orbits'' 
described in \cite{BerryTabor76}.

In separable two degrees of freedom systems 
all the combinations of the four possibilities might occur
in principle.
Actually there is some arbitrariness in how to combine
the possible real and non-real motions from the separated
degrees of freedom. 
However, in the present case it seems to be natural to form pairs from 
tunneling and scattering orbits, giving rise to 
imaginary complex tori. These tori are denoted 
by $\tTO$ and $\tTR$ in \fig~\ref{fig:effpot}.
\NEU
Continuing the $\tTO$ torus below the potential minimum 
gives rise to the $\tTOc$ torus.
The real complex orbit is the natural continuation of an elliptic
orbit, which disappears at the minimum of the potential.
The torus denoted by $\tOc$ in \fig~\ref{fig:effpot} is therefore
real in its $\rad$-part and real complex in its $\ang$-part.
These real complex tori lead to an energy surface that extends beyond
the positive quadrant, as discussed by Berry and Tabor.
There are no 
continuations of the $\tR$ and $\tTR$ tori, since for the $\rad$-motion 
the hard billiard wall does not create any complex zeroes for low energy. 
Therefore the energy surface cannot be continued at $\kappa^2 = 1$.

For type $\tOc$ $(\kappa^2 < 0)$ the actions and winding number 
of the desymmetrized billiard are given by
\bege
\begin{array}{lllrl}
 \label{eq:IWIIc}
 \Idrad & = & \frac{\sqrt{2E}}{\pi}\;\left( \sqrt{1-\kappa^2} \sin \tau  
         + \sqrt{a^2-\kappa^2} \;(\Fell(\tau,k) 
         - \; \Eell(\tau,k)) \right) & =
         & \frac{1}{2}\Ifrad \ , \nonumber \\
 \Idang & = & \frac{\sqrt{2E}}{\pi}\; \left( \sqrt{a^2-\kappa^2} \;(\Eell(k)
           - \;\Kell(k)) \right)
           & = & \frac{1}{2}\Ifang \ , \\
 \wrd & = & \Fell(\tau,k)/\Kell(k) & = & \wrf\ , \nonumber 
\end{array}
\ende
with $\sin^2 \tau = \ds 1-a^2$ and $k^2 = -\kappa^2/({a^2-\kappa^2})$.
The formulas~(\ref{eq:IWIIc}) equal the
formulas~(\ref{eq:IWII}) for imaginary $\kappa$. It is important
to notice that although the actions are real, the tori are
classically not allowed, because both position and momentum are complex.
These formulas define the continuation of the energy surface shown
in \fig~\ref{fig:esurface}a.
The winding number of the tori of type $\tOc$ 
decreases very slowly to zero for $\kappa^2 \to -\infty$.
Resonant real complex tori can also be defined, because their real
frequencies can fulfill a resonance conditions. 
The action $S$ and the period of a non-isolated orbit on a resonant
complex torus are always positive, although the angular component
$\Ifang$ is negative.

\section{Exact quantum mechanics}
\label{sec:qmechanic}

The Schr\"odinger equation including the boundary condition separates 
in elliptic coordinates.
Using the angle-radius parametrization $(\ang,\rad)$ gives the
standard form of the Mathieu equation
\nootnote{In AS the standard form reads $z''+(a-2q\cos 2t)z=0$, which
gives the relation 
$\SeKo - \ec/2(1+\cos 2t) = a - 2q\cos 2t$, such that
$\SeKo - \ec/2 = a$ and $\ec/2 = 2 q$. The notation with $\SeKo$ and
$c$ is motivated from the equivalence with the spheroidal wave equation
as given in NR.
}
\begin{eqnarray}
\label{eqn:matang}
        \DD\ang \Fang + (\SeKo - \ec\cos^2 \ang) \Fang & = & 0 \\
\label{eqn:matrad}
        \DD\rad \Frad - (\SeKo - \ec\cosh^2\rad) \Frad & = & 0,
\end{eqnarray}
which follows directly from (\ref{eq:separation2},\ref{eq:separation1}).
The relation to the physical parameters is
\begin{eqnarray}
\ec             & = & 2 E a^2 / \hbar^2                   \\
\SeKo           & = & 2 E \kappa^2 / \hbar^2 . 
\end{eqnarray}

For Dirichlet boundary conditions the eigenfunction 
$\Psi(\ang,\rad) = \Fang(\ang)\Frad(\rad)$ 
must be zero on the billiard boundary, which gives 
$\Frad(\rad_{\text{max}}) = 0$. 
The solution in the angular variable must be periodic with period $2\pi$,
to give a physical solution. 
Floquet theory guarantees the existence of solutions with period a multiple
of $\pi$, because (\ref{eqn:matang}) is linear with $\pi$-periodic 
coefficients. 
In the classical terminology the special values of the 
parameters, for which $\pi$ or $2\pi$ periodic solutions of 
(\ref{eqn:matang}) exist are called 
characteristic values. For even solutions, i.e.\ $\Fang(0)=0$,  they are
denoted by $a = \SeKo - \ec/2$, and similarly by $b$ for odd solutions.
Since the Mathieu equation is an equation of Sturm-Liouville type, 
these eigenvalues are all real, ordered as $a_0 < b_1 < a_1 \cdots$ for
fixed $\ec$ and the corresponding eigenfunctions 
have $i$ zeroes in the interval $\ang \in [0,\pi)$. 
Solutions with even $i$ have period $\pi$, 
those with odd $i$ have period $2\pi$.

If $\Fang(\phi)$ is even, 
then $\Psi$ is symmetric with respect to the $x$-axis:
$\Psi(x,y) = \Psi(x,-y)$, see (\ref{eq:radang}).
We denote this symmetry by $\pi_y = +1$, 
respectively by $\pi_y = -1$ for odd $\Fang$ with $\Psi(x,y) = -\Psi(x,-y)$.
Similarly $\Psi$ can be even or odd with respect to $x$, which is denoted
by $\pi_x = \pm1$. 
The four possible parity combinations are given in 
\tab~\ref{tab:bc}, together with the sign of the value of 
$\Fang(\pi/2 k)$, $k=0,1,2,3$. In order to obtain a smooth wave function $\Psi$
in the part of the $x$-axis connecting the foci,
the radial solution must satisfy $\Frad(0) = 0$ if $\Fang$ is odd.
If both parities are the same, the angular solution has period $\pi$.
In the last column we indicate which coordinate axis becomes a
nodal line for the wave function of the corresponding parity.
The given signs are only defined up to a global factor.

\begin{table}[!h]
\tabstart
\begin{center}
\begin{tabular}{|c|c|c|c|c|c|c|c|c|}\hline
  $\pi_x$ & $\pi_y$ & $\psixi(0)$ & $\psieta(0)$ &
  $\psieta(\frac{\pi}{2})$ & $\psieta(\pi)$ &
  $\psieta(\frac{3\pi}{2})$ & period $\psieta(\alteta)$ & node \\ \hline
  $+$ & $+$ & + &$\pm$& + &$\pm$& + & $\pi$ & \\ 
  $-$ & $+$ & + & + & 0 &$-$& 0 & $2\pi$ & $x=0$ \\
  $+$ & $-$ & 0 & 0 & + & 0 &$-$& $2\pi$ & $y=0$ \\
  $-$ & $-$ & 0 & 0 & 0 & 0 & 0 & $\pi$  & $x=0,\,y=0$ \\ \hline
\end{tabular}
\caption[]{\label{tab:bc}%
\capsty Symmetry properties of the wave function depending on the
four parities. The four states which have the same quantum
number are listed in the order of increasing energy.
}
\end{center}
\tabend
\end{table}

\begin{Filesave}{bilder}
\def\FIGwaves{%
\centerline{ 
        \mbox{$(0,0)_{++}$}
        \psfig{figure=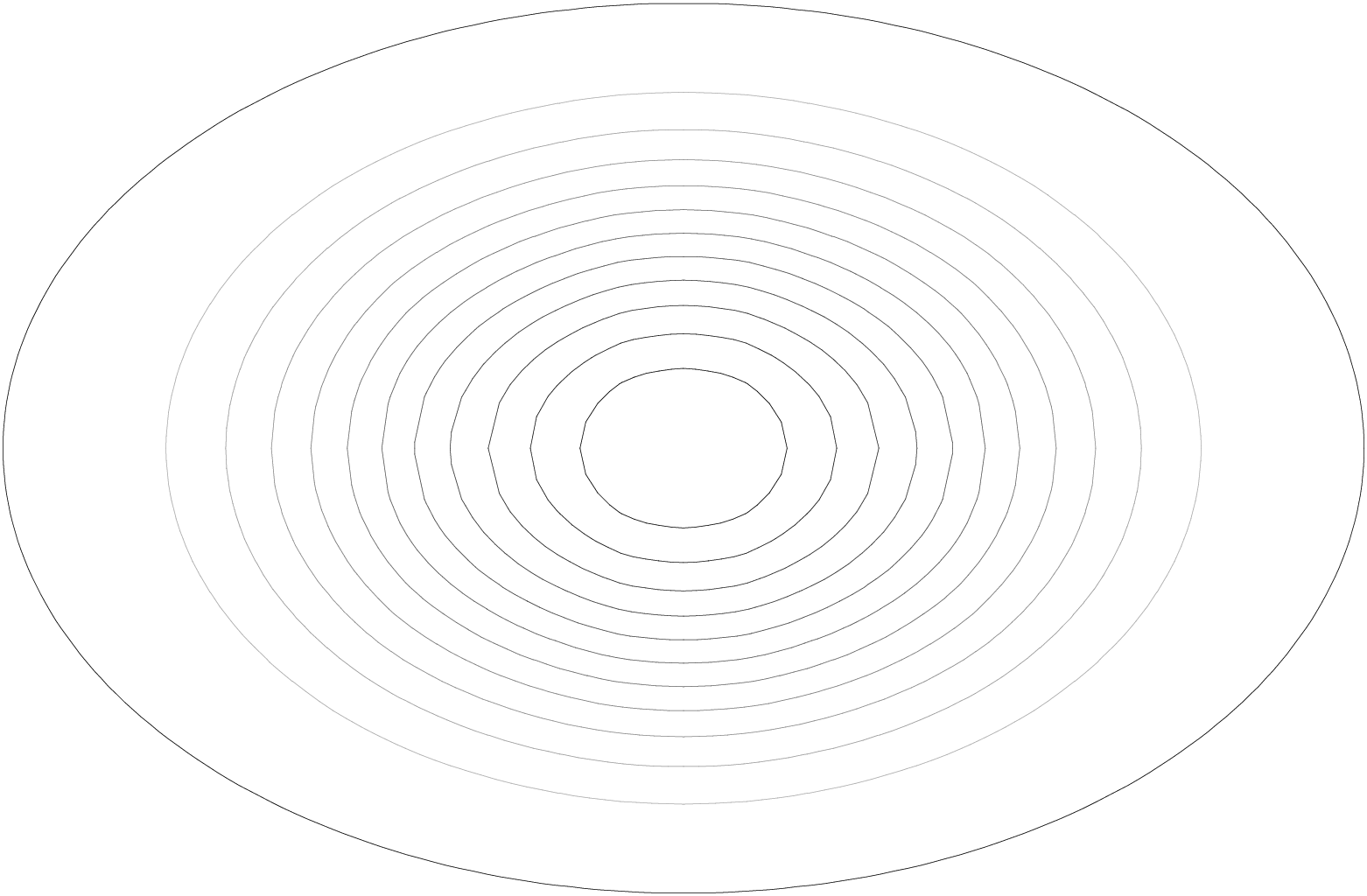,width=4.0cm,angle=-90}
        \hspace*{1cm}
        \mbox{$(0,0)_{-+}$}
        \psfig{figure=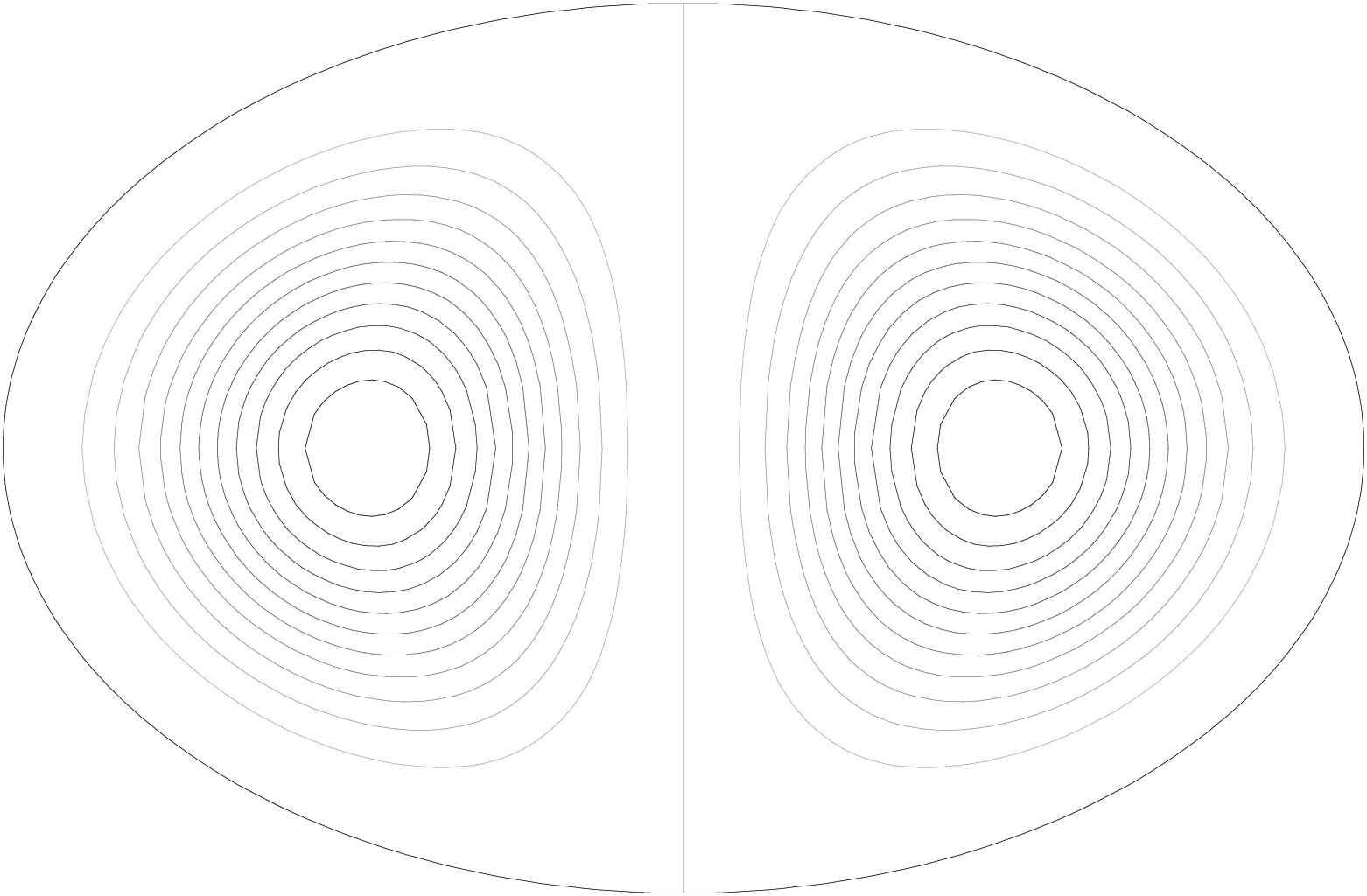,width=4.0cm,angle=-90}
} 
  \centerline{
        \mbox{$(0,0)_{+-}$}
        \psfig{figure=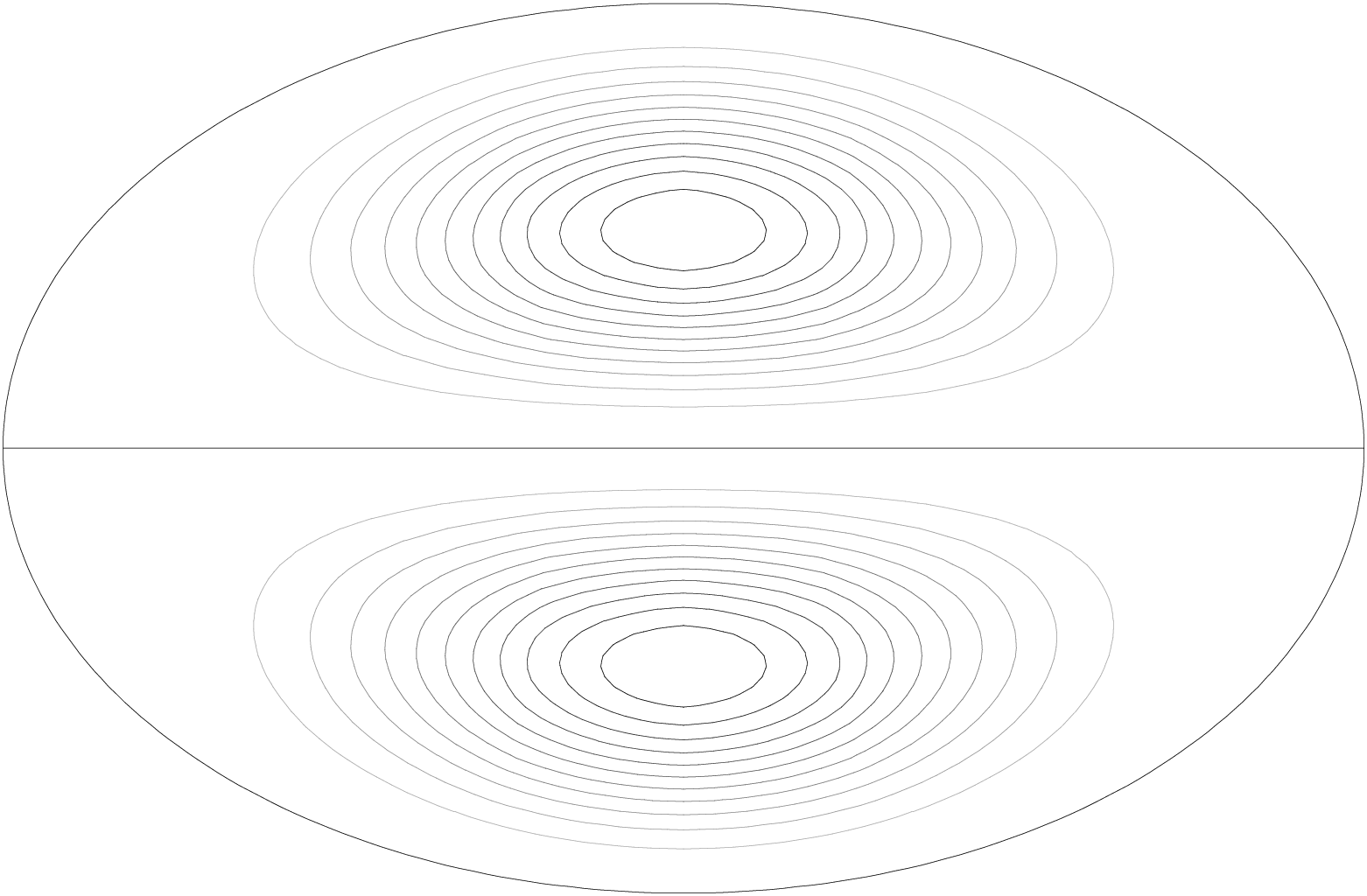,width=4.0cm,angle=-90}
        \hspace*{1cm}
        \mbox{$(0,0)_{--}$}
        \psfig{figure=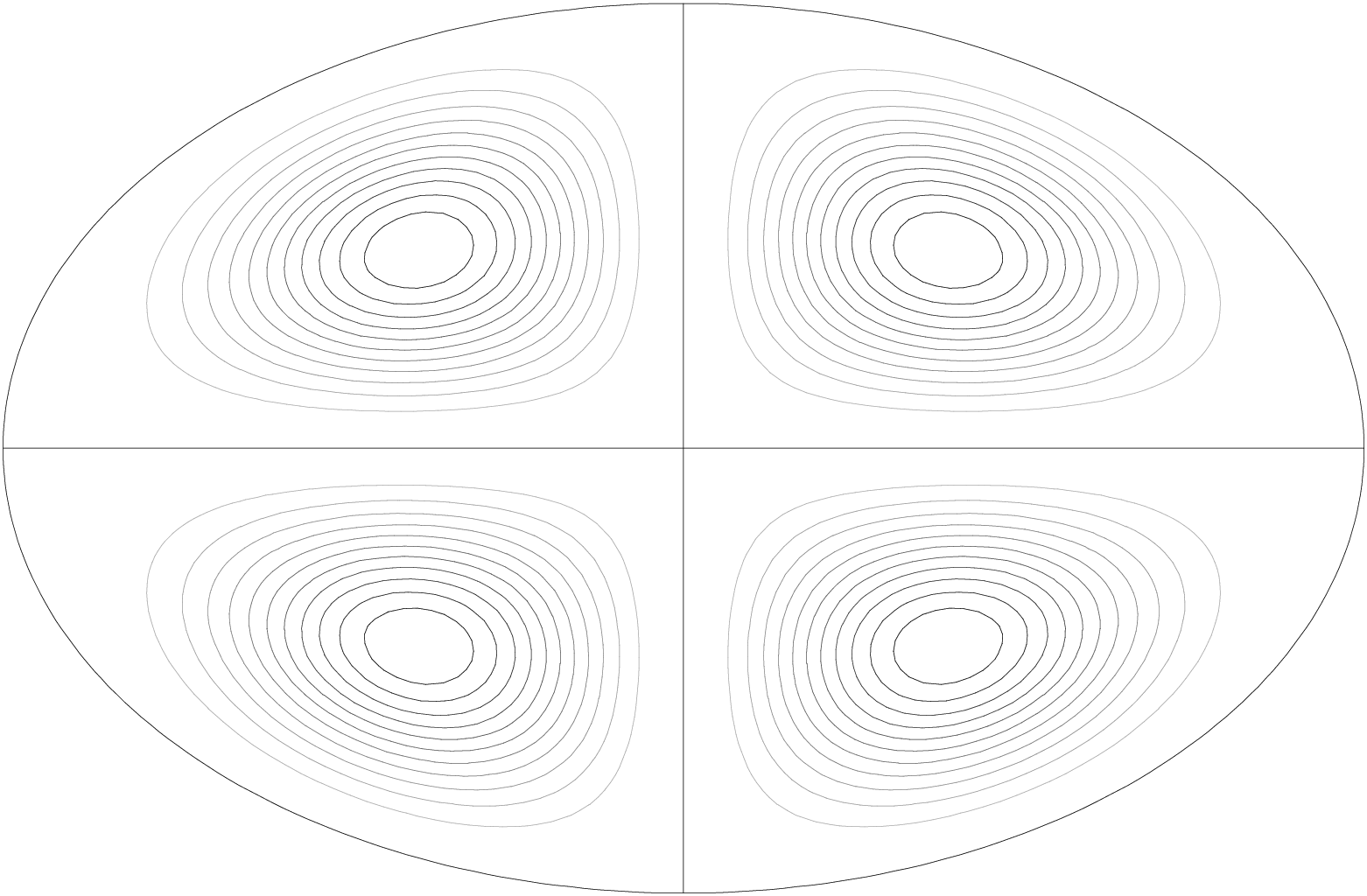,width=4.0cm,angle=-90}
} }
\def\figwaves{%
Contourplots of the probability density of the wave functions for the four 
different parity states $(\pi_x,\pi_y)$ for quantum numbers 
$(\nrad,\nang)=(0,0)$ and with $a=0.7$. 
The contourlines are equally spaced from 0 to the maximum probability 
density of the wave function.
}
\FIGo{fig:waves}{\figwaves}{\FIGwaves}
\end{Filesave}

We define the radial quantum number $\nrad$ 
as the number of zeroes of $\Frad(\rad)$ in the range 
$\rad \in (0,\rho_{\text{max}})$,
i.e.\ not counting the zeroes at the boundaries. 
Similarly the angular quantum number $\nang$ 
gives the number of zeroes of $\Fang(\ang)$ in the range 
$\phi \in (0,\pi/2)$, again not counting possible zeroes at the boundaries. 
This choice of quantum numbers is consistent with the \EBK\
quantization (\ref{eq:EBK}) for the symmetry reduced system.
Therefore it does not directly give the number of nodes of the wave
functions in the full system: depending on the
parity there are 0, 1, 2 or 3 additional zeroes in the range $[0,\pi]$
for the states with parity $++$, $-+$, $+-$, and $--$, respectively, 
such that the total number of angular nodes is $4 \nang + 2 - \pi_x - \pi_y$.
The energy eigenvalues for the four states with the same quantum 
numbers are in the same order.
In general we denote a state by $(r,l)_{\pi_x \pi_y}$.
Note that the four different parity combinations correspond to the
description of the symmetry reduced quarter billiard, where on the
coordinate axes Dirichlet (parity $-1$) or Neumann 
(parity $+1$) boundary conditions are required.
To illustrate the symmetries, the probability density of the
$(0,0)$ state for each parity is shown in \fig~\ref{fig:waves}.

The transformation $\cos\ang = \eta$ and $\cosh\rad = \xi$
gives the algebraic form of the Mathieu equation
\begin{eqnarray} 
\label{eqn:matalgang}
(1-\eta^2)\DD\eta\Feta - \eta\D\eta\Feta + (\SeKo - \ec\eta^2)\Feta & =
& 0  \\
\label{eqn:matalgrad}
(1-\xi^2) \DD\xi \Fxi  - \xi \D\xi \Fxi  + (\SeKo - \ec\xi^2) \Fxi & = & 0\ ,
\end{eqnarray}
from which it is obvious that the angular and the radial equation are actually
the same, only evaluated on different ranges of the independent variable,
$\eta\in[-1,1]$, $\xi\in[1,1/a]$.
This equation, although it has regular
singular points at $\xi,\eta = \pm 1$, is better suited for the numerical
solution of the eigenvalue problem.
The requirement for the solution to be smooth at these points replaces
the periodic boundary condition.

The Mathieu equation is a special case of the spheroidal wave equation, 
\begin{equation} \label{eqn:spheroidal}
(1-\sxi^2)\DD\sxi\sFxi - 2(m+1)\sxi\D\sxi\sFxi + 
        (\SeKo - 1/4 - m(m+1)-\ec\sxi^2)\sFxi=0 \ ,
\end{equation}
which appears in the case of the billiard inside the rotational symmetric 
ellipsoid; $m$ is the quantum number of the angular momentum of 
rotation around the axis of symmetry. 
With $m=-1/2$ in (\ref{eqn:spheroidal}) we reobtain (\ref{eqn:matalgang})
and (\ref{eqn:matalgrad}).
We will see that $m=+1/2$ also produces solutions of the Mathieu equation.

In order to obtain solutions with $\pi_y=-1$, which must be zero at 
$\phi = 0,\pi$, we must construct solutions which are zero at the 
regular singular points. 
It is necessary to factor out this behavior by the ansatz
$g(\sxi) = \sqrt{1-\sxi^2} h_{1/2}(\sxi)$, 
$f(\sxi) = \sqrt{\sxi^2-1} h_{1/2}(\sxi)$, 
which transforms (\ref{eqn:matalgang}) 
and (\ref{eqn:matalgrad}) into the equation
\begin{equation}
        (1-\sxi^2) \DD\sxi h_{1/2} - 3 \sxi\D\sxi h_{1/2} + (\SeKo - 1 - \ec \sxi^2) h_{1/2} = 0,
\end{equation}
which again is a special case of the spheroidal wave equation
(\ref{eqn:spheroidal}), in this case corresponding to $m=+1/2$, however. 

In \fig~\ref{fig:xietagraph} four solutions of (\ref{eqn:spheroidal}) 
are shown. The radial and the angular part smoothly join at
the regular singular point at $x=1$.
We conclude that the spectrum of the 2-dimensional billiard is
obtained from the spectrum of the 3-dimensional rotational symmetric
billiard if the angular quantum number $m$ in the spheroidal wave
function is set to a half integer ``spin'' number $\pm1/2$ 
instead of to an integer number as in the 3-dimensional problem.

\begin{Filesave}{bilder}
\def\FIGxietagraph{%
  \centerline{\psfig{figure=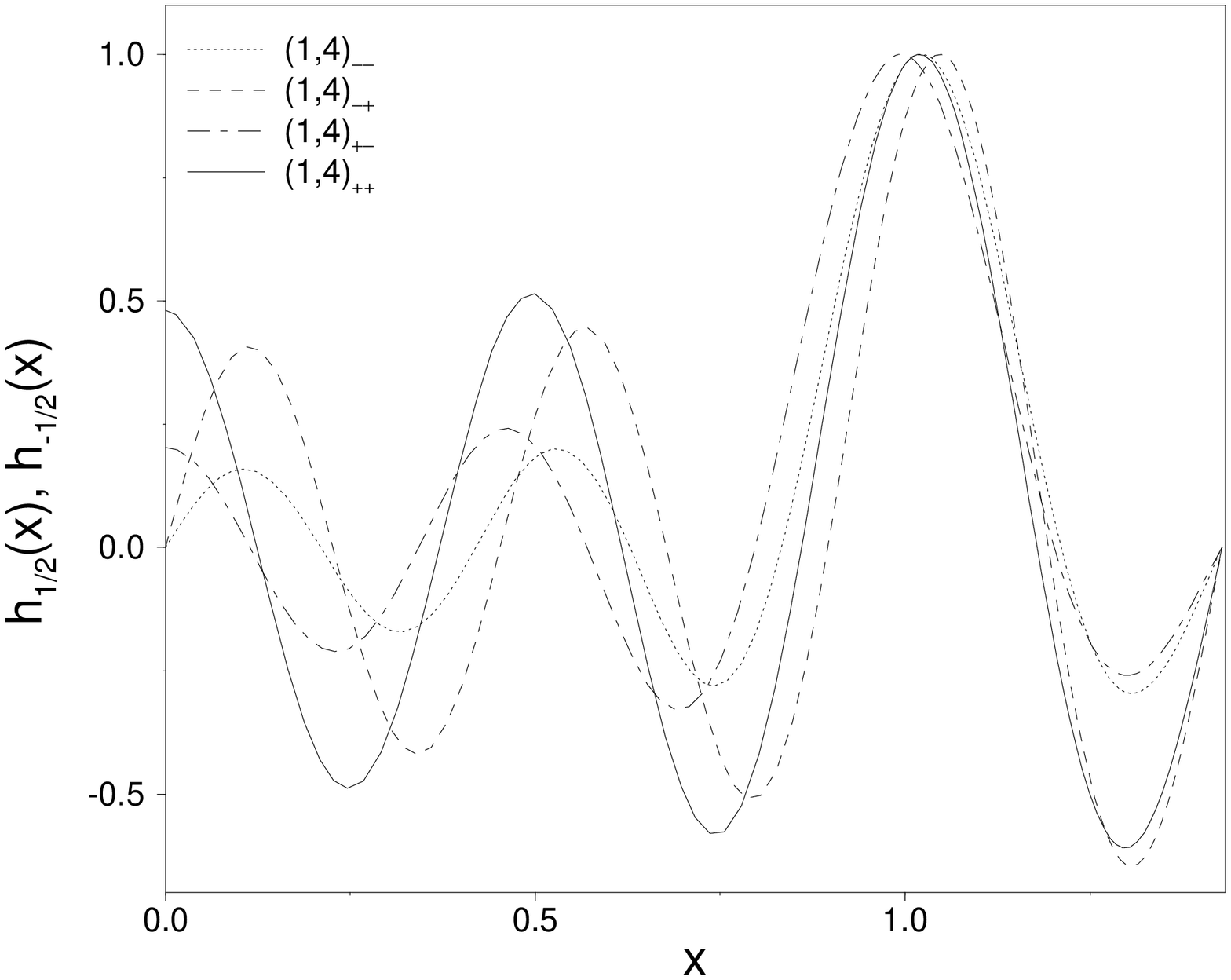,width=9.0cm,angle=-90}}
}
\def\figxietagraph{%
The solutions of (\ref{eqn:spheroidal}) for the state $\nrad=1$, $\nang=4$
with the four possible parities, i.e.\ even or odd at $x=0$ ($\pi_x$)
and $m=-1/2$ for $\pi_y=1$ or $m=1/2$ for $\pi_y=-1$; $a=0.7$.}
\FIGo{fig:xietagraph}{\figxietagraph}{\FIGxietagraph}
\end{Filesave}

In the standard theory of the Mathieu equation $c$ is fixed and the
eigenvalue $\SeKo$ is determined. In the billiard problem we have to
simultaneously satisfy also the boundary condition for the radial 
equation. Since both separation constants $\SeKo$ and $\ec$ appear
in each equation, although the variables are separated, the separation
constants are not separated, see e.g.~\cite{Morse53}, which requires
a nonstandard approach to the numerical solution of this 
Sturm-Liouville eigenvalue problem. 
For the angular equation (\ref{eqn:matalgang}) 
we have boundary conditions at
$\eta=-1$ and at $\eta=0$, corresponding to $\ang=\pi$ and $\ang=3\pi/2$,
for the radial equation at $\xi = 1$ and $\xi = 1/a$, corresponding to
$\rad = 0$ and to the billiard boundary.
Introducing a new independent variable by $\xi = 1/a + \zeta (1/a-1)$,
the ranges of $\zeta$ and $\eta$ coincide, and we can use a standard
shooting method as, e.g., described in \cite{Press88}, to solve both coupled
eigenvalue problems simultaneously. Even though the final solution is
smooth, as shown in \fig~\ref{fig:xietagraph}, it is not possible
to numerically integrate through the singularity. Instead the
integration always starts an epsilon away from the singularity with
an analytically calculated regular initial velocity \cite{Press88}.
In order to find a specific state it is necessary to have a fairly
good initial guess for the eigenvalues, otherwise the shooting
method might converge to another state. We calculate the initial
guess via the uniform \WKB\ approach described in the next section,
and have found that this always works.
The results of these computations are shown in \fig~\ref{fig:statesX},
and some eigenvalues are listed in \tab~\ref{tab:states}.

\begin{Filesave}{bilder}
\def\FIGstatesX{%
  \centerline{\psfig{figure=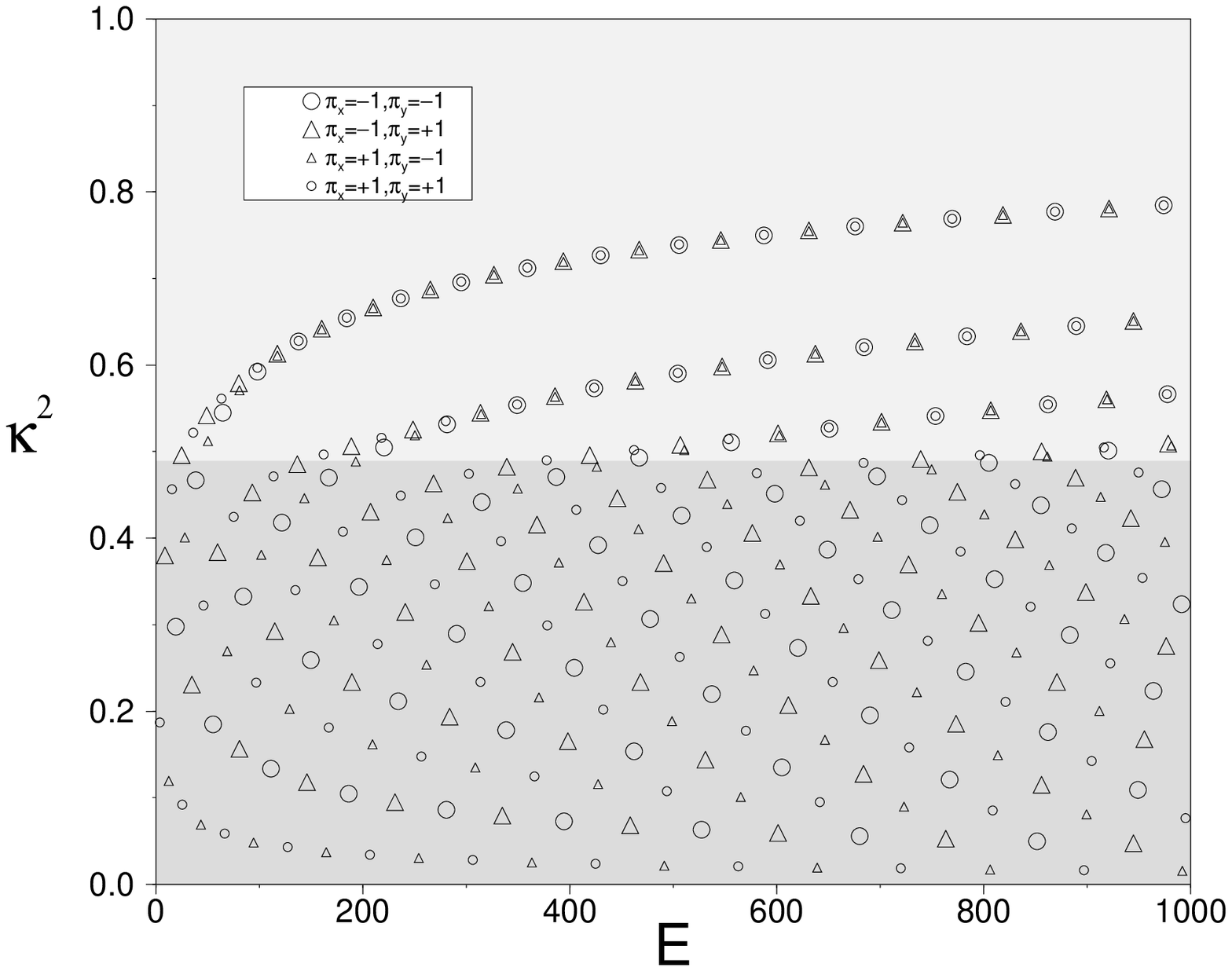,width=11.0cm}}
}
\def\figstatesX{%
The eigenvalues $(E,\kappa^2)$ of the eigenstates of the elliptic 
billiard with $a=0.7$.
The region of type $\tR$ states (rotational, $\kappa^2 > a^2$) 
is shown in light grey, 
type $\tO$ states (oscillating, $\kappa^2 < a^2$) in dark grey.
}
\FIGo{fig:statesX}{\figstatesX}{\FIGstatesX}
\end{Filesave}

\begin{Filesave}{bilder}
\def\FIGstatesI{%
  \centerline{\psfig{figure=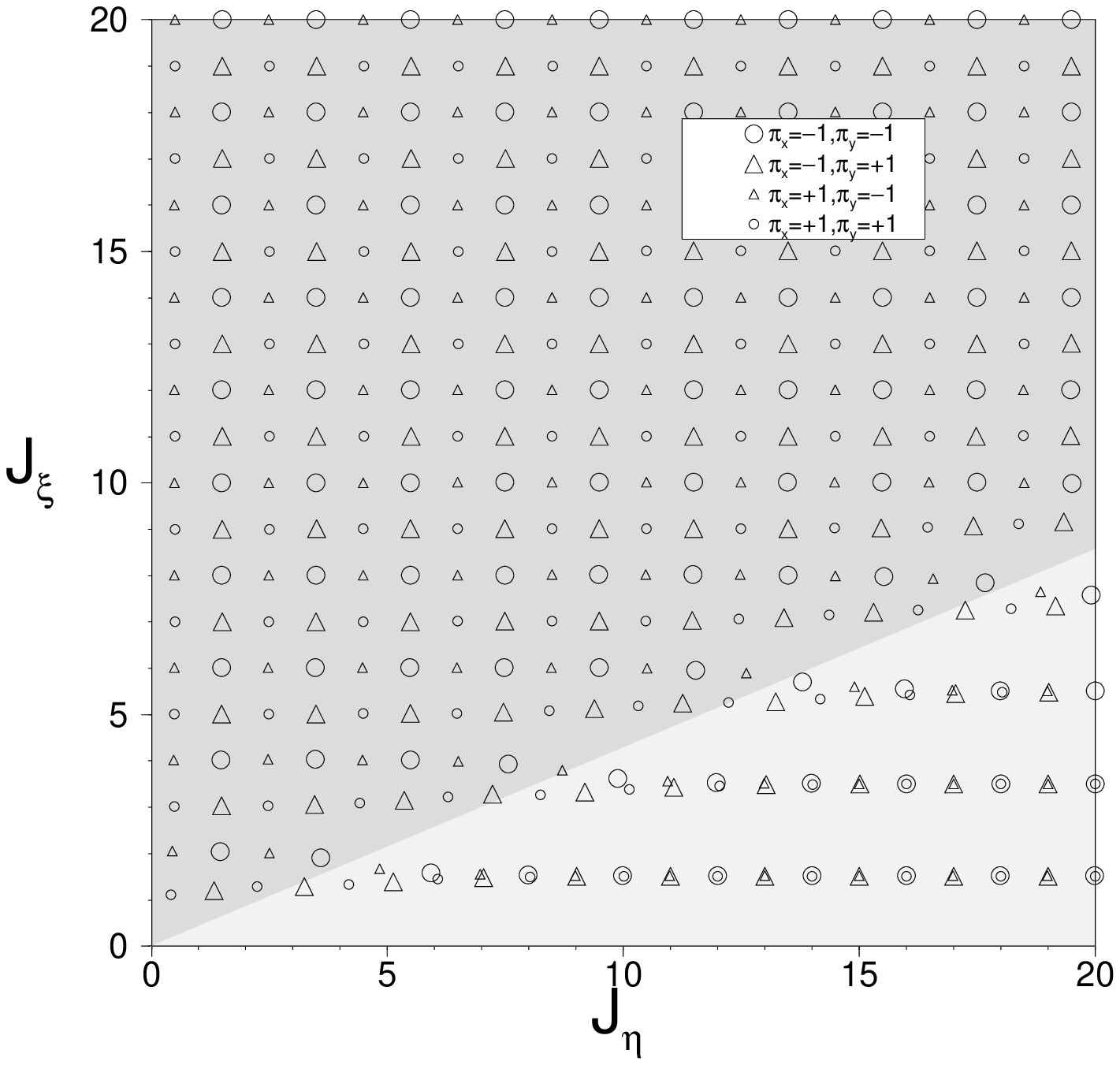,width=11.0cm}}
}
\def\figstatesI{%
The classical actions 
$\BJf = 2 \BId$ in units of $\hbar$ corresponding to the exact
quantum mechanical eigenstates in the elliptic billiard for $a=0.7$. 
Except for states close to the separatrix,
they are located on a lattice given by \EBK\ quantization. 
The structure near the separatrix can be explained by
uniform \WKB\ quantization.
Grey code as in \fig~\ref{fig:statesX}.
} 
\FIGo{fig:statesI}{\figstatesI}{\FIGstatesI}
\end{Filesave}

\begin{Filesave}{bilder}
\def\figMwaves{%
Contourplots of the probability densities of the 
wavefunctions $(\nrad,\nang)_{++}$ of the elliptic billiard
with $a=0.7$.
The vertical axis shows $r = 0,\ldots,5$ and the 
horizontal axis shows $l = 0,\ldots,5$.
The classical caustic corresponding to the second eigenvalue
is also shown.
}
\def\FIGMwaves{\centerline{\psfig{figure=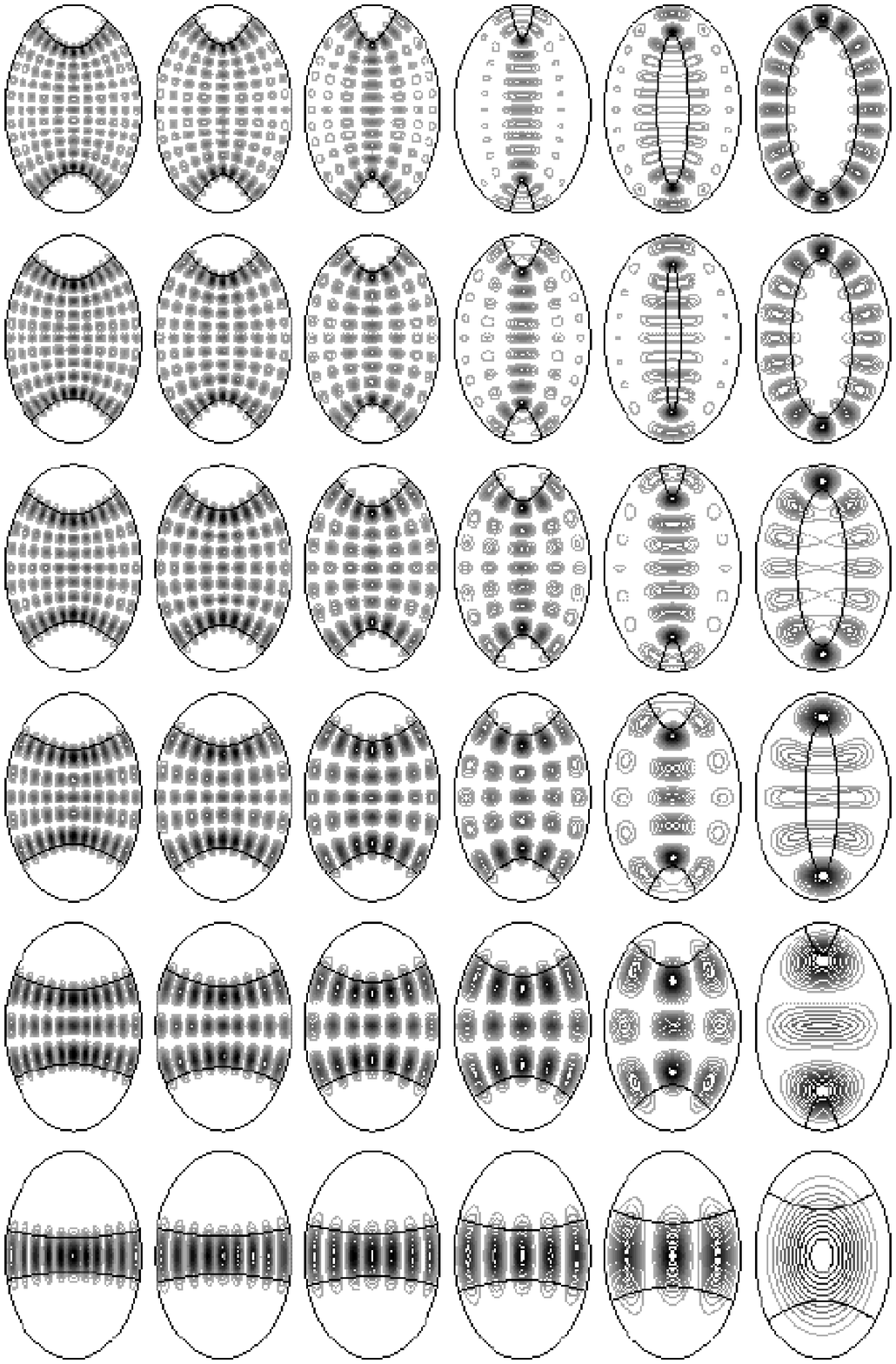,angle=-90,width=15.0cm}}}
\FIGo{fig:Mwaves}{\figMwaves}{\FIGMwaves}
\end{Filesave}

In \fig~\ref{fig:Mwaves} a few symmetric wavefunctions are illustrated.
The wave functions for the other parities are related to the $++$ states
show in \fig~\ref{fig:Mwaves} as illustrated in \fig~\ref{fig:waves}.
In the left column the localization around the stable
isolated periodic orbits is clearly visible. It becomes stronger when $\nrad$ 
is increased. 
In the bottom row the same happens for the orbit $\ob$. Both cases 
can be explained by considering the \WKB\ wavefunctions for the corresponding
states, which is exponentially small outside the classical caustics 
and nonzero inside. 
Although the \WKB\ wavefunction diverges at the 
classical caustic it is correct insofar as the quantum probability density
is relatively high close to (and inside) the classical caustic, 
as can be seen e.g.\ for the $(4,4)_{++}$ state.
The surprising fact, which can not be explained by the above reasoning, is the 
localization around the focus points, e.g.\ in the $(1,4)_{++}$ state.
This phenomenon occurs 
at the transition from the rotational states (classical caustic type $\tR$)
to oscillating states (classical caustic type $\tO$), i.e.\ close
to the classical separatrix.
Eigenstates with higher quantum numbers can show even stronger 
localization around the foci, because their second eigenvalue can be found 
closer to the critical value $\kappa=a=0.7$.
\NEU
Besides this very strong localization on the focus points these
states also show a ``scar'' \cite{Heller84}
of the unstable orbit, e.g.~see the $(2,5)_{++}$ state. Although the
probability density is low compared to the one in the focus points,
it is high compared to the region further away from the $x$-axis.
The behaviour outside the foci can again be explained by the 
\WKB\ wave function, which diverges for $y=0$ for the critical 
value $\kappa=0.7$.
We conclude that in the integrable elliptic billiard 
states localize around stable periodic orbits, 
and they also show ``scars'' along the unstable periodic orbit,
with an additional strong localization on the focus points of the ellipse.

The oscillating states of type $\tO$ are always non-degenerate.
The rotational states of type $\tR$ become more and more degenerate when the
distance from the separatrix is increased. The classical reason for this
increasing degeneracy is the fact that there are two tori for fixed constants 
of motion, connected via time reversal symmetry. In a simple \EBK\ 
quantization these states would be exactly degenerate.
Two type $\tR$ states with the same $\nrad$ and $\nang$ become 
(approximately) degenerate 
if they have period $2\pi$, i.e.\ if $\pi_x\pi_y = -1$. 
If, however, the period is $\pi$, $\pi_x\pi_y = 1$, two states with
the same $\nrad$ but $\nang$ differing by 1 become degenerate. 
This is consistent with the fact that the total number of nodes
on the circle $\ang \in [0,2\pi)$ is $4\nang + 2 - \pi_x - \pi_y$, 
which gives the same number for the degenerate states.
The shift from degenerate states of type $\tR$ to non-degenerate states of 
type $\tO$ can most clearly be seen if the exact quantum eigenvalues 
$E$, $\kappa^2$ are transformed into the classical action space
using equations (\ref{eq:IWI}) and (\ref{eq:IWII}), see \fig~\ref{fig:statesI}.
In order to represent the actions of the full system for both types of motion 
in one picture in such a way that on average one state occupies a box
of size $\hbar^2$, we use the doubled action of the symmetry reduced
billiard $\BJf = 2 \BId$.
\NEU
This amounts to doubling the action $I_\rad$ for type $\tR$ tori, 
which can be understood from two facts: 
firstly there are two classical tori with the same
action for type $\tR$ tori and secondly the corresponding quantum
mechanical states are degenerate in the \EBK\ approximation.
The type of movement that each parity state can perform 
in its ``semiclassical quantum cell'' will become clear in the next section.

\section{Uniform \WKB\ quantization}
\label{sec:semiclassic}

Consider a Hamiltonian $H=p^2/2+V(q)$ with a potential as e.g.\ 
given by the effective potentials of \Sec~\ref{sec:classic},
see \fig~\ref{fig:effpot}.
Let $V(q)<E$ for $q<q_1$ and
$q>q_2$ and $V(q)>E$ in the range $q_1<q<q_2$.
The \WKB\ solutions to the
left ($j=1$) and to the right ($j=2$) of the potential barrier are
\bege \label{eq:wavef}
\psi_j(q) =
\left(A_j^+\exp(iS_j(q)/\hbar)+A_j^-\exp(-iS_j(q)/\hbar)\right)/\sqrt{p(q)}\,,
\ende
where 
\begin{equation}
p(q)=\sqrt{2\left(E-V(q) \right)}, \qquad
S_j(q)=\int_{q_j}^q p(q')\,dq'\, .
\end{equation}
The matrix  connecting  the constants $A_1^+$ and $A_1^-$ to $A_2^+$
and $A_2^-$ (see e.g.\ the review of Berry and Mount \cite{BM72} and
the references therein) is given by 
\bege\label{eq:connection}
\begin{array}{cc}
\left(
 \begin{array}{c}
  A_2^+\\
  A_2^-
 \end{array}
\right) = M
\left(
 \begin{array}{c}
  A_1^+\\
  A_1^-
 \end{array} 
\right), \quad &
M = e^{\Theta/\hbar}\left(\begin{array}{cc}
  \sqrt{1+e^{-2\Theta/\hbar}} & -i \\
  i & \sqrt{1+e^{-2\Theta/\hbar}}\\
  \end{array}\right) \,,
\end{array}
\ende
with the penetration integral $\Theta=-i\int_{q_1}^{q_2}p(q')\,dq'$.
In the terminology of \Sec~\ref{sec:classic} this 
is $i$ times the action of an imaginary complex tunneling orbit.
This formula is also valid in the case where the energy $E$
lies everywhere above the potential $V(q)$.
Then the classical turning 
points  become complex ($q_1$ complex conjugate to $q_2$) and the 
penetration integral $\Theta$ becomes negative, corresponding
to an imaginary complex scattering orbit.
In a potential with several turning points in each region away from the 
classical turning points a \WKB\ wave function is reasonable. 
In classically forbidden regions the phase $S$  becomes complex leading 
to a real exponential. Uniform semiclassical quantization conditions may 
be obtained by a piecewise connection of these wave functions, and by imposing 
the correct boundary conditions, e.g.\ an exponential decay in classically 
forbidden regions, periodic boundary conditions in the case of a rotor with 
a potential, or Dirichlet conditions in the case of a hard potential wall,
e.g.\ in a billiard. 
The resulting quantization conditions take into account real tori and 
imaginary complex orbits.

The quantization conditions for the planar elliptic billiard
are obtained from the effective potentials shown in \fig~\ref{fig:effpot}.
 We allow for  negative values of $\rad$, which can formally be achieved
 by letting $\phi$ vary between $0$ and $\pi$ only.
 This is helpful in order to incorporate the different parities (see the 
 boundary  conditions in \tab~\ref{tab:bc}) in the semiclassical wave 
 functions $\psiwkbrad_j(\rad)$ (\equ~(\ref{eq:wavef}) for the
 radial degree of freedom). 
The wave functions $\psiwkbrad_j(\rad)$ are symmetric or antisymmetric 
with respect to $\rad = 0$, corresponding to $\pi_y = +1$ or $\pi_y = -1$ 
(see \tab~\ref{tab:bc}), consequently  
\bege\label{eq:symmetry}
  \begin{array}{ccc}
    \left(\begin{array}{c}
      A^+_2 \\
      A^-_2\\
    \end{array}\right)
  & = & 
  \pi_y     \left(\begin{array}{cc}
      0 & 1 \\
      1 & 0 \\
    \end{array}\right)

  \left(\begin{array}{c}
      A^+_1 \\
      A^-_1\\
    \end{array}\right) \ .
  \\
  \end{array}
\ende
The matrix $P$ describes the phase shift along the
classically allowed region
\NEU
\bege
P = \left(\begin{array}{cc}
  e^{i\frac{\pi}{2} \Jfrad/\hbar} & 0 \\
  0 & e^{-i\frac{\pi}{2} \Jfrad/\hbar} \\
  \end{array}\right) \,,
\ende
where it is important not to use the action $\Ifrad$ of the
full billiard, because this quantity jumps by a factor of 2 as the
effective energy increases beyond the top of the barrier. 
Instead we use $\BJf = 2 \BId$ as introduced in the last
section, such that the phase shift is continuous at the
transition from type $\tO$ to type $\tR$ motion.
Inserting the Dirichlet boundary conditions 
$\psiwkbrad_1(\rad_0) = \psiwkbrad_2(\rad_3) = 0$ 
into (\ref{eq:wavef}) leads to 
\bege\label{eq:bcond}
  \begin{array}{cc}
    \left(\begin{array}{c}
      A^+_1 \\
      A^-_1\\
    \end{array}\right)
   \propto P
  \left(\begin{array}{c}
      1 \\
      -1\\
    \end{array}\right), \qquad
    \left(\begin{array}{c}
      A^+_2 \\
      A^-_2\\
    \end{array}\right)
  \propto P^{-1}
  \left(\begin{array}{c}
      1 \\
      -1\\
    \end{array}\right)
  \\
  \end{array}  \,.
\ende

The fourth relation among the constants $A_i$ is given by 
(\ref{eq:connection}), where the barrier penetration integral 
$\Theta = \Theta_\rad$ is calculated from one turning point to 
the other and is given by
\bege
\Theta_\rad = \left\{ \begin{array}{cl} 
-2\sqrt{2E}\kappa
\left(\Eell(\sqrt{1-{a^2}/{\kappa^2}})-\Kell(\sqrt{1-{a^2}/{\kappa^2}})\right)
    & {\rm type\;} \tTR \\
-2\sqrt{2E}a
\left(\Eell(\sqrt{1-{\kappa^2}/{a^2}})-\frac{\kappa^2}{a^2}\Kell(\sqrt{1-{\kappa^2}/{a^2}})\right)
    & {\rm type\;} \tTO \\
-2\sqrt{2E}\sqrt{a^2-\kappa^2}\Eell(a/\sqrt{a^2-\kappa^2})
    & {\rm type\;} \tTOc \ .
\end{array}\right.
\ende
The action of imaginary complex orbits of type $\tTOc$ will be needed in 
the next section.

Composing the connection formula~(\ref{eq:connection}), the symmetry
(\ref{eq:symmetry}) and the boundary conditions~(\ref{eq:bcond}) 
gives 
\bege\label{eq:qcond1}
  \begin{array}{ccc}
    PMP\left(\begin{array}{c}
      1 \\
      -1\\
    \end{array}\right)
  & = & 
  \pi_y
  \left(\begin{array}{c}
      -1 \\
      1\\
    \end{array}\right) \ .
  \\
  \end{array}
\ende
Decomposing this complex Equation into its real and
imaginary part we obtain the quantization conditions 
for the radial degree of freedom,
\bege
\label{eq:cond1a}
\cos{(\pi \Jfrad/\hbar)} =
\frac{-\pi_y}{\sqrt{1+e^{2\Theta_\rad/\hbar}}} 
\ende
and
\bege
\label{eq:cond1b}
\sin{(\pi \Jfrad/\hbar)} =
\frac{-1}{\sqrt{1+e^{-2\Theta_\rad/\hbar}}} \ ,
\ende
which have to be fulfilled simultaneously.
The second Equation is not independent of the first one, it just
selects half the number of the solutions of the first Equation.

For the \WKB\ wave function $\psiwkbang(\ang)$ we use $\Veff (\ang)$
(\fig~\ref{fig:effpot}b), and we impose periodic boundary conditions.
Following the calculations of Miller \cite{Miller68} 
for the different parities we obtain the quantization conditions
\bege
\label{eq:cond2a}
\cos{(\pi \Jfang/\hbar)} = \frac{\pi_x
  \pi_y}{\sqrt{1+e^{2\Theta_\ang/\hbar}}} 
\ende
and 
\bege
\label{eq:cond2b}
\sin{(\pi \Jfang/\hbar)} =
\frac{\pi_x}{\sqrt{1+e^{-2\Theta_\ang/\hbar}}} \ ,
\ende
with $\Theta_\ang = -\Theta_\rad$.
The equivalence of the
  absolute values of $\Theta_\rad$ and $\Theta_\ang$ is a result of the
  especially symmetric separation of the Hamiltonian (\ref{eq:HamEll}).
The condition in \equ~(\ref{eq:cond2b}) just selects
the $x$-parity of the solutions of (\ref{eq:cond2a}).

\NEU
In Table~\ref{tab:limitcases} the limiting cases for large $|\Theta|$ 
of the right hand sides of the Equations~(\ref{eq:cond1a}-\ref{eq:cond2b})
are given. The resulting \EBK\ quantization conditions for the full billiard
with quantum numbers $(n_\rad, n_\ang)$
and the connection to the quantum numbers $(\nrad,\nang)$ of the
symmetry reduced billiard are indicated. 
For type $\tR$ motion 
these conditions are equivalent to the \EBK\ quantization conditions 
for the circular billiard.  
\begin{table}[!h]
\begin{center}
{ \small
\begin{tabular}{|c|lll|lll|}\hline
  \equ & 
\multicolumn{3}{|c|}{type $\tO$: $\Theta_\rad = -\Theta_\ang \ll 0$} & 
\multicolumn{3}{|c|}{type $\tR$: $\Theta_\rad = -\Theta_\ang \gg 0$} \\ \hline
  (\ref{eq:cond1a}) & $-\pi_y$ & & $\Ifrad = \Jfrad =
  (n_\rad+4/4)\hbar$ & $0$ & & $\Ifrad = \Jfrad/2 =
  (n_\rad+3/4)\hbar$ \\
  (\ref{eq:cond1b}) & $0$ & \raisebox{1.5ex}[-1.5ex]{$\bigg\}$} &
  $n_\rad = 2r+(1-\pi_y)/2$ & $-1$ & \raisebox{1.5ex}[-1.5ex]{$\bigg\}$}
  & $n_\rad = r$ \\ \hline
  (\ref{eq:cond2a}) & $0$ & & $\Ifang = \Jfang =
  (n_\ang+2/4)\hbar$ & $\pi_x \pi_y$ & & $\Ifang = \pm \Jfang =
  \pm n_\ang\hbar$ \\
  (\ref{eq:cond2b}) & $\pi_x$ &
  \raisebox{1.5ex}[-1.5ex]{$\bigg\}$} & $n_\ang = 2l+(1-\pi_x)/2$ & $0$
  & \raisebox{1.5ex}[-1.5ex]{$\bigg\}$} & $n_\ang = 2l+(2-\pi_x-\pi_y)/2$ \\ \hline
\end{tabular}
}
\caption[]{\label{tab:limitcases} \capsty The limiting cases of the
  right hand sides of the
  Equations~(\ref{eq:cond1a}-\ref{eq:cond2b}). The quantum numbers for the corresponding \EBK\ quantization
  conditions are $n_\rad, n_\ang \in \N\cup \{0\}$ for the full system
  and $r,l \in \N\cup \{0\}$ for the desymmetrized billiard.}
\end{center}
\end{table}

%
\rem{
In the limiting case $a \to 0$ where the ellipse degenerates into a
circle, we can set $\Theta_\ang \ll 0$ always 
and the conditions~(\ref{eq:cond1a}), (\ref{eq:cond1b}),
(\ref{eq:cond2a}) and (\ref{eq:cond2b}) reduce to 
\begin{eqnarray}
\cos{(\pi \Jfrad/\hbar)} & = & \sin{(\pi \Jfang/\hbar)} = 0 \nonumber \\
\sin{(\pi \Jfrad/\hbar)} & = & -1 \\
\cos{(\pi \Jfang/\hbar)} & = & \pi_x\pi_y \ . \nonumber
\end{eqnarray}
These Equations are equivalent to the \EBK\ quantization conditions for the 
circular billiard which read $\Ifrad = \Jfrad/2 = (n_\rad+3/4)\hbar$ and 
$\Ifang = \pm \Jfang = n_\ang\hbar$, where $n_\rad \in \N$ and $n_\ang \in \Z$.
} 

How to calculate the $(E,\kappa^2)$- or the $(\Jfrad,\Jfang)$-spectrum from
the conditions in \equ~(\ref{eq:cond1a}), (\ref{eq:cond1b}),
(\ref{eq:cond2a}) and (\ref{eq:cond2b})? The actions
and the barrier penetration integrals are functions of
$(E,\kappa^2)$. With Newton's method we can numerically obtain
$(E,\kappa^2)$ as functions of $(\Jfrad,\Jfang)$ and therefore
$\Theta_\rad$ and $\Theta_\ang$ as functions of  $(\Jfrad,\Jfang)$. 
Ignoring (\ref{eq:cond1b}) and (\ref{eq:cond2b}) for the moment,
the remaining Equations~(\ref{eq:cond1a}) and (\ref{eq:cond2a}) can 
thus be rewritten in the form
\bege\label{eq:condn}
\begin{array}{lllrl}
f^{\pi_y}_1(\Jfrad,\Jfang) & := & \cos{(\pi \Jfrad/\hbar)} +
\pi_y g_1(\Theta_\rad(\Jfrad,\Jfang)) & = & 0  \\
f_2^{\pi_x,\pi_y}(\Jfrad,\Jfang) & := & \cos{(\pi \Jfang/\hbar)} -
\pi_x \pi_y g_2(\Theta_\ang(\Jfrad,\Jfang)) & = & 0 \ ,
\end{array}
\ende
where $g_1$ and $g_2$ are smooth functions onto the intervall $(0,1)$. 
A solution of (\ref{eq:condn}) can be viewed as an
intersection of the lines $f_1^{\pi_y}= 0$ and $f_2^{\pi_x,\pi_y} = 0$ 
in action space with the same corresponding parities. 
The solutions lie inside a box with the edges defined by
the extremal values of $g_1$ and
$g_2$. Because of the periodicity of the left hand sides of
Equations~(\ref{eq:cond1a}-\ref{eq:cond2b}) these boxes are always arranged in the same way
inside a cell in 
the $(\Jfrad,\Jfang)$-space. We call these quadratic cells 
``semiclassical quantum cells''
with quantum numbers $(\nrad,\nang)$. Their edges have length $\Delta
\Jfrad = \Delta \Jfang = 2\hbar$ and they tessellate the whole 
$(\Jfrad,\Jfang)$-space. Inside a quantum cell there are
four quantum states, one for each parity. Each state of fixed parity
is confined to a ``parity box'' of width $\hbar/2$, shown by bold lines in
\fig~\ref{fig:qcell}. The size of the parity box is a result 
of the fact that the Maslov indices change by two 
upon the transition of the separatrix, see \tab~\ref{tab:limitcases}
These parity boxes take care of the
remaining conditions (\ref{eq:cond1b}) and (\ref{eq:cond2b}):
inside a box they are automatically fulfilled when 
(\ref{eq:condn}) holds. To find the solution guaranteed to lie in the
interior of a box, we use the bisection method described in
\cite{Vrahatis95}. Having got all solutions in action space,
we compute again the solutions in $(E,\kappa^2)$-space with Newton's
method to obtain the semiclassical eigenvalue spectrum.
Introducing the semiclassical quantum cells in action space ensures
that always the right state is found, even when two states are
almost degenerate. 
The method can easily be extended to the case of more than two degrees 
of freedom.  
\begin{Filesave}{bilder}
\def\figqcell{ Schematic view of the four states in 
    a single semiclassical quantum cell 
    $(\Delta \Jfrad = \Delta \Jfang = 2\hbar)$ for (a) the region $\tR$, (b) the region 
    $\tO$ and (c) close to the separatrix corresponding to the states with
    quantum numbers $(0,1)$. 
    The intersections of the lines $f_1^{\pi_y}= 0$ and $f_2^{\pi_x,\pi_y} = 0$
    with equal parities (dots) are the solutions of the quantization
    conditions~(\ref{eq:cond1a}), (\ref{eq:cond1b}), (\ref{eq:cond2a})
    and (\ref{eq:cond2b}) if they are located inside a parity box. 
    The squares and the triangles are the
    states obtained from \EBK\ quantization in regions $\tR$ and $\tO$,
    respectively. 
    Parity boxes that are crossed by the bold dashed separatrix  
    may contain two \EBK\ states (bold squares and triangles)
    or no \EBK\ states, because both corners (the thin squares and
    triangles) are in the wrong region.
    The \EBK\ Maslov index is $(3,0)$ in region $\tR$ 
    (where $J_\rad = 2I_\rad$) and $(4,2)$ in region $\tO$.}
\def\FIGqcell{\centerline{\psfig{figure=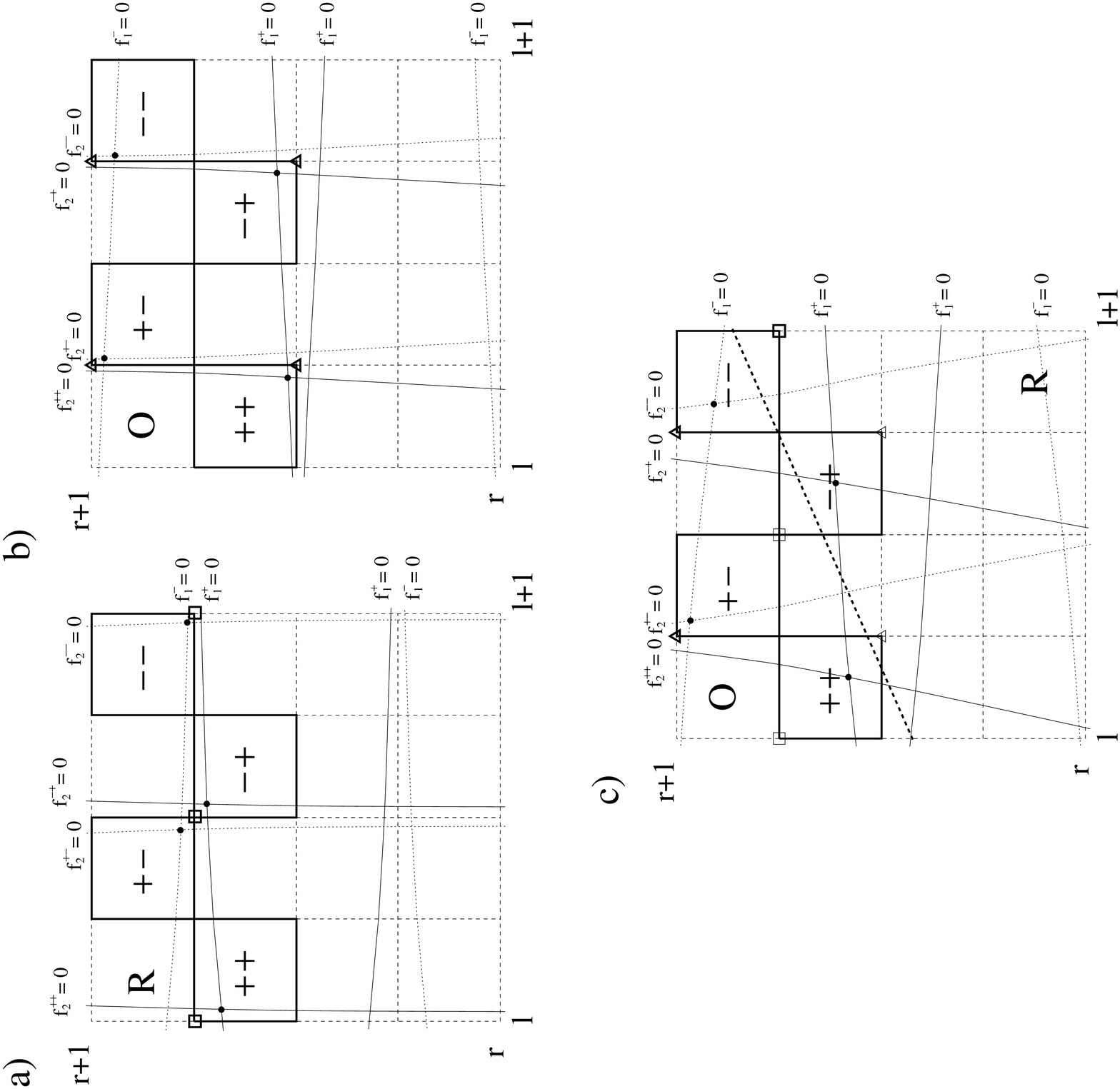,angle=-90,width=15cm}}}
\FIGo{fig:qcell}{\figqcell}{\FIGqcell}
\end{Filesave}

\rem{
\begin{Filesave}{bilder}
\begin{figure} 
  \centerline{\psfig{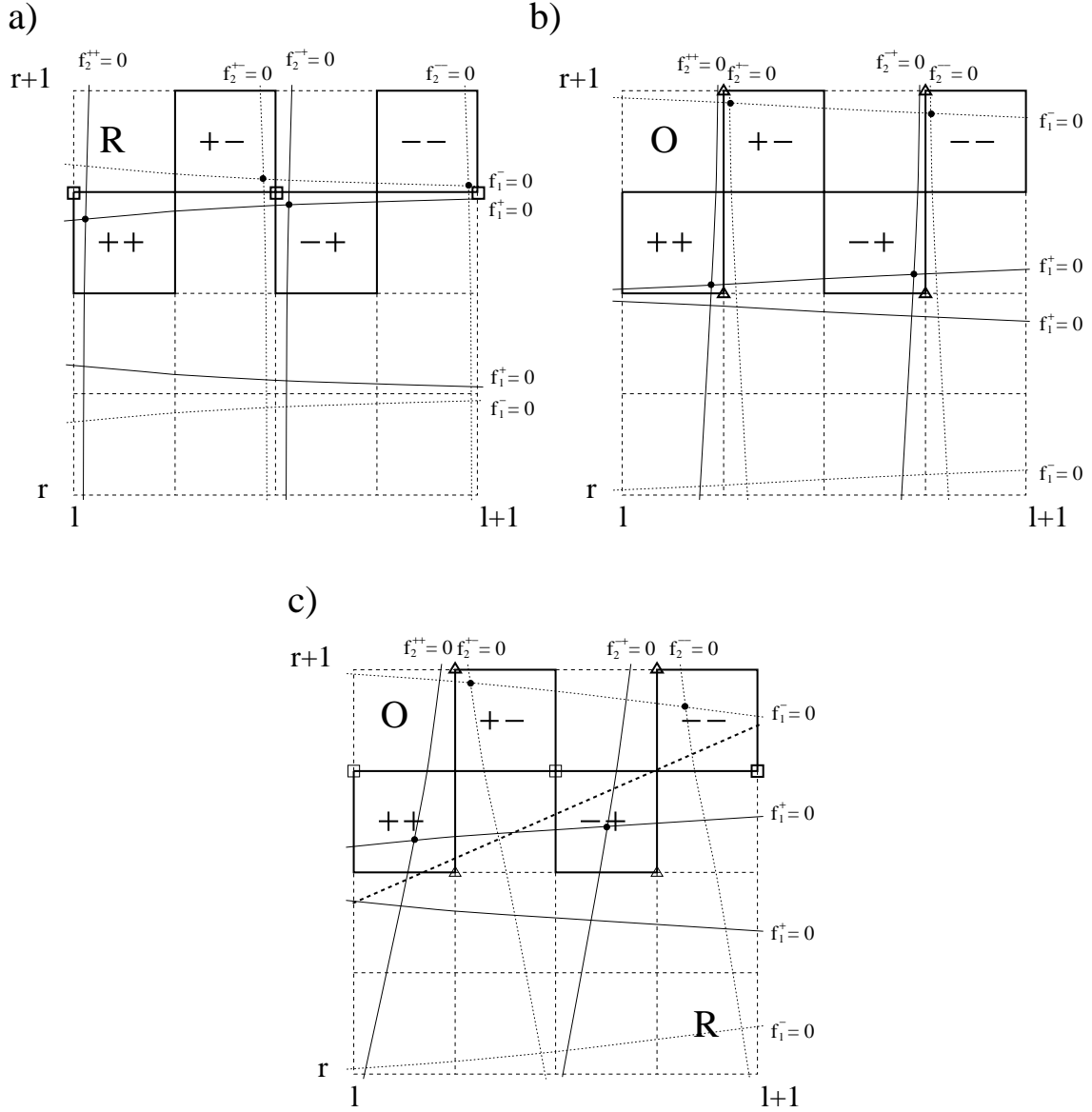}}
  \caption[]{\label{fig:qcell} \capsty Schematic view of the four states in 
    a single semiclassical quantum cell 
    $(\Delta \Jfrad = \Delta \Jfang = 2\hbar)$ for (a) the region $\tR$, (b) the region 
    $\tO$ and (c) close to the separatrix corresponding to the states with
    quantum numbers $(0,1)$. 
    The intersections of the lines $f_1^{\pi_y}= 0$ and $f_2^{\pi_x,\pi_y} = 0$
    with equal parities (dots) are the solutions of the quantization
    conditions~(\ref{eq:cond1a}), (\ref{eq:cond1b}), (\ref{eq:cond2a})
    and (\ref{eq:cond2b}) if they are located inside a parity box. 
    The squares and the triangles are the
    states obtained from \EBK\ quantization in regions $\tR$ and $\tO$,
    respectively. 
    Parity boxes that are crossed by the bold dashed separatrix  
    may contain two \EBK\ states (bold squares and triangles)
    or no \EBK\ states, because both corners (the thin squares and
    triangles) are in the wrong region.
    The \EBK\ Maslov index is $(3,0)$ in region $\tR$ 
    (where $J_\rad = 2I_\rad$) and $(4,2)$ in region $\tO$.}
\end{figure}
\end{Filesave}
}

Figure~\ref{fig:statescmp} and \tab~\ref{tab:states} show the 
result of the semiclassical calculation compared to the exact 
quantum mechanical results. The
distinction between the parities is omitted in
\fig~\ref{fig:statescmp} because this becomes already clear from
\fig~\ref{fig:statesI}. 
In \fig~\ref{fig:errors} the relative errors of the semiclassical eigenvalues 
$(E_{\text{qm}}-E_{\text{sc}})/E_{\text{qm}}$  and
$(\kappa_{\text{qm}}-\kappa_{\text{sc}})/\kappa_{\text{qm}}$ 
are plotted versus $\kappa_{\text{qm}}$, such that the
behavior of the error with respect to the position of the state on
the classical energy surface can be seen. 
The semiclassical energy eigenvalues are almost always too low, 
see also \tab~\ref{tab:states}. The opposite holds for the values 
of the second eigenvalue, which usually are too high.
The only exception (for both eigenvalues) occurs in the neighborhood
of the classical separatrix. 
The fact that \fig~\ref{fig:errors} looks rather symmetric indicates
that the relative errors of the two eigenvalues are strongly correlated.
Concerning the semiclassical limit we see series of states with
increasing $\nrad$ quantum number and decreasing error for small
$\kappa$, and series with increasing $\nang$ quantum number and
decreasing error for large $\kappa$. 
However, increasing $\nang$ with $\kappa < a$ does not decrease the error.
The error of the whispering gallery states $(0,\nang)$ is 
particularly large.
In general, however,
the agreement between the exact eigenvalues
and the semiclassical values is very good. 
\def\figerrors{%
The relative error of semiclassical energy eigenvalue
(circles) and $\kappa$ eigenvalue (pluses) plotted
versus $\kappa$. The obvious patterns result from series of
states with the indicated quantum numbers. The arrows point into
the direction of increasing quantum number.
}
\def\FIGerrors{\centerline{\psfig{figure=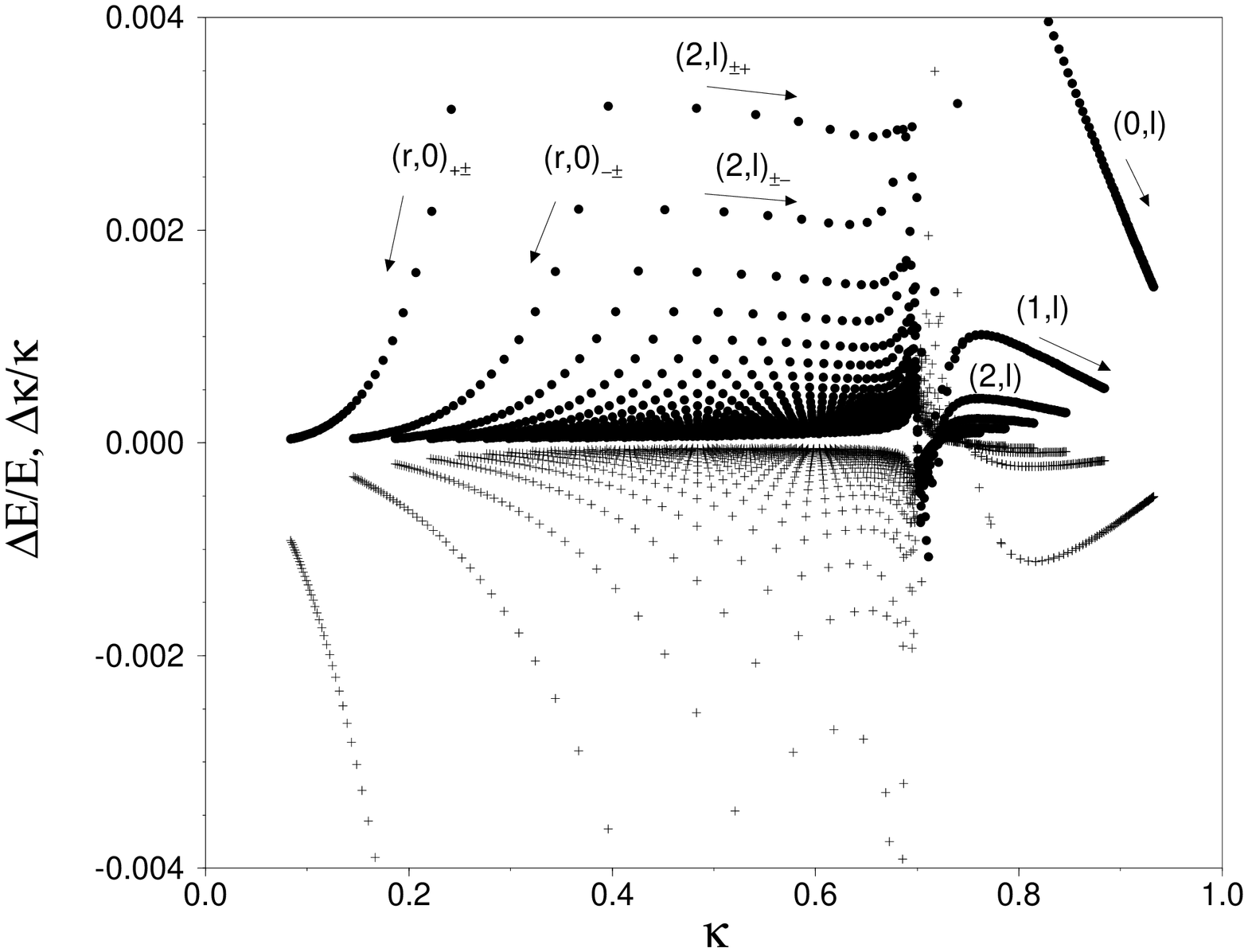,width=11cm,angle=-90}}}
\FIGo{fig:errors}{\figerrors}{\FIGerrors}
\begin{Filesave}{bilder}
\def\figstatescmp{Comparison of the exact states
    (circles) to the semiclassical states (intersections of the lines $f_1^{\pi_y} = 0$ and
    $f_2^{\pi_x,\pi_y} = 0$ with the same corresponding
    parities). The dashed line is the energy surface $E=200$. $\Jfrad$ and $\Jfang$ are measured in units of $\hbar$.}
\def\FIGstatescmp{\centerline{\psfig{figure=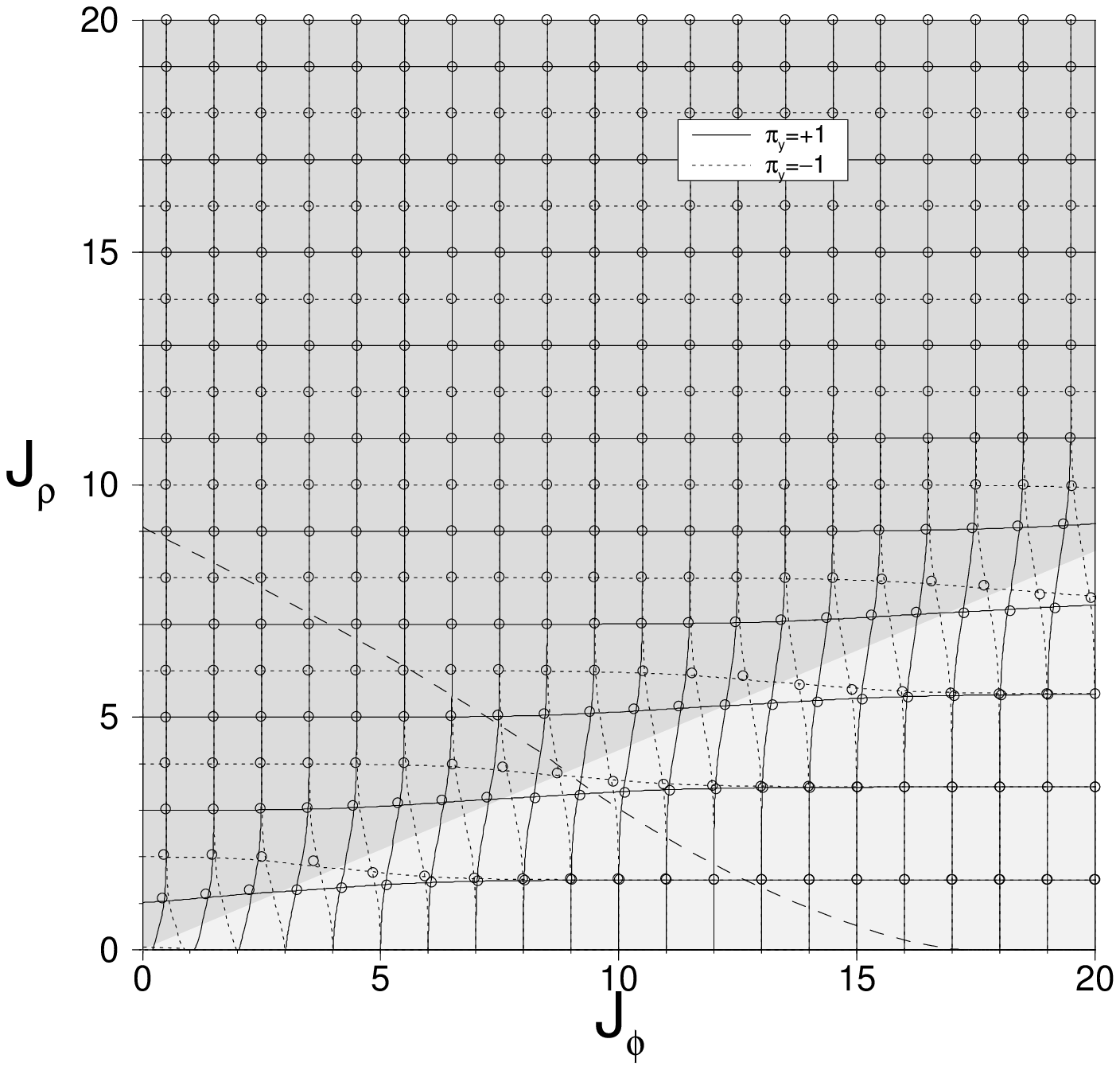,angle=0,width=13.0cm}}}
\FIGo{fig:statescmp}{\figstatescmp}{\FIGstatescmp}
\end{Filesave}
\rem{
\begin{Filesave}{bilder}
\begin{figure}
  \centerline{\psfig{figure=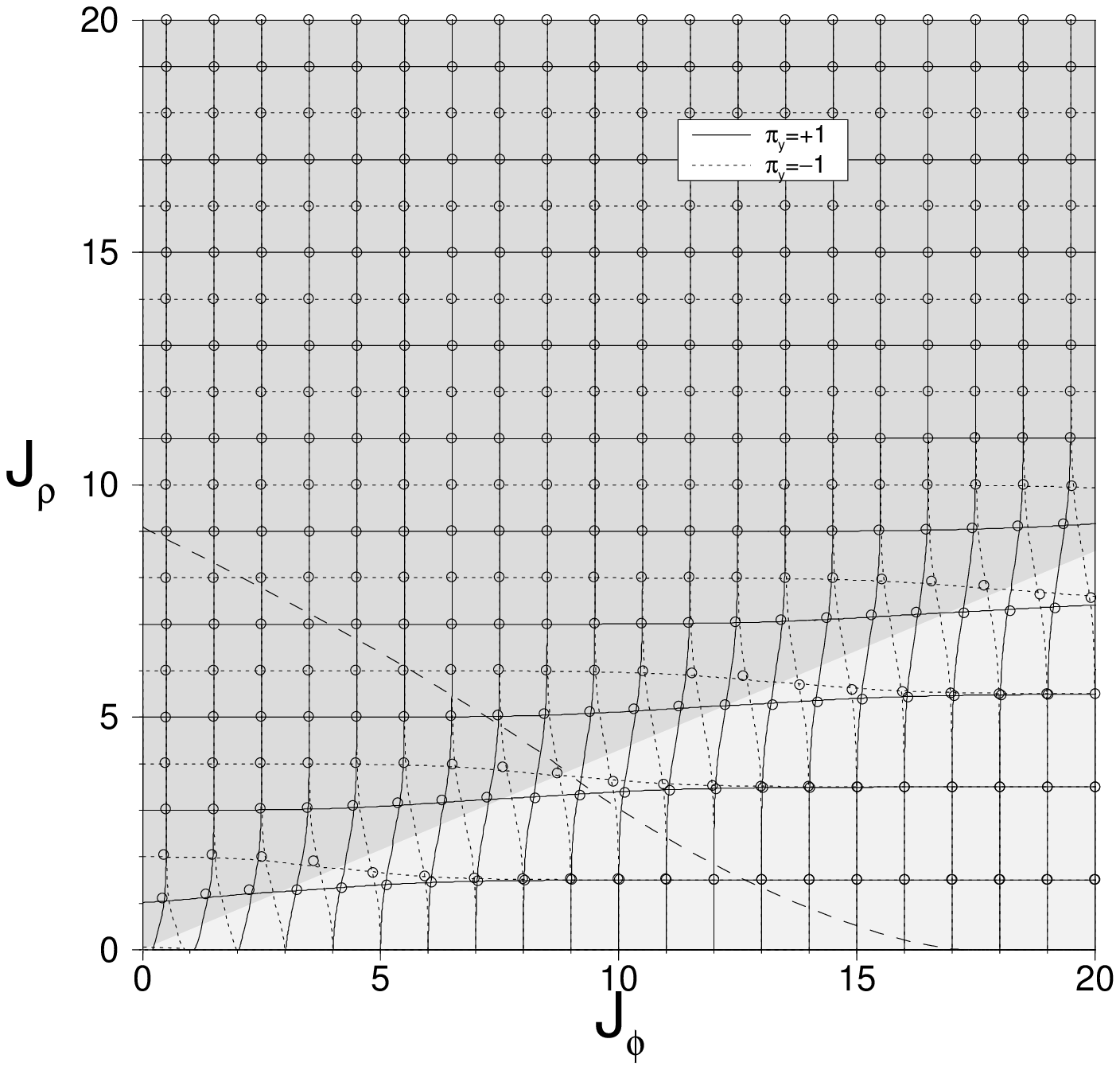,angle=0,width=13.0cm}} 
  \caption[]{\label{fig:statescmp} \capsty Comparison of the exact states
    (circles) to the semiclassical states (intersections of the lines $f_1^{\pi_y} = 0$ and
    $f_2^{\pi_x,\pi_y} = 0$ with the same corresponding
    parities). The dashed line is the energy surface $E=200$. $\Jfrad$ and $\Jfang$ are measured in units of $\hbar$.} 
\end{figure}
\end{Filesave}
}
\begin{table}[!h]
\tabstart
\begin{center}
{ \small
\begin{tabular}{|c|c|c|c|c|c|c|c|}\hline
  $E_{{\text{qm}}}$ & $\kappa^2_{{\text{qm}}}$ & 
  $E_{{\text{sc}}}$ & $\kappa^2_{{\text{sc}}}$ & $\nrad$ & $\nang$ & $\pi_x$ & $\pi_y$ \\ \hline
 4{.}26746 & 0{.}18714 &  3{.}99539 & 0{.}21120 & 0 & 0 & + & + \\
 9{.}05834 & 0{.}37981 &  8{.}65488 & 0{.}39821 & 0 & 0 &$-$& + \\
12{.}5768 & 0{.}12960 & 12{.}3295 & 0{.}13666 & 0 & 0 & + &$-$\\
15{.}9933 & 0{.}45667 & 15{.}5268 & 0{.}46827 & 0 & 1 & + & + \\
19{.}3576 & 0{.}30380 & 18{.}9734 & 0{.}31216 & 0 & 0 &$-$&$-$\\
25{.}0613 & 0{.}49625 & 24{.}7502 & 0{.}49755 & 0 & 1 &$-$& + \\
25{.}8947 & 0{.}09212 & 25{.}6683 & 0{.}09516 & 1 & 0 & + & + \\
\vdots & &  &  &  &  &  &  \\ 
1000{.}46 & 0{.}13551 & 1000{.}21 & 0{.}13560 & 8  & 2 & + &$-$ \\ 
1001{.}65 & 0{.}65621 & 1000{.}78 & 0{.}65650 & 1 & 13 &$-$&$-$ \\ 
1001{.}65 & 0{.}65621 & 1000{.}78 & 0{.}65650 & 1 & 14 & + & + \\ 
1002{.}01 & 0{.}43314 & 1001{.}40 & 0{.}43342 & 5  & 8 & + & + \\ 
1008{.}07 & 0{.}19149 & 1007{.}80 & 0{.}19159 & 8  & 3 & + & + \\ 
1010{.}97 & 0{.}36819 & 1010{.}52 & 0{.}36838 & 6  & 6 &$-$& + \\ 
1013{.}91 & 0{.}48076 & 1012{.}72 & 0{.}48148 & 4  & 9 &$-$& + \\ 
1018{.}48 & 0{.}24440 & 1018{.}17 & 0{.}24451 & 7  & 4 & + &$-$ \\ 
1029{.}35 & 0{.}78842 & 1026{.}76 & 0{.}78970 & 0  & 16 & + &$-$ \\ 
1029{.}35 & 0{.}78842 & 1026{.}76 & 0{.}78970 & 0  & 16 &$-$& + \\ \hline
\end{tabular}
}
\caption[]{\label{tab:states} \capsty The quantum mechanical
  eigenvalues $(E_{{\text{qm}}},\kappa^2_{{\text{qm}}})$ and the
  semiclassical eigenvalues $(E_{{\text{sc}}},\kappa^2_{{\text{sc}}})$ 
  of the billiard in the ellipse for $a=0.7$ with $E_{{\text{qm}}} < 26$ 
  or $1000 < E_{{\text{qm}}} < 1030$.}
\end{center}
\tabend
\end{table}

Now we study the change of the regular
lattice obtained from \EBK\ quantization induced by the 
uniformization.
There are three typical situations in action space
represented in \fig~\ref{fig:qcell}: region $\tR$ (a) and region $\tO$ (b)
both far away from the separatrix and the region close to 
the separatrix (c). In region $\tR$ the \EBK\ states
$(\nrad,\nang+1)_{++}, (\nrad,\nang)_{--}$ (e.g.\
\tab~\ref{tab:states} rows 9 and 10) and $(\nrad,\nang)_{+-},
(\nrad,\nang)_{-+}$ (e.g.\ rows 16 and 17) are almost degenerate. 
This corresponds to the degeneration of clockwise and counterclockwise
rotations in the planar circular billiard. The \EBK\ states 
form regular lattices, with mesh size $\Delta \Jfrad =
2\hbar$ and $\Delta \Jfang = \hbar$ with two states on each corner in region 
$\tR$ and $\Delta \Jfrad =\Delta \Jfang = \hbar$ in region $\tO$.
Far away from the separatrix the uniform semiclassical states lie near
the \EBK\ states with Maslov index $(3,0)$ for type $\tR$ and
$(4,2)$ for type $\tO$. Approaching the separatrix a transition between
the two regular \EBK\ lattices takes place. Crossing the
separatrix from region $\tR$ to region $\tO$ the 
shift of the \EBK\ states
is $\delta \Jfrad = -\hbar/2$ and $\delta \Jfang = \hbar/2$ for 
$\pi_y = +1$ and $\delta \Jfrad = \hbar/2$ and $\delta \Jfang = -\hbar/2$ 
for $\pi_y = -1$. 
The uniform quantization smoothly joins the \EBK\ lattices
along the separatrix.
It is natural to interpret the location of the 
semiclassical states as two separate meshes: 
one for $\pi_y = +1$ and one for $\pi_y = -1$, indicated
by full and dotted lines in \fig~\ref{fig:statescmp}.

Figure~\ref{fig:qcell}c shows that close to the separatrix 
not only the accuracy of the \EBK\ states is unsatisfactory but also that \EBK\
quantization sometimes yields a wrong number of states. 
In \fig~\ref{fig:qcell}c the situation for the $(0,1)$-quantum cell is shown. 
\NEU
The \EBK\ states are marked by bold squares in region $\tR$
and by bold triangles in region $\tO$. A square represents two \EBK\ states
because of the degeneracy in region $\tR$. 
The $+-$ parity box contains one \EBK\ state, marked by the bold triangle.
The opposite corner of this parity box could be reached with the Maslov 
index defined for region $\tR$. However, it is located on the wrong side
of the separatrix, and therefore not a true \EBK\ state indicated 
by a thin square. 
By the same reasoning the parity boxes $++$ and $-+$ contain no 
true \EBK\ state. On the contrary the $--$ parity box contains
two true \EBK\ states.
\rem{
Choosing the Maslov indices for both regions separately, it is obvious that
there are two $--$ states (the bold square and triangle) and one $+-$ state 
(the bold triangle) but no $-+$ and $++$ states\footnote{In the 
\EBK\ approximation each bold square accounts for two degenerate 
states with parity $--$ and $++$. The bold square in c) thus 
stands for a $++$ state, which, however, does not belong to the 
quantum cell $(0,1)$.}
in the quantum cell. 
}
This indicates the main deficiency of \EBK\ 
quantization for systems that exhibit a separatrix.

\rem{
The \WKB\ wave functions $\psiwkbrad_j(\rad)$ in the regions
$j=1$ and $j=2$ are given by
\bege
\label{eq:wavef}
\psiwkbrad_j(\rad) =
(A^+_j\exp{(iS_j(\rad)/\hbar)}+A^-_j\exp{(-iS_j(\rad)/\hbar)})/
\sqrt{|p_{\rad} (\rad)|}\ ,
\ende
with
\bege
S_j(\rad) = \int^{\rad}_{\rad_j}p_{\rad} (\rad')d\rad'\ .
\ende
Note that $S_2(\rad)$ is complex when the energy lies below the
potential barrier.
} 

\section{The Berry-Tabor trace formula}
\label{sec:bt}

In the last section we have seen how the imaginary complex orbits
lead to a Maslov index smoothly varying across the separatrix.
The real complex orbits can be taken into account by a completely
different approach introduced by Berry and Tabor \cite{BerryTabor76}. 
The goal of this section is to incorporate the imaginary complex
orbits into the Berry-Tabor trace formula, such that all kinds
of classically forbidden tori are taken into account. 

Starting from \EBK\ quantization Berry and Tabor
derived a formula for the semiclassical density of states
\rem{
\begin{eqnarray}
\sds = \sum_{n=1}^{\infty}\delta\left(E-E_n\right)
\end{eqnarray}
}
\begin{eqnarray}\label{eq:sds}
\sds = \sum_{{\bm n}}\delta\left(E-H(({\bm n}+\frac{\Bmaslovf}{4})\hbar)\right)
\end{eqnarray}
in terms of resonant tori for an integrable system with $\dof$ degrees of 
freedom. They transformed \equ~(\ref{eq:sds}) via the
Poisson summation formula and performed the integrals in stationary-phase approximation. The result is
\begin{equation}\label{eq:bt}
\sds = \overline{n}(E) + \frac{2}{\hbar^{\frac{\dof+1}{2}}}\sum\limits_\Brtopo           
\frac{1}{|\Bomf(\BIf^\rtopo)||\Brtopo|^{\frac{\dof-1}{2}}\sqrt{|\curv (\BIf^\rtopo)|}} 
 \sum\limits^\infty_{q=1}\frac{\cos{(q(\action(\BIf^\rtopo)/\hbar-\frac{\pi}{2}\Bmaslovf\Brtopo)+ 
\frac{\pi}{4}\beta)}}{q^{\frac{\dof-1}{2}}}.
\end{equation}

The first summation is over all relatively prime non-negative integers
$\Brtopo$, i.e. over families of prime orbits, while the second
summation runs over all their repetitions. 
The Maslov indices are always the same since the 
energy surface is assumed to have no separatrix.
The term $\Brtopo = 0$ is
excluded from the sum and denoted by $\overline{n}(E)$; it gives the
mean density of states, the so-called Thomas-Fermi term. For the
elliptic billiard we have $\overline{n}(E)=A_b/(2\pi\hbar^2)$
according to {\em Weyl's law} where $A_b=\pi \sqrt{1-a^2}$ gives the 
billiard's area. In \cite{Sieber96} it is shown that next order
correction to Weyl's law, proportional to the circumference of 
the billiard, is contained in the whispering gallery orbits in the
remaining sum.

To improve the convergence of the series (\ref{eq:bt}) it is advantageous to
introduce a smoothed density of states $\sdsg$ by giving $E$ a small
imaginary part $i\gamma$. Then a $\delta$-peak changes into a Lorentz
function with half width energy $\gamma$. The semiclassical formula for
$\sdsg$ differs from formula~(\ref{eq:bt}) by a decay factor $D =
\exp{(-\gamma q T(\BIf^\rtopo)/\hbar)}$ before the cosine term in the second
summation.

Berry and Tabor show that formula~(\ref{eq:bt}) can be improved by
taking into account resonant real complex tori. This removes unphysical
discontinuities in the density of states resulting from contributions
of resonant tori which are suddenly classically realized as $E$
changes. For a system that scales with respect to the energy this
cannot occur, nevertheless the system parameter $a$ can take 
over the role of the energy, e.g.\ for the elliptic billiard all 
real complex orbits become realized for $a \to 1$. 
Even without a parameter these 
corrections are sensible because they improve the situation
when the stationary phase approximation is bad because there
is a stationary point close to, but outside the range of integration.
The resulting trace formula incorporating real complex orbits for two
degrees of freedom systems reads
\begin{equation}\label{eq:btuniform}
\begin{array}{r}
\sdsg = \overline{n}(E) + \frac{2}{\hbar^{3/2}}\sum\limits_\Brtopo           
\frac{1}{|\Bomf(\BIf^\rtopo)|\sqrt{|\Brtopo||\curv (\BIf^\rtopo)|}} 
 \sum\limits^\infty_{q=1}D\;A\frac{\cos{(q(\action(\BIf^\rtopo)/\hbar-\frac{\pi}{2}\Bmaslovf\Brtopo)+\theta)}}{\sqrt{q}} \\
-\frac{2}{\sqrt{2\pi}\beta\hbar}\sum\limits_\Brtopo           
\frac{1}{|\Bomf(\BIf^\rtopo)|\sqrt{|\Brtopo||\curv (\BIf^\rtopo)|}} 
 \sum\limits^\infty_{q=1}D\left(\frac{\sin{(q(\action_2/\hbar-\frac{\pi}{2}\Bmaslovf\Brtopo))}}{\Lambda_2\sqrt{q}}-\frac{\sin{(q(\action_1/\hbar-\frac{\pi}{2}\Bmaslovf\Brtopo))}}{\Lambda_1\sqrt{q}}\right) \\
+\frac{1}{\pi\hbar}\sum\limits_\Brtopo \sum\limits^\infty_{q=1}
 D\left(\frac{\sin (q(\action_2/\hbar-\frac{\pi}{2}\Bmaslovf_2\Brtopo))}{q|\Bomf_2|\Brtopo \BIf'_2}-\frac{\sin (q(\action_1/\hbar-\frac{\pi}{2}\Bmaslovf_1\Brtopo))}{q|\Bomf_1|\Brtopo \BIf'_1}\right) ,
\end{array}
\end{equation}
where $\action_{1,2}$ are the actions, $\Bmaslovf_{1,2}$ the Maslov 
indices and $\BIf'_{1,2}$ the normalized derivatives with respect to
a second constant of motion $\xi$, which parametrizes the energy
surface in action space; 
all quantities with indices $1,2$ are evaluated at the 
boundaries $\xi_1$ and $\xi_2$ of the energy surface. 
$A$ is the amplitude and $\theta$ the argument
of the complex Fresnel integral
\bege\label{eq:fresnelintegral}
F = \frac{1}{\sqrt{2\pi}}\int^{\Lambda_2/\sqrt{\hbar}}_{\Lambda_1/\sqrt{\hbar}} dx e^{i\frac{\beta}2
  x^2} \ ,
\ende
where $\Lambda_{1,2} \in \R$ are defined by
$\Lambda^2_{1,2} = 2q/\beta(\action_{1,2}-\action)$ and
$\Lambda_{1,2} \stackrel{\scriptscriptstyle >}{\scriptscriptstyle <} 0$ if
$\xi_{1,2} \stackrel{\scriptscriptstyle >}{\scriptscriptstyle <}
\xi^\rtopo$, where $\BIf^\rtopo = \BIf(\xi^\rtopo)$.
 The series~(\ref{eq:btuniform}) holds for real and real complex
tori.  

\nono{Could we put this to the next chapter:
In a later paper P.J.~Richens~\cite{Richens82} derives an approximate
form of the last two boundary terms in (\ref{eq:btuniform}) and
interpretes them as contributions of the stable isolated periodic
orbits. He shows that in the limiting case these contributions are equal to the
corresponding terms in the Gutzwiller's trace
formula~\cite{Gutz71}. This will become important in the next
section. 
}

Formulas~(\ref{eq:bt}) and (\ref{eq:btuniform}) can be applied to
simple systems \cite{BerryTabor76,AtkinsEzra95} but not to systems
with separatrices like the elliptic billiard. In the derivation of
Berry and Tabor the energy surface is assumed to be strictly convex 
or strictly concave and to have a smooth curvature. 
This is not the case for a system with
separatrices, e.g. the elliptic billiard, as can be seen in 
\fig~\ref{fig:esurface}. 
We are going to study three versions of the Berry-Tabor trace formula
with sucessive improvements concerning states close to the separatrix.
A first approach, similar to \cite{Joyeux95},
would be to divide the energy surface into smooth patches 
and to consider (\ref{eq:bt}) or (\ref{eq:btuniform}) for each patch.
We call this the \EBKBT\ quantization. 
The manifestation of the separatrix in the \EBK\ approximation as
a discontinuous quantization condition is carried over to the 
\EBKBT\ quantization by the appearance of the wrong number of terms in 
the summation~(\ref{eq:bt}) or (\ref{eq:btuniform}). 
In the \WKBBT\ quantization we correct this deficiency
by taking the Maslov index in (\ref{eq:sds}) as a uniform Maslov phase,
smoothly varying across the separatrix. However, the new term is considered
constant in the stationary phase approximation. This improves the results 
and still gives a sum over periodic orbits.
The best results are obtained if the Maslov phase is fully taken into
account in the stationary phase approximation, leading to summation
over (in general) non-periodic orbits in a \uniWKBBT\ trace formula.

The resonant tori of an integrable two degrees of freedom system are 
characterized by the winding number $\wrf(\kappa^\rtopo) =
\rtopo_\ang/\rtopo_\rad$.
For a billiard system without potential $\wrf$ is independent of the energy. 
Thus we can obtain all resonant tori for one reference energy. 
The actions, the frequencies and the reciprocal curvature scale with
$\sqrt{E}$. Hence all these quantities must only be calculated for the
reference energy. 
For the elliptic billiard the winding number $\wrf$ is restricted 
to the interval $(0,1)$. Thus for resonances up to a fixed order we have 
$0 < \rtopo_\rad \le \rtopo_{\text{max}}$ and $0 < \rtopo_\ang < \rtopo_\rad$.
For the summation over $q$, i.e. over the repetitions, we incorporate all 
orbits with period $q T$ less than a cutoff time 
$T_c = \frac{\hbar}{\gamma}\ln{\frac{2}{\epsilon}}$
to guarantee that the error in the second summation of formula
(\ref{eq:bt}) is smaller than $\epsilon$. 
In the following we choose $a=0.7$, $\hbar = 1$, $\gamma
= 0.3$ and $\epsilon = 10^{-8}$. Thus the cutoff time is $T_c \approx 63.7$.
For numerical reasons it is not possible to approach the cusp of
$\wrf$ (see \fig~\ref{fig:esurface}b for $\wrd$) arbitrarily
close. We can only include all resonant tori with $|\kappa^2-a^2| >
10^{-14}$, i.e. $\wrd < 0.472$ for type $\tR$ and $\wrd < 0.943$ for 
type $\tO$. But the contribution of tori with larger winding numbers are 
very small because of the divergence of the curvature at $\kappa^2 = a^2$. 
Taking $\rtopo_{\text{max}} = 250$, we include $26\,395$ and neglect $2\,026$ 
resonant tori in the summation. 
The remaining interval $[0,2\arccos{(a)}/\pi) \approx [0,0.506)$ carries
the real complex tori (type $\tOc$). Since for $a=0.7$ 
the real complex torus with $\wrf = 1/2$ 
has the predominant effect we found it sufficient to consider
resonant real complex tori with $\kappa^2 \geq \kappa^2_{\text{min}} =
-15$, i.e. $\wrf(\kappa) \geq \wrf(\kappa_{\text{min}}) =
0.285$. Resonant tori with $\kappa^2 < \kappa^2_{\text{min}}$ give
negligible constributions, even though the curvature is very small, due to
the destructive interference in the 
Fresnel integral (\ref{eq:fresnelintegral}). 
The number of incorporated real complex tori is $4\,205$, $5\,1$ are left out.

In the \EBKBT\ approach we set $\Bmaslovf = (3,0)$ and $\beta = 1$ 
for the patch $\tR$ and $\Bmaslovf = (4,2)$ and $\beta = -1$ for patch $\tO$. 
On patch $\tR$ the \EBK\ states are exactly degenerate, thus all the
contributions are multiplied by 2.
Since there are only real complex tori of type $\tOc$, in
\equ~(\ref{eq:btuniform}) for patch $\tO$ we only consider terms corresponding
to the boundary $\xi_1$ ($\kappa^2 = 0$) and no boundary terms for patch 
$\tR$. 
Figure \ref{fig:EBKBT} presents the spectrum $\sdsg$ calculated
with the \EBKBT\ quantization in the range $12.5 \leq E \leq 47.5$. 
For comparison the exact spectrum smoothed in a Lorentzian manner is shown. 
We focus on the states shown in the quantum cell $(0,1)$ close to 
the separatrix shown in \fig~\ref{fig:qcell}. The other states
in \fig~\ref{fig:EBKBT} are reproduced fairly well.
First we observe that the $(0,1)_{+-}$ state is also reproduced quite well,
which is due to the fact that the corresponding parity cell 
correctly contains one \EBK\ state only.
The peak for the parity box $(0,1)_{++}$ 
splits into two peaks with approximately half amplitude. The reason is
that there is no true \EBK\ state in this parity box. However the
two corners of the box carry \EBK\ states located in the wrong 
region (the thin square and triangle in \fig~\ref{fig:qcell}),
and they still give a small contribution in the trace formula.
The situation is similar for the $(0,1)_{-+}$ state. 
Here the \EBK\ peak also splits into two small peaks; 
one accidently overlaps with the $(1,0)_{++}$ \EBK\ state and the
other one is seen as a small bump on the left side.  
In both cases the \EBKBT\ quantization almost fails to
produce a state.
The parity box $(0,1)_{--}$ in \fig~\ref{fig:qcell} 
contains two \EBK\ states. Since both of them 
are in the correct region (the bold square and triangle)
they both produce peaks with almost the usual amplitude,
which can be clearly seen by considering $--$ parity only,
i.e.\ by looking at the billiard in the quarter of an ellipse.
In \fig~\ref{fig:EBKBT} we instead observe that the
$(0,2)_{++}$ peak has almost doubled amplitude:
this is the result of the assumed degeneracy of states 
in the region $\tR$.
So here the \EBKBT\ quantization produces one state too
much. The corresponding peaks in the spectrum are called 
``large spurious peaks'' in \cite{Joyeux95}.
\begin{Filesave}{bilder}
\def\figebkbt{Comparison of the exact
    quantum mechanical density of states (dotted line) to the semiclassical 
    density $\sdsg$ calculated from \EBKBT\ quantization (solid line).
    The labels refer to the exact peaks. 
    \EBKBT\ peaks connected to parity boxes with the wrong number 
    of states (see \fig~\ref{fig:qcell}) do not give satisfactory results.}
\def\FIGebkbt{\centerline{\psfig{figure=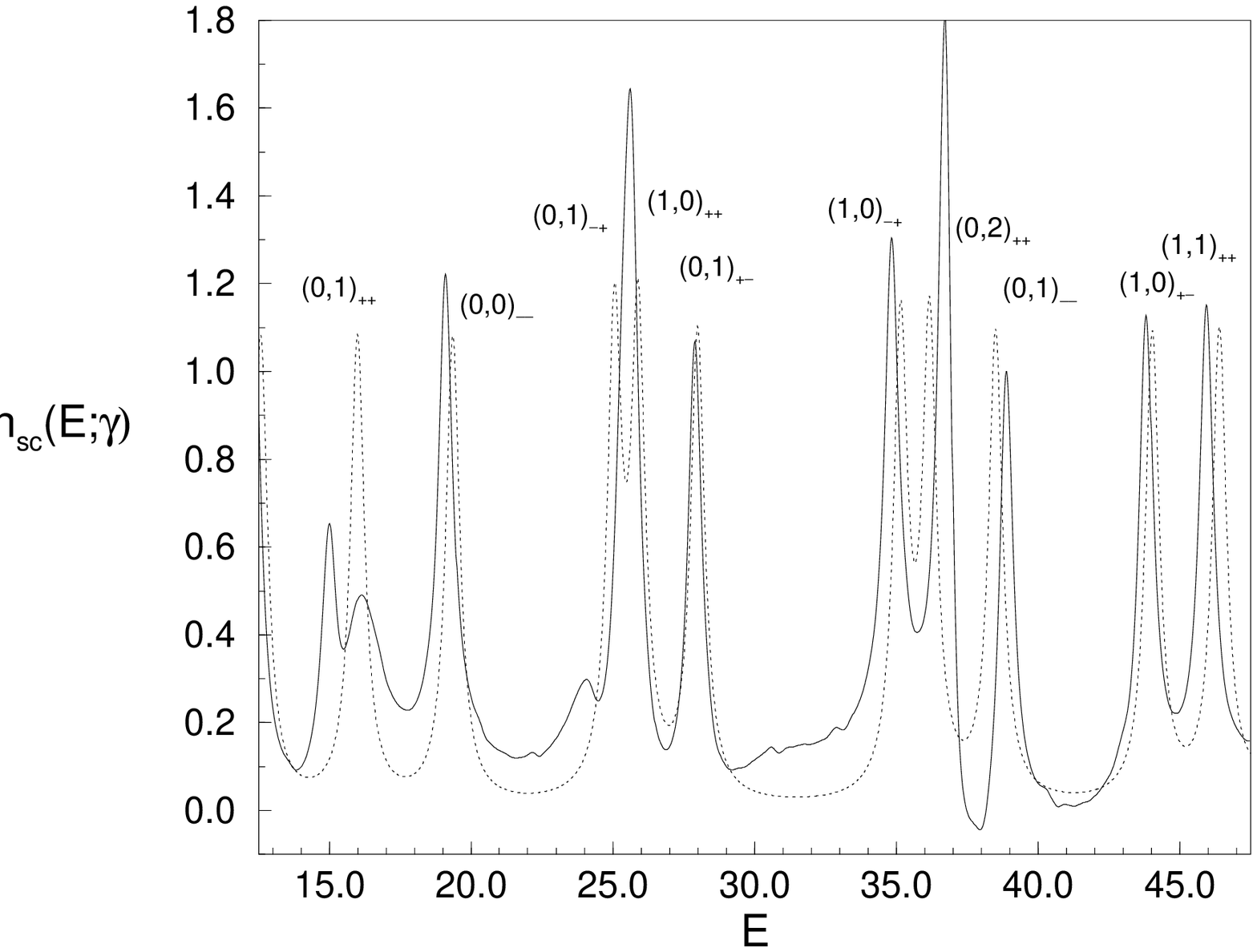,width=13.0cm}}}
\FIGo{fig:EBKBT}{\figebkbt}{\FIGebkbt}
\end{Filesave}
\rem{
\begin{Filesave}{bilder}
\begin{figure}
  \centerline{\psfig{figure=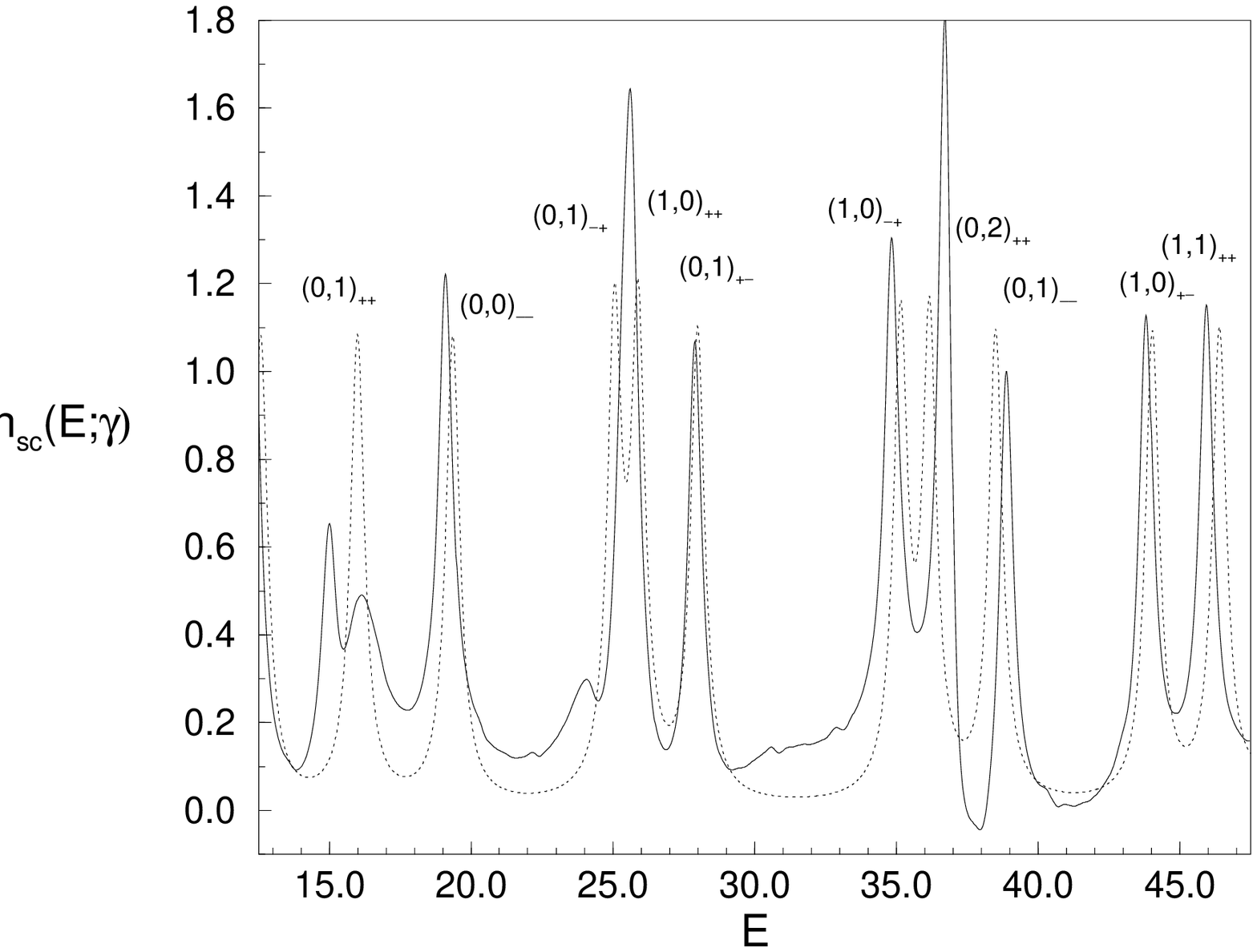,width=13.0cm}}
  \caption[]{\label{fig:EBKBT} \capsty Comparison of the exact
    quantum mechanical density of states (dotted line) to the semiclassical 
    density $\sdsg$ calculated from \EBKBT\ quantization (solid line).
    The labels refer to the exact peaks. 
    \EBKBT\ peaks connected to parity boxes with the wrong number 
    of states (see \fig~\ref{fig:qcell}) do not give satisfactory results. 
}
\end{figure}
\end{Filesave}
}
\begin{Filesave}{bilder}
\def\figWKBBT{Comparison of the exact
    quantum mechanical density of states (dotted line) to the semiclassical 
    density $\sdsg$ calculated with 
    \WKBBT\ quantization (solid line). \WKBBT\ peaks corresponding
    to parity boxes with no \EBK\ state at all ($(0,1)_{++}$ and $(0,1)_{-+}$)
    are improved as compared to \fig~\ref{fig:EBKBT}. The
    parity box $(0,1)_{--}$ with two \EBK\ states still produces two peaks.}
\def\FIGWKBBT{\centerline{\psfig{figure=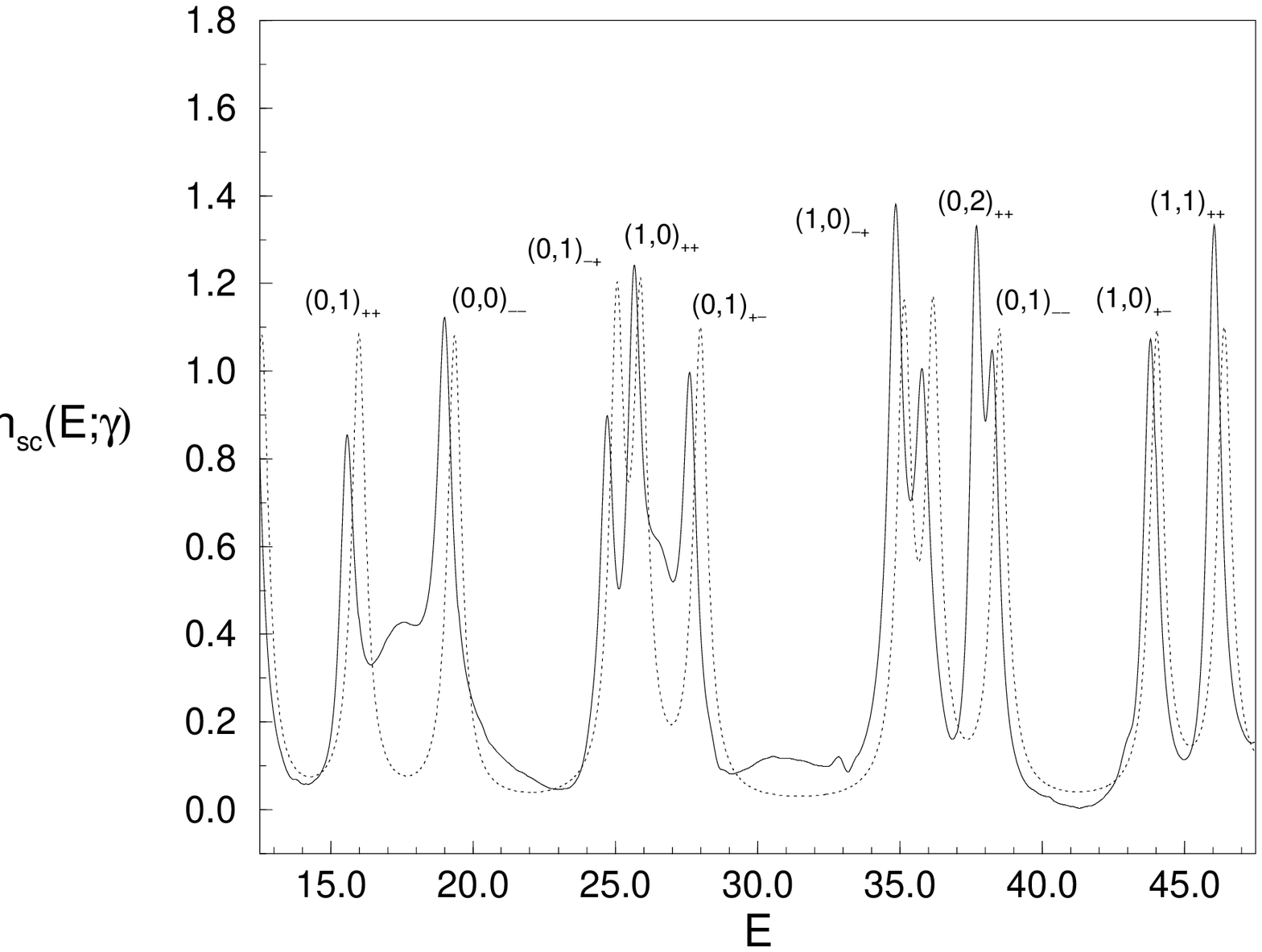,width=12.0cm}}}
\FIGo{fig:WKBBT}{\figWKBBT}{\FIGWKBBT}
\end{Filesave}
\rem{
\begin{Filesave}{bilder}
\begin{figure}
  \centerline{\psfig{figure=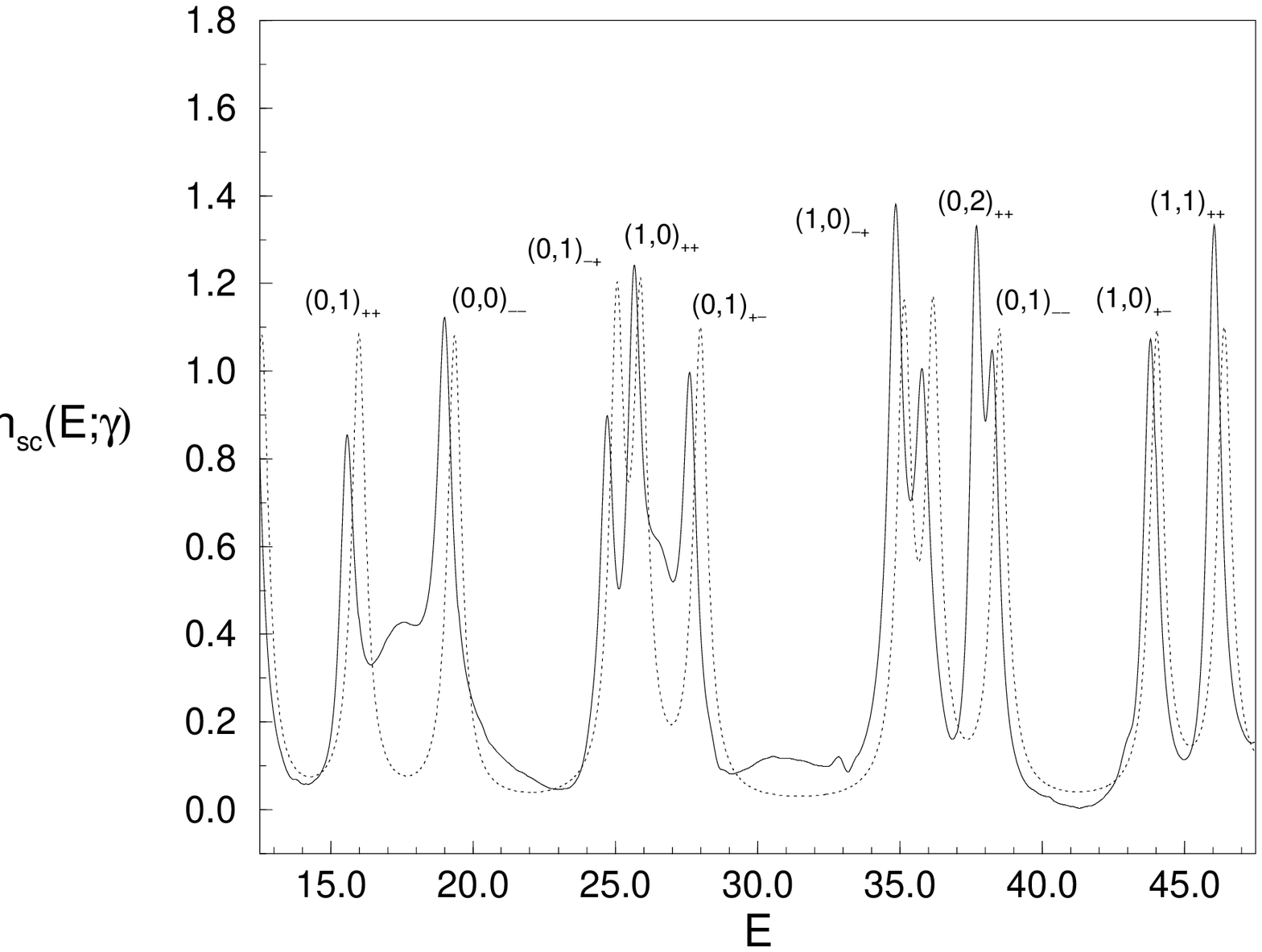,width=12.0cm}} 
  \caption[]{\label{fig:WKBBT} \capsty Comparison of the exact
    quantum mechanical density of states (dotted line) to the semiclassical 
    density $\sdsg$ calculated with 
    \WKBBT\ quantization (solid line). \WKBBT\ peaks corresponding
    to parity boxes with no \EBK\ state at all ($(0,1)_{++}$ and $(0,1)_{-+}$)
    are improved as compared to \fig~\ref{fig:EBKBT}. The
    parity box $(0,1)_{--}$ with two \EBK\ states still produces two peaks.
    } 
\end{figure}
\end{Filesave}
}%

To remove these problems the tunneling through the potential barriers 
should be incorporated by a summation over imaginary complex orbits. 
For one-degree-of-freedom-systems
the consideration of tunneling orbits can be done in a simple way
\cite{Miller79,RCL89}. For systems with more degrees of freedom the
problem is rather involved. In particular a Green's function approach in
terms of action angle variables is still out of reach except for
special systems \cite{Gutz90}. 
Thus we again start with \equ~(\ref{eq:sds}) as Berry and Tabor did. 
At the heart of their approach are the \EBK\ quantization conditions of 
the form~(\ref{eq:EBK}) which allow for a resummation of the
semiclassical density of states via Poisson summation. At this stage
we allow for a varying $\Bmaslovf$ in \equ~(\ref{eq:EBK}), i.e.\ we
base our calculations on a uniform quantization condition. Note that
this uniformization is completely different from the uniformization of
Berry and Tabor, which involves only real complex tori. We incorporate 
both approaches, with the result that all kinds of classically forbidden
tori described in \Sec~\ref{sec:classic} are taken into account.

An effective Maslov phase $\Bmaslovd(\BId;\pi_x,\pi_y)$ 
varying smoothly on 
the energy surface can be extracted from Equations~(\ref{eq:cond1a}-\ref{eq:cond2b}). 
Inserting the \EBK\ quantization condition (\ref{eq:EBK}) into these
Equations the quantum number $n$
drops out because it appears as $2\pi n$ in the argument of the trigonometric 
functions. Thus we obtain the effective Maslov phases as
\bege\label{eq:alpharad}
\tilde{\alpha}_\rad(\BId;\pi_x,\pi_y) = \pi_y \frac{2}{\pi}
\arctan{(e^{\Theta_\rad(\Idrad,\Idang)/\hbar})}+3-\pi_y
\ende
and
\bege\label{eq:alphaang}
\tilde{\alpha}_\ang(\BId;\pi_x,\pi_y) = \pi_y \frac{2}{\pi}
\arctan{(e^{\Theta_\ang(\Idrad,\Idang)/\hbar})}+2-\pi_y-\pi_x
\ .
\ende
Then we introduce new variables, behaving like continuous quantum numbers
\begin{equation}\label{eq:newI}
\Bn = \frac{1}{\hbar}\BId-\frac{1}{4}\Bmaslovd(\BId)
\end{equation}
for each parity. 
The energy surfaces in these variables (one for each parity), 
which now depend on the energy through $\Bmaslovd$, 
are shown in \fig~\ref{fig:effesurf}. The energy surfaces with
equal $\pi_y$ have the same shape, but differ in a
horizontal shift by the constant $1/2$. For $\hbar \to 0$
$\Bmaslovd$ becomes a step function such that except for a shift
and the parity splitting the classical energy surfaces are reobtained 
in the semiclassical limit.
The quantization conditions for these variables 
read $\Bn \in \N^2$ and are trivial.
In \Sec~\ref{sec:semiclassic} we studied how the lattice of states
changes across the separatrix, while the energy surface was the
same for all energies. Now we turn the point of view and introduce
new variables $\Bn$ giving a trivial lattice but a more complicated
energy surface instead. The lines of constant $\tilde{n}_\rad,\tilde{n}_\ang$ in the usual
action variables are shown in \fig~\ref{fig:statescmp}.

Since the quantization conditions in the new variables are of \EBK\ type,
we can repeat the derivation of the trace formula by Berry and Tabor.
The phase being approximated in this derivation is $\Brtopo \Bn$.
Without separatrix the only non constant part in $\Bn$ is $\BId$,
and the stationary phase condition leads to resonant tori.
In order to keep the summation over resonant tori we must
assume that $\Bmaslovd(\BId;\pi_x,\pi_y)$ varies slowly in action space,
which is also the approach followed in \cite{Sieber96}.
This gives a trace formula with two minor modifications:
Firstly each sum is subdivided into separate sums for each parity. 
Secondly $\Bmaslovd$ now is different for every resonant torus. 
We call this method \WKBBT\ quantization. 
Figure~\ref{fig:WKBBT} shows the \WKBBT\ spectrum for 
the same parameters as above. The states $(0,1)_{++}$ and
$(0,1)_{-+}$ (parity cells without \EBK\ state) are reconstructed in a 
satisfactory manner. 
The $(0,1)_{--}$ state (parity cell with two \EBK\ states) 
has improved, but it still contains two contributions, which is
due to the fact that we assume $\Bmaslovd$ to be constant in
the stationary phase approximation, which is a particularly bad
assumption in the neighborhood of the separatrix.
\def\figeffesurf{The energy surfaces in
    the variables $\Bn$ for the energies $25$, $300$, $1026.76$. In
    contrast to \fig~\ref{fig:statescmp} here the shape of the energy
    surfaces depends on the energy but the semiclassical states form a
    regular lattice. For the energy $E=1026.76$ the almost
    degeneracy of the semiclassical states $(0,16)_{+-}$ and
    $(0,16)_{-+}$ is shown.}
\def\FIGeffesurf{\centerline{\psfig{figure=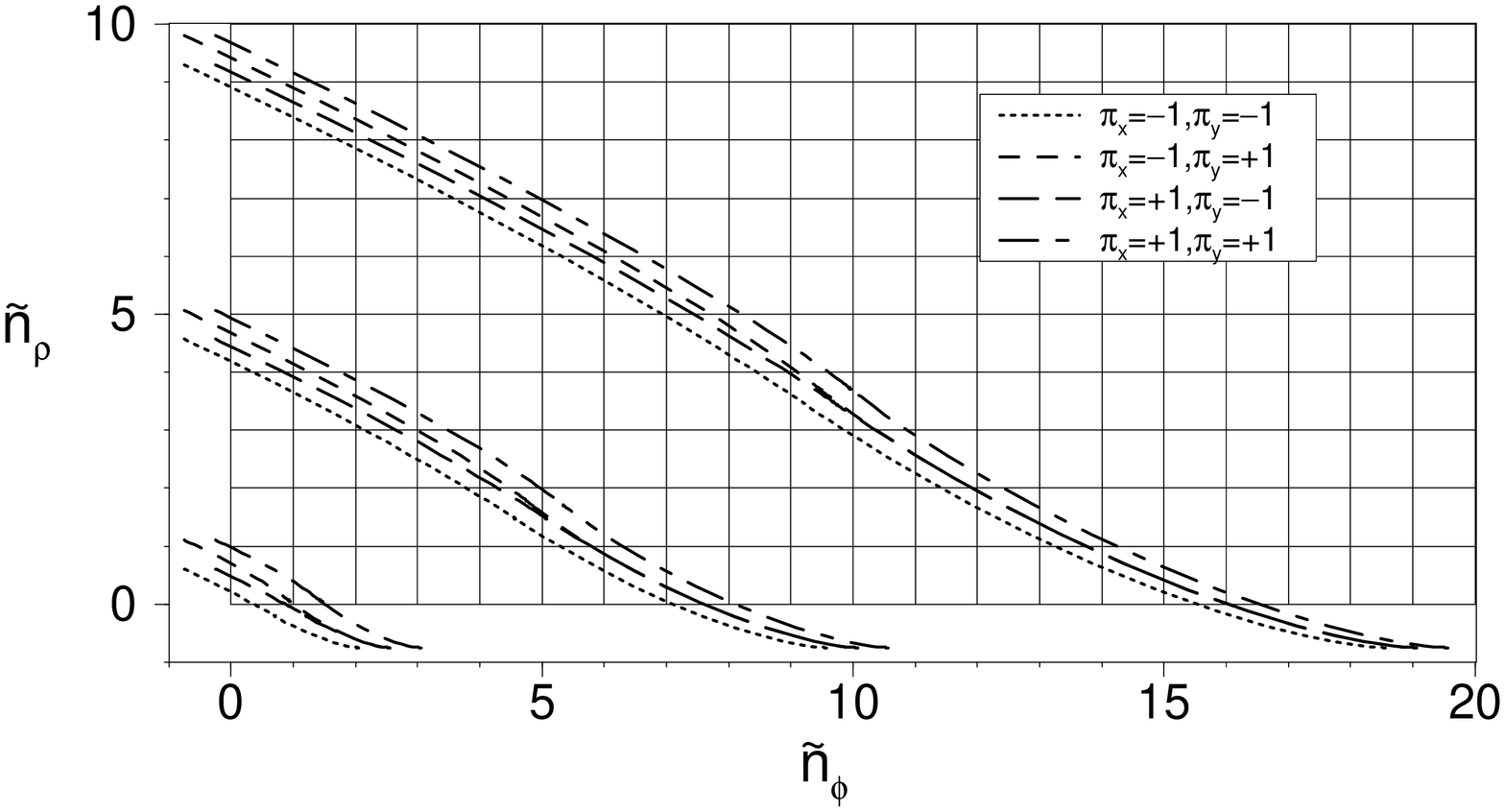,angle=-90,width=13.0cm}}}
\FIGo{fig:effesurf}{\figeffesurf}{\FIGeffesurf}
\rem{
\begin{figure}
  \centerline{\psfig{figure=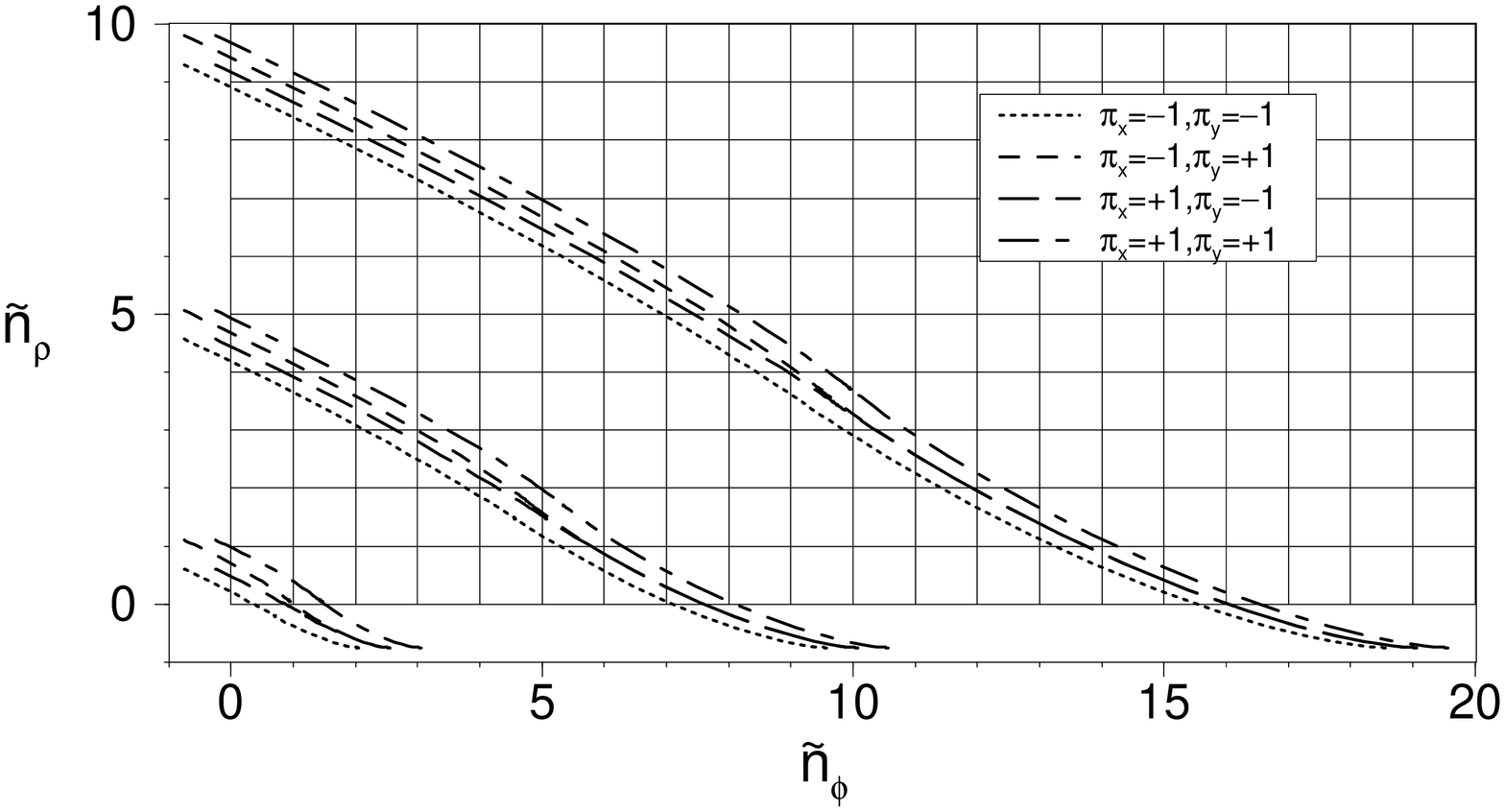,angle=-90,width=13.0cm}} 
  \caption[]{\label{fig:effesurf} \capsty The energy surfaces in
    the variables $\Bn$ for the energies $25$, $300$, $1026.76$. In
    contrast to \fig~\ref{fig:statescmp} here the shape of the energy
    surfaces depends on the energy but the semiclassical states form a
    regular lattice. For the energy $E=1026.76$ the almost
    degeneracy of the semiclassical states $(0,16)_{+-}$ and
    $(0,16)_{-+}$ is shown.}
\end{figure}
}%

Thus we are finally led to the necessity to fully take into account
the variation of $\Bmaslovd$. In essence this means to take the
surfaces in \fig~\ref{fig:effesurf} as new energy surfaces, and
define all quantities (most notably the winding number) with respect
to them. Most important, the stationary points are now given by
\begin{equation}\label{eq:newstationary}
\Brtopo \frac{\partial \BId}{\partial \kappa^2} = \frac{\hbar}{4}\Brtopo
\frac{\partial \Bmaslovd}{\partial \kappa^2} 
= \frac{\Theta'}{\cosh \Theta} (\rtopo_\ang - \rtopo_\rad)
\,.
\end{equation}
Although far away from the separatrix $|\Theta|$ is large and
this condition almost reduces to
the ordinary resonance condition, in general
Equations~(\ref{eq:bt}) and (\ref{eq:btuniform}) are no
longer summations over resonant tori and consequently the resulting 
semiclassical density of states is not determined by periodic orbits.
Instead the tori to be taken into account in the sum are given
by the rational values of a new effective winding number
\begin{equation}
  \wrfn(E,\kappa) = -\frac{\partial \tilde{n}_\rad}{\partial \kappa^2} \Big/ 
        \frac{\partial \tilde{n}_\ang}{\partial \kappa^2} \,,
\end{equation}
such that the solutions of \equ~(\ref{eq:newstationary}) are given by
$\wrfn(E,\kappa) = \rtopo_{\ang}/\rtopo_{\rad}$. 
The overall structure of the trace formula does not change, however,
the frequencies and the curvature have to be replaced by the respective
expressions obtained from the surface $\Bn(\kappa)$.
An important difference between the winding numbers $\wrf$ and $\wrfn$ is 
that the latter depends on the energy. 
This leads to an enormous numerical effort
for the calculation of $\sdsg$, because it is necessary to determine
the stationary points for every energy separately.
We call this method \uniWKBBT\ quantization in contrast to the
\WKBBT\ quantization, 
since it neglects the varying $\Bmaslovd$ 
in the stationary phase approximation. 
The \uniWKBBT\ spectrum calculated with the same parameters as above
is shown in \fig~\ref{fig:uniWKBBT}. 
As expected the splitting of the $(0,1)_{--}$ state is now removed. 
In the evaluation of the sum we define the boundaries
of the energy surfaces as $\Bn(\kappa = 0)$ and $\Bn(\kappa =
1)$. This choice is somewhat arbitrary since one could also
define the boundaries as $\tilde{n}_\rad = 0$ and $\tilde{n}_\ang = 0$. 
But we want to focus onto states close to the separatrix 
where this arbitrariness is not relevant.  
\NEU 
The little peaks in \fig~\ref{fig:uniWKBBT} are artefacts which could
be removed by including larger $\Brtopo$ in the sum.
\begin{Filesave}{bilder}
\def\figuniWKBBT{Comparison of the 
     \WKBBT\ density of states (dotted line) to the density
    calculated with the \uniWKBBT\ quantization
    (solid line). The density obtained from
    the energy eigenvalues calculated in \Sec~\ref{sec:semiclassic}
    (dashed) gives the expected location of the peak, which is
    slightly displaced from the peak in the exact quantum mechanical density 
    (long dashed). The main point is that  
    the splitting of the $(0,1)_{--}$ peak is removed.}
\def\FIGuniWKBBT{\centerline{\psfig{figure=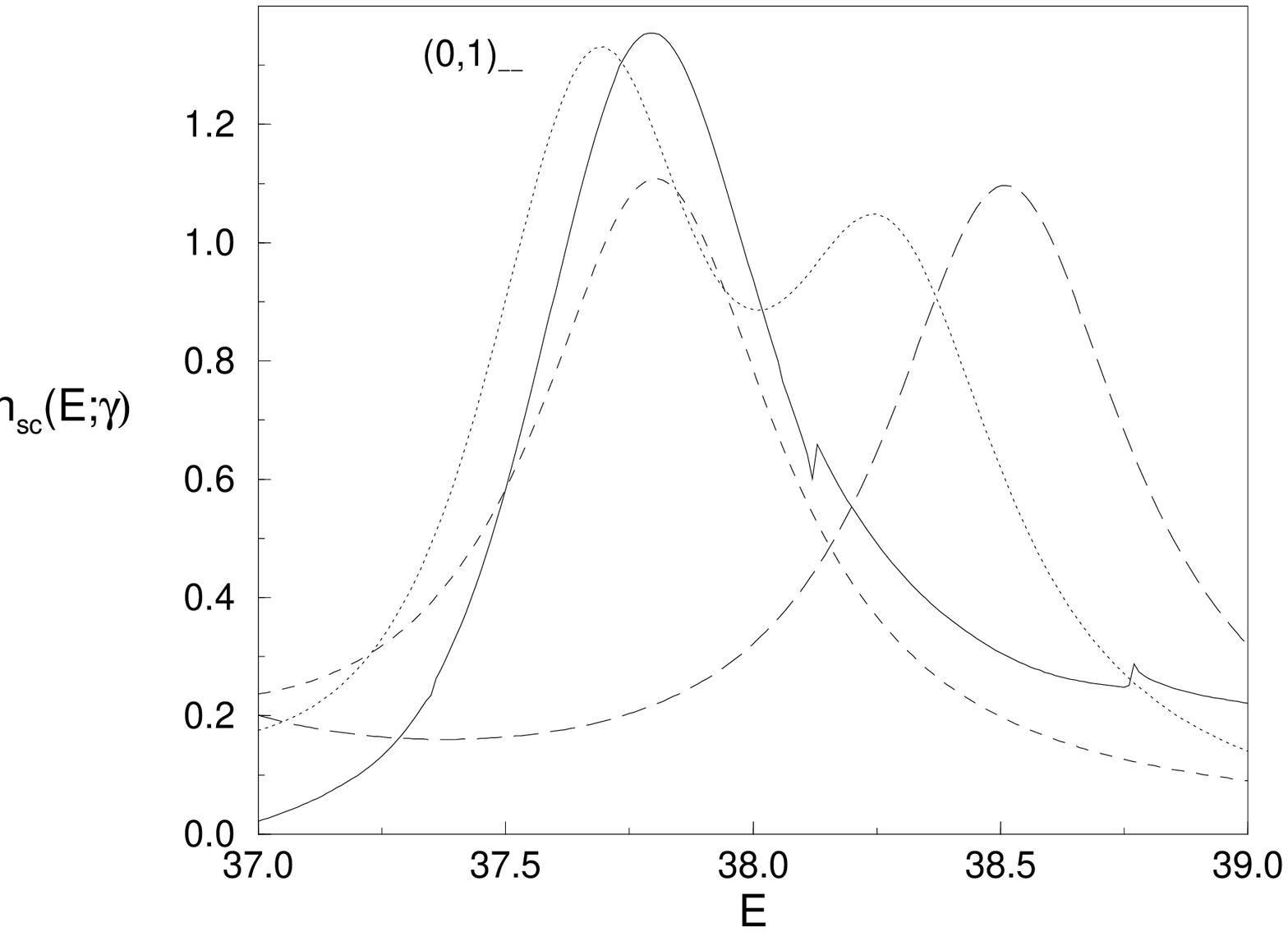,width=12.0cm}}}
\FIGo{fig:uniWKBBT}{\figuniWKBBT}{\FIGuniWKBBT}
\end{Filesave}
\rem{
\begin{Filesave}{bilder}
\begin{figure}
  \centerline{\psfig{figure=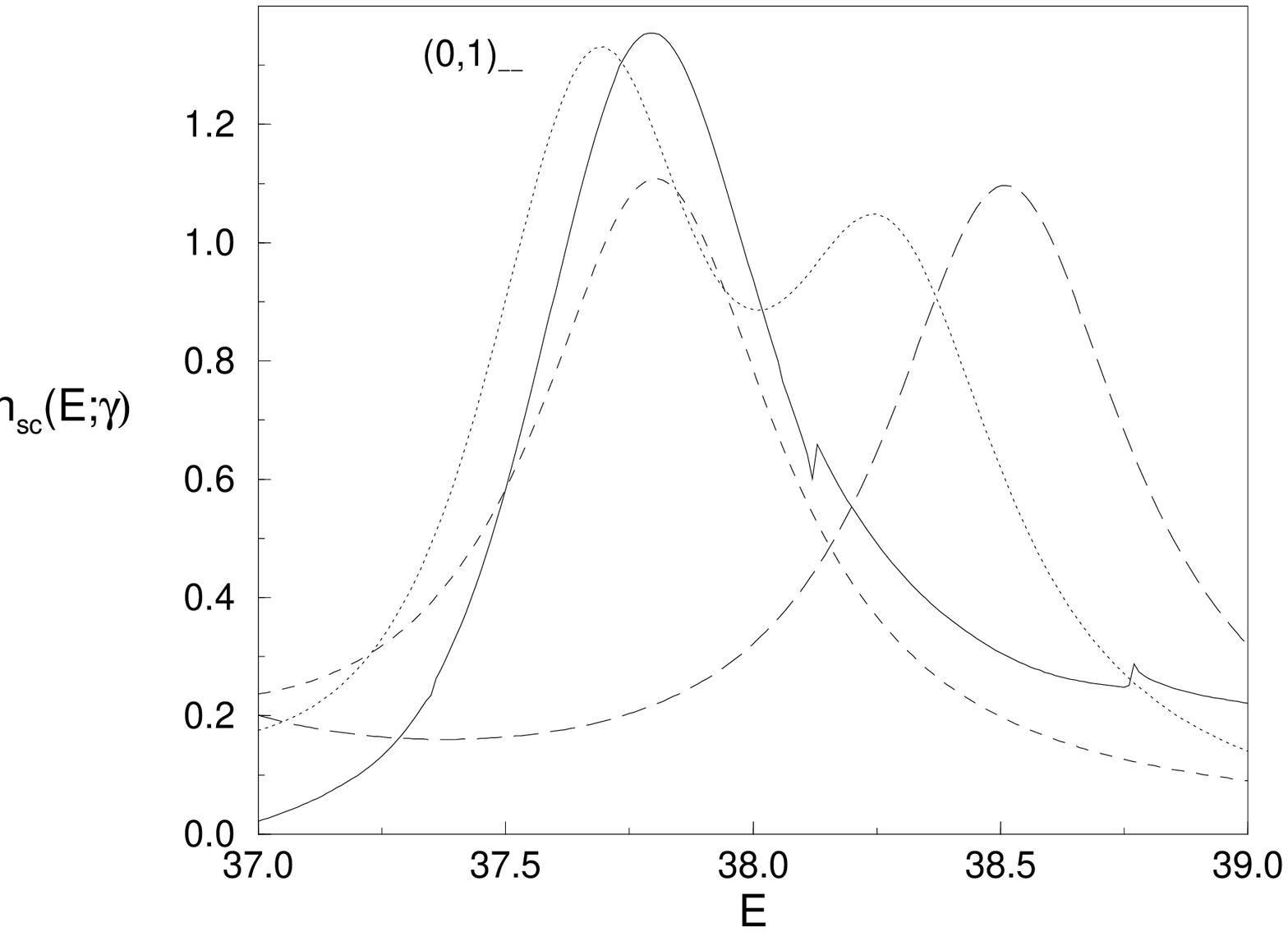,width=12.0cm}} 
  \caption[]{\label{fig:uniWKBBT} \capsty Comparison of the 
     \WKBBT\ density of states (dotted line) to the density
    calculated with the \uniWKBBT\ quantization
    (solid line). The density obtained from
    the energy eigenvalues calculated in \Sec~\ref{sec:semiclassic}
    (dashed) gives the expected location of the peak, which is
    slightly displaced from the peak in the exact quantum mechanical density 
    (long dashed). The main point is that  
    the splitting of the $(0,1)_{--}$ peak is removed.} 
\end{figure}
\end{Filesave}
}
The derivation of the effective Maslov phase 
(\ref{eq:alpharad}) and (\ref{eq:alphaang}) 
makes it obvious that 
in the uniform $\Bmaslovd$ the effects of tunneling and scattering orbits 
are incorporated into the Berry-Tabor sum.
Now we want to show that the consideration of an
effective $\Bmaslovd$ is equivalent to an explicit incorporation
of imaginary complex orbits with constant $\Bmaslovd$.


We consider two crossed double well potentials, which is slightly more 
general than the elliptic billiard since the barrier penetration
integrals $\Theta_1$ and $\Theta_2$ are then unrelated.
The effective $\Bmaslovd$ obtained by the method described in
\Sec~\ref{sec:semiclassic} can always be written as
 $\tilde{\alpha}_i = -2\zeta_i/\pi+\tilde{\alpha}_{ci}$, with $\zeta_i =
\arctan{\exp{(-\Theta_i)}}$ and $\Bmaslovd_c = (\tilde{\alpha}_{c1}, \tilde{\alpha}_{c2})$
is a constant integer vector. We
rewrite the cosine term in \equ~(\ref{eq:bt}) with
$x=q(\action/\hbar-\pi/2{\tilde{\bm \alpha}_c}{\Brtopo})+\pi/4\beta$ and $q_i
= q\rtopo_i > 0$ as the real part of
\begin{equation}\label{eq:ghostsum}
  e^{i(x+q_1\zeta_1+q_2\zeta_2)} =
  e^{ix}(r_1+it_1)^{q_1}(r_2+it_2)^{q_2}
\end{equation}
with 
\begin{equation}\label{eq:riti}
  r_i = \cos\zeta_i = \frac{1}{\sqrt{1+e^{-2\Theta_i}}} \;\; \mbox{ and }\;\;  
  t_i = \sin\zeta_i = \frac{1}{\sqrt{1+e^{2\Theta_i}}} \,.
\end{equation}
Now $r_i$ and $t_i$ can be interpreted as the absolute values of the
refection and transmission coefficients of the potential barriers for
one reflection or transmission (\cite{Ozorio84,RCL90}),
respectively.
Expanding the last term of \equ~(\ref{eq:ghostsum}) gives
\begin{equation}\label{eq:ghostsum2}
  e^{ix}\sum_{\stackrel{k_1,l_1=1}{\scriptscriptstyle{k_1+l_1=q_1}}}^{q_1} {q_1 \choose k_1} i^{l_1} r_1^{k_1}
  t_1^{l_1}\sum_{\stackrel{k_2,l_2=1}{\scriptscriptstyle{k_2+l_2=q_2}}}^{q_2} {q_2 \choose k_2}i^{l_2} r_2^{k_2} t_2^{l_2} \,,
\end{equation}
where the summations go above all permutations of $r_1$, $t_1$ and
$r_2$, $t_2$, i.e. we suppress the binomial coefficients. 
Finally we expand \equ~({\ref{eq:ghostsum2}) and take the real part
  to give 
\begin{equation}\label{eq:ghostsum3} 
  \sum_{\stackrel{k_1+l_1=q_1}{\scriptscriptstyle {k_2+l_2=q_2}}}
  {q_1 \choose k_1}{q_2 \choose k_2} \cos{(x+\frac{\pi}{2}l_1+\frac{\pi}{2}l_2)} \, r_1^{k_1}
  t_1^{l_1} r_2^{k_2} t_2^{l_2} \, .
\end{equation}
Again the summation is above all permutations of $r_1$, $t_1$ and
$r_2$, $t_2$. 
The phase in the cosine term in \equ~(\ref{eq:ghostsum3}) can be
interpreted as accumulation of phase shifts of $\pi/2$
resulting from single transmissions. Thus
\equ~(\ref{eq:ghostsum3}) tell us how to extend the
summation~(\ref{eq:bt}) to a summation over
imaginary complex tori. One must incorporate all combinations of real 
and imaginary complex orbits with the same resulting period by
replacing the cosine term in \equ~(\ref{eq:bt}) by \equ~(\ref{eq:ghostsum3}).

\rem{ = R\cos{(x)}-T\sin{(x)}
\end{equation}
with $R = R_1 R_2 - T_1 T_2$, $T = T_1 R_2 + T_2 R_1$, $R_i =
\cos{(q_i\zeta_i)}$ and $T_i = \sin{(q_i\zeta_i)}$. We expand $R_i$ and $T_i$ to
\begin{equation}\label{eq:Ri}
  R_i = \sum_{\stackrel{m_i=0}{\scriptscriptstyle\text{even}}}^{q_i} {q_i \choose m_i}
  r_i^{q_i-m_i} t_i^{m_i} (-1)^{m_i/2}
\end{equation}
and 
\begin{equation}\label{eq:Ti}
  T_i = \sum_{\stackrel{m_i=1}{\scriptscriptstyle\text{odd}}}^{q_i} {q_i \choose m_i}
  r_i^{q_i-m_i} t_i^{m_i} (-1)^{(m_i-1)/2}
\end{equation}
} 

}

\section{The length spectrum}
\label{sec:invproblem}

Instead of calculating the density of states by a summation over
resonant tori one can turn the tables and look at the length spectrum,
i.e.\ the Fourier transform of the oscillating part of the density
of states $n_{\text{osc}}(E)=n(E)-\overline{n}(E)$. In the context of
hyperbolic systems this viewpoint is referred to as {\em inverse
quantum chaology}, see, e.g.~\cite{Wintgen87,BSS95}.

The familiar way to discover the appearance of periodic orbits 
in the quantum mechanical spectrum is to calculate its power spectrum. 
Similar to the phases in the Gutzwiller trace formula, the phases in 
the Berry-Tabor summation over resonant tori 
are proportional to the action of periodic orbit representatives
of the tori. 
The action in a billiard scales with $\sqrt{E}$, 
such that we take the wavenumber $k :=\sqrt{2E}/ \hbar$ as the
integration variable and determine $\left|p(L)\right|^2$ from
\begin{eqnarray}
p(L):= 
   \int_0^{\infty}dk\, \varrho(k) n_{\text{osc}}(E(k))\exp \left( ikL \right)\exp\left(-tk\right)\,.
\label{eqnfourier}
\end{eqnarray}
Here the factor $\varrho(k)=\hbar^2k$ gives the measure with respect
to the wavenumber.
The fading function $\exp(-tk)$ is introduced to
reduce the significance of higher eigenvalues and it has been
chosen in order to make the analytical calculations feasable.
$t$ has to be chosen 
appropriately in order to incorporate the finiteness of the available energy
range. Taking the exact eigenvalues up to $E_{\text{max}}=100\,000$
($\hbar=1$) calculated according to the method described in 
\Sec~\ref{sec:qmechanic} one deals with $n_{\text{max}}=35\,169$ levels 
for the ellipse parameter $a=1/\sqrt{2}$.
Throughout this section we choose $a=1/\sqrt2$ 
for reasons that will become clear below. 
The condition  to be imposed  on $t$ is $t\gg\ln2/k_{\text{max}}$ 
with $k_{\text{max}}=\sqrt{2mE_{\text{max}}}/\hbar$. We found it adequate
to set $t=0.025$. 
\begin{Filesave}{bilder}
\def\figpowerspectrum{Power spectrum
$|p(L)|^2$ for $a=1/\sqrt{2}$. The smaller figure shows a
magnification of the range $[4.0,5.4]$ enclosed by the arrows. The
tick marks above mark resonant tori, the tick marks and labels below mark 
isolated periodic orbits.}
\def\FIGpowerspectrum{\centerline{\psfig{figure=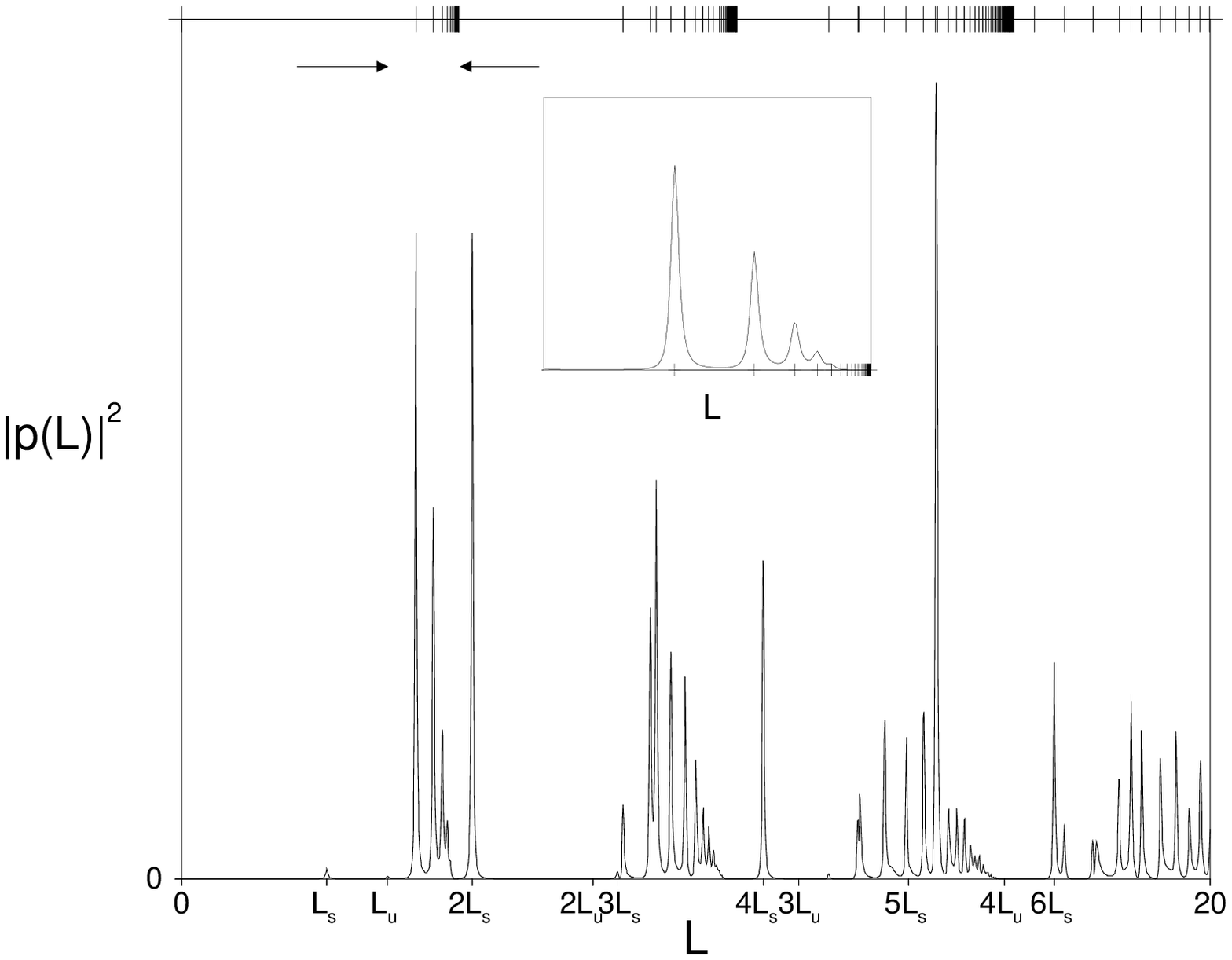,width=12.0cm}}}
\FIGo{fig:powerspectrum}{\figpowerspectrum}{\FIGpowerspectrum}
\end{Filesave}
\rem{
\begin{Filesave}{bilder}
\begin{figure}
  \centerline{\psfig{figure=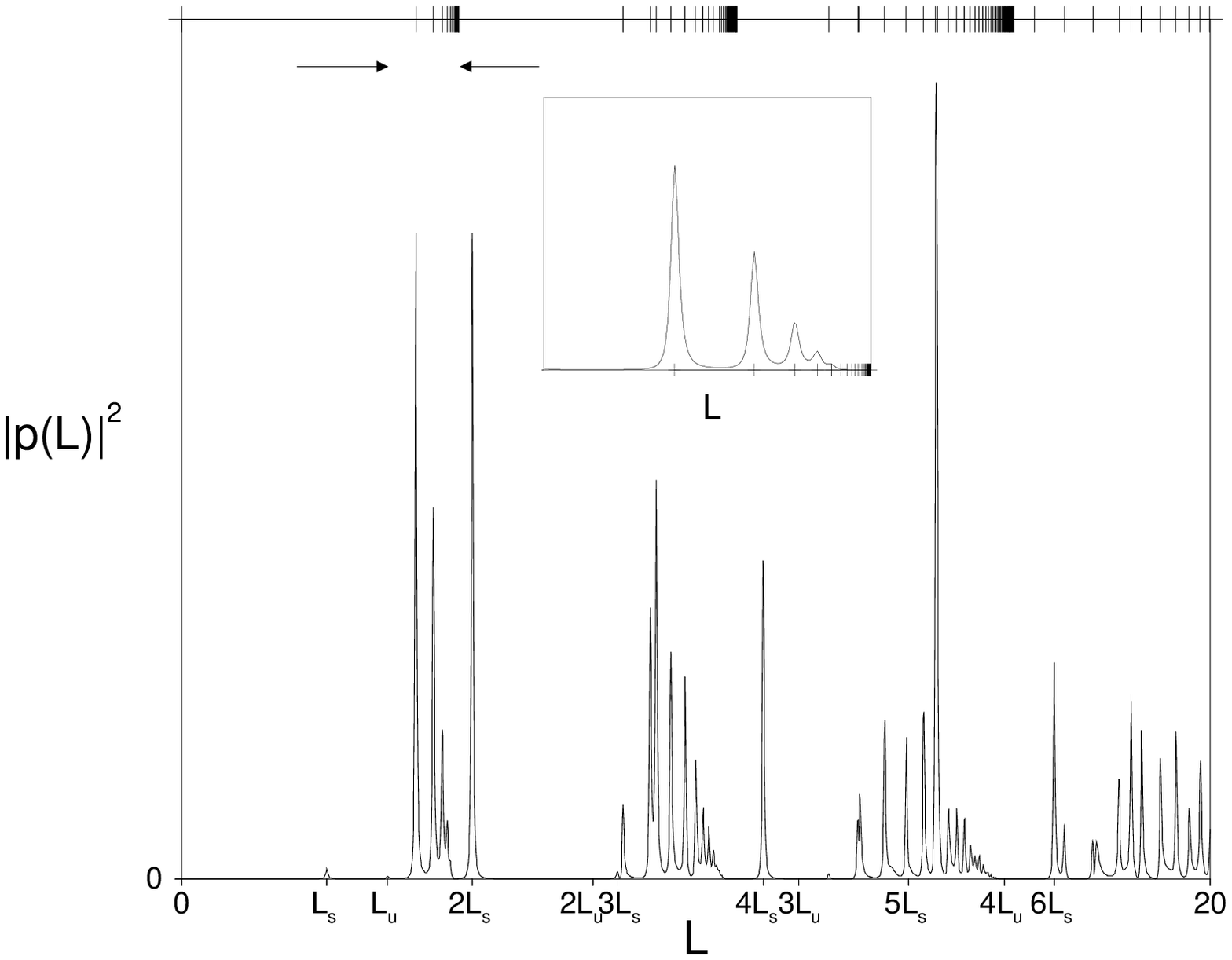,width=12.0cm}}
  \caption[]{\label{powerspectrum} \capsty Power spectrum
$|p(L)|^2$ for $a=1/\sqrt{2}$. The smaller figure shows a
magnification of the range $[4.0,5.4]$ enclosed by the arrows. The
tick marks above mark resonant tori, the tick marks and labels below mark 
isolated periodic orbits.}
\end{figure}
\end{Filesave}
}

Figure~\ref{fig:powerspectrum} shows $|p(L)|^2$ versus the length $L$.
The tick marks above give the lengths of periodic orbit representatives of 
resonant tori. The spectrum shows equally spaced clusters of
contributions. The small enclosed Figure gives a magnification of the
range $[4.0,5.4]$ around the first cluster. Here the peaks
correspond to type $\tR$ resonant tori, with winding number $1/\rtopo_\rad$,
the whispering gallery orbits.
Their lengths
accumulate in the length $\lengthb$ of the sliding orbit along the
billiard boundary $\ob$ that can be considered as the
$\rtopo_\rad \rightarrow\infty$ limit of these orbits. The amplitudes of the
contributions from these orbits decrease with growing number of reflections 
$\rtopo_\rad$, damped by 
$(1+\rtopo_\rad^2)^{-1/2}$,
the diverging curvature, and the diverging frequency, see~(\ref{eq:bt}).
The other clusters in \fig~\ref{fig:powerspectrum} lie
around integer multiples of $\lengthb$. 
Here multiple traversals of the whispering gallery orbits 
with winding number $1/\rtopo_\rad$ and 
new ones with winding number $2/\rtopo_\rad$, $3/\rtopo_\rad$ (coprime) 
etc.\ accumulate. 
Contributions of type $\tO$ tori occur only sparsely 
in the spectrum as compared to the type $\tR$ tori.
They have no accumulation orbit like the type $\tR$ tori. 
The shortest orbits lying on a type $\tO$ resonant torus have
length $L=9.23$ followed by $L=13.18$ and $L=14.67\,$. 
This means that the low part of the  spectrum is dominated by type 
$\tR$ tori and the clusters they produce.
The whispering gallery orbits lead to an infinite number of (families of) 
periodic orbits with finite action, thus violating the generic 
growth behavior of the number of periodic orbits in integrable systems. 
In addition to resonant tori contributions \fig~\ref{fig:powerspectrum}
also  shows peaks  at lengths corresponding to the unstable orbit
$\ou$  and to the stable periodic $\os$ and multiples thereof. 

In the following we will have a closer look at the amplitudes
corresponding to isolated periodic orbits, and to the amplitudes of the
low resonant tori. For this purpose we consider the real part,
i.e.\ the cosine transform, of $p(L)$ 
because it reveals much more information than the absolute value.
Richens showed in \cite{Richens82} that as a limiting case 
the uniform version of the Berry-Tabor summation contains contributions 
of stable isolated periodic orbits equal to the corresponding terms in
Gutzwiller's trace formula \cite{Gutz90}. He also suggests that the
unstable isolated orbits should  appear in a similar way,
which is rigorously shown for the ellipse billiard in \cite{Sieber96}. 
Thus the $q$th traversal of  $\os$  and $\ou$ contribute 
\begin{eqnarray}
\frac{T(E)}{\pi\hbar\sqrt{\left|\det\left({\bf M}^{q}-{\bf 1}\right)\right|}}\cos\left(q\frac{S(E)}{\hbar}-\mu_q \frac{\pi}{2}\right)
\label{eqnperiodcont}
\end{eqnarray}
to the density of states. Here $T(E)$ is the period, $S(E)$ the
action of the periodic orbit, ${\bf M}$ the reduced $2\times2$
monodromy matrix describing its stability 
and $\mu_q$ the Morse index of its $q$th traversal.
For the calculation of Morse indices see e.g.~\cite{CRL90}.
Since for the stable orbit we have $\mu_q = 4q+1+2[qw]$, 
where $[\,]$ denotes the integer part 
it is not immediately possible to factor out $q$.
%
For $\os$ and $\ou$
$ 
\trace {\bf M}=2-4L\left(\kappa_1+\kappa_2\right)+4\kappa_1\kappa_2L^2\,,
$ 
where $L$ is the flight length between two consecutive reflections and
$\kappa_1$ and $\kappa_2$ are the curvatures at the reflection points
with a positive sign in case of a convex billiard boundary
\cite{Dullin96b}. 
For $\os$ we find $L=\lengths/2$ and 
$\kappa_1=\kappa_2=\sqrt{1-a^2}$, 
while for $\ou$ the trace is always larger than $2$ because
$L=\lengthu/2 = 2$ and $\kappa_1=\kappa_2=\left(1-a^2\right)^{-1}$.  
The winding number $w\in [0,1)$ for $\os$ and the stability exponent  $u$ 
for $\ou$ are given by
\begin{eqnarray}
\left.\begin{array}{r}
\exp (\pm i2\pi w)\\
\exp(\pm u)
\end{array}
\right\}
:= \frac{1}{2}\left[\pm \sqrt{(\trace {\bf M})^2-4}+\trace{\bf M}  \right]
\label{eq:stabilityexponents}
\end{eqnarray}
where the corresponding matrix ${\bf M}$ has to be inserted. 
The winding number $w$ obtained from the eigenvalues of the
monodromy matrix ${\bf M}$ is the same as in \equ~(\ref{eq:IWII}). 
Now we can write the stability term in (\ref{eqnperiodcont})
as $\sqrt{\left|\det\left({\bf M}^q-{\bf 1}\right)\right|}=2|\sin(q\pi w)|$ 
or 
$2\sinh(qu/2)$, respectively.
Matching the signs of the two sine functions for $\os$ one can
rewrite (\ref{eqnperiodcont}) as
\renewcommand{\arraystretch}{1.6}
\begin{equation}
\begin{array}{cl}
\displaystyle
\frac{T(E)}{2\pi\hbar}\frac{\sin(qS/\hbar)}{\sin(q\pi w)} 
        & \text{for $\os$, elliptic,} \\
\displaystyle
\frac{T(E)}{2\pi\hbar}\frac{\cos(q(S/\hbar - \mu_{u} \pi/2))}{\sinh(qu/2)} 
        & \text{for $\ou$, hyperbolic,} 
\end{array}
\end{equation}
\renewcommand{\arraystretch}{1.3}
where $\mu_q = q \mu_u$ has been used for $\ou$, with 
$\mu_u = 4+2$.
It is important to notice that the amplitudes of the contributions 
of isolated periodic orbits and of
resonant tori differ in the power of $\hbar$, the former is
proportional to $1/\hbar$, the latter to $1/\hbar^{3/2}$.

A problem arises when the winding number $w$ becomes rational, leading
to a divergent amplitude for the contribution of the corresponding
number of traversals where neighboring trajectories of $\os$ are
closed in phase space. 
In order to study this phenomenon we take $a=1/\sqrt{2}$ in this section, 
slightly different from $a=0.7$ as before. Then the winding ratio for
the stable orbit \equ~(\ref{eq:stabilityexponents}) becomes $w=1/2$.
As worked out by Richens, in this case the thin
resonant torus surrounding the periodic orbit rather than the periodic
orbit alone determines the contribution to the density of
states. Equation  (\ref{eqnperiodcont}) then has to be replaced by 
\begin{equation}
\frac{1}{\hbar ^{3/2}}\frac{T(E)}{2\pi\left(1+w^2\right)^{3/4}\sqrt{|\curv(E)|}}\frac{\cos \left(q\left( S(E)/\hbar-\pi w-
  \frac{\pi}{2} \alpha_\rad\right) +\frac{\pi}{4}\beta\right)}{\sqrt{q}}\,.
\label{thintoruscontribution}
\end{equation}
The amplitude is again proportional to $1/\hbar^{3/2}$ signaling a
torus contribution.\\ 
Inserting the semiclassical results for $n_{\text{osc}}(E)$ into
\equ~(\ref{eqnfourier}) and taking the real part we obtain 
\begin{eqnarray}
\int_0^{\infty}dk\,\varrho(k)n_{\text{osc}}(E(k))\cos(kL)\exp(-tk)\approx\sum_{\text{p.obj.}}A_{\text{p.obj.}}(L)
\label{equationfouriertranscomp}
\end{eqnarray}
where the summation on the right hand side runs over all ``periodic objects'',
i.e.\ resonant tori, isolated periodic orbits and thin resonant tori in
cases where the isolated stable periodic orbits become
resonant. Taking the fixed \EBK\ phases in \equ~(\ref{eq:bt})  all
the different kinds of contributions $A_{\text{p.obj.}}$ can be
calculated analytically. 
The scaling properties of the action variables in 
equations (\ref{eq:IWI}) and (\ref{eq:IWII}), the amplitudes in equations
(\ref{eq:bt}), (\ref{eqnperiodcont}), and (\ref{thintoruscontribution}) 
allow for a scaling with respect 
to the wavenumber $k$. Hence the semiclassical results
together with (\ref{eqnfourier}) 
lead to a summation over integrals of the form
\begin{eqnarray}
\tilde{A}(L):= A \int_0^{\infty}dk\,\cos\left(\tilde{L}k+n\frac{\pi}{4}\right)\cos\left( L k\right)\exp\left(-tk\right)k^{\sigma}
\label{eqnintegralfunction}
\end{eqnarray}
with $n\in\Z,\, A,\tilde{L},L,t\in \R,\,t>0$ and $\sigma=0,1/2$.
Defining the functions
\begin{eqnarray} \label{eqnfunctiondef}
\fcos(L):=&\quad&\frac{\Gamma(\sigma+1)}{\left((\tilde{L}-L)^2+t^2\right)^{\frac{\sigma +1}{2}}}\cos\left((\sigma+1)\arctan\right(\frac{\tilde{L}-L}{t}\left)\right)\\
&+&\frac{\Gamma(\sigma+1)}{\left((\tilde{L}+L)^2+t^2\right)^{\frac{\sigma +1}{2}}}\cos\left((\sigma+1)\arctan\right(\frac{\tilde{L}+L}{t}\left)\right)
  \nonumber
\end{eqnarray}
and similarly $\fsin(L)$ by replacing cosine by sine in 
\equ~(\ref{eqnfunctiondef}) the integrals 
for $\tilde{A}(L)$ 
can be solved analytically \cite{GradRyzh65} and 
are listed in \tab~\ref{tableintegralfunction}.
\begin{table}[ht]
\tabstart
\centering
\begin{tabular}{|c|c||c|c|}
\hline
$n$ & $\tilde{A}(L)$ & $n$ & $\tilde{A}(L)$ \\ \hline
$0$ & $A\fsin(L)$ & $\pm 4$ & $-A\fsin(L)$ \\  
$\pm1$ & $\frac{A}{\sqrt{2}}\left(\fsin(L)\mp \fcos(L)\right)$ & $\pm 5$ & $\frac{A}{\sqrt{2}}\left(-\fsin(L)\pm  \fcos(L)\right)$ \\  
$\pm2$ & $\mp A\fcos(L)$ & $\pm 6$ & $\pm A\fcos(L)$ \\  
$\pm3$ & $\frac{A}{\sqrt{2}}\left(-\fsin(L)\mp \fcos(L)\right)$ & $\pm 7$ & $\frac{A}{\sqrt{2}}\left( \fsin(L) \pm \fcos(L)\right)$ \\ \hline
\end{tabular}
\tabend
\caption[]{\capsty Table of the results for $\tilde{A}(L)$.}
\label{tableintegralfunction}
\end{table}
The classical quantities $\BIf^\rtopo$, $\bm \omega$, and $\curv$ are 
understood to be calculated for $E=1$. 
Then the mapping of the parameters is given by:
\begin{itemize}
\item 
\begin{equation}
\left(\tilde{L},A,n,\sigma\right) = \left(
2\pi q \Brtopo \BIf^\rtopo,
\frac{2\epsilon}{\sqrt{|{q \Brtopo}|}|{\bm \omega}|\sqrt{|\curv|}},
\left(-2q {\bm \alpha}\Brtopo+\beta\right) \,\,\mbox{mod}\,8
,1/2 \right)
\end{equation}
for the resonant torus contributions. The degeneracy 
factor $\epsilon$ is $2$ for type $\tR$ tori and $1$ for type $\tO$ tori.
\item
\begin{equation}
\left( \tilde{L},A,n,\sigma \right)
=\left(
q\lengths,
\frac{\lengths}{2\pi\sqrt{q}(1+w^2)^{3/4}\sqrt{|\curv|}},
\left(-4qw-2q\alpha_\rad +\beta\right) \,\,\mbox{mod}\,8
,1/2 \right)
\end{equation}
for the thin resonant torus case where $q w$ is an integer.
\item
\begin{equation}
\left( \tilde{L},A,n,\sigma \right)
=\left(
q\lengths,
\frac{\lengths}{2\pi\sin(q\pi w)}, -2 
,0 \right)
\end{equation}
for the stable isolated periodic orbit case with $qw$ not an integer.
\item
\begin{equation}
\left( \tilde{L},A,n ,\sigma\right)
=\left(
q\lengthu,
\frac{\lengthu}{2\pi\sinh(q u / 2)},
\left(-2q \mu_u\right) \,\,\mbox{mod}\,8
,0 \right)
\end{equation}
for the unstable isolated periodic orbit case with positive trace. 
\end{itemize}
\begin{Filesave}{bilder}
\def\figfouriertransformed{Comparison of
the exact to the semiclassical result for $\Re(p(L))$ ($a=1/\sqrt{2}$). 
The solid line is the cosine transform of the exact $n_{\text{osc}}(E)$, the
dotted line marks the semiclassical curve. (b),(c) and (d) give
magnifications of  the
length  ranges $[2.5,5.4]$, $[5.4,10.8]$ and $[10.8,16.2]$, respectively.}
\def\FIGfouriertransformed{\centerline{\psfig{figure=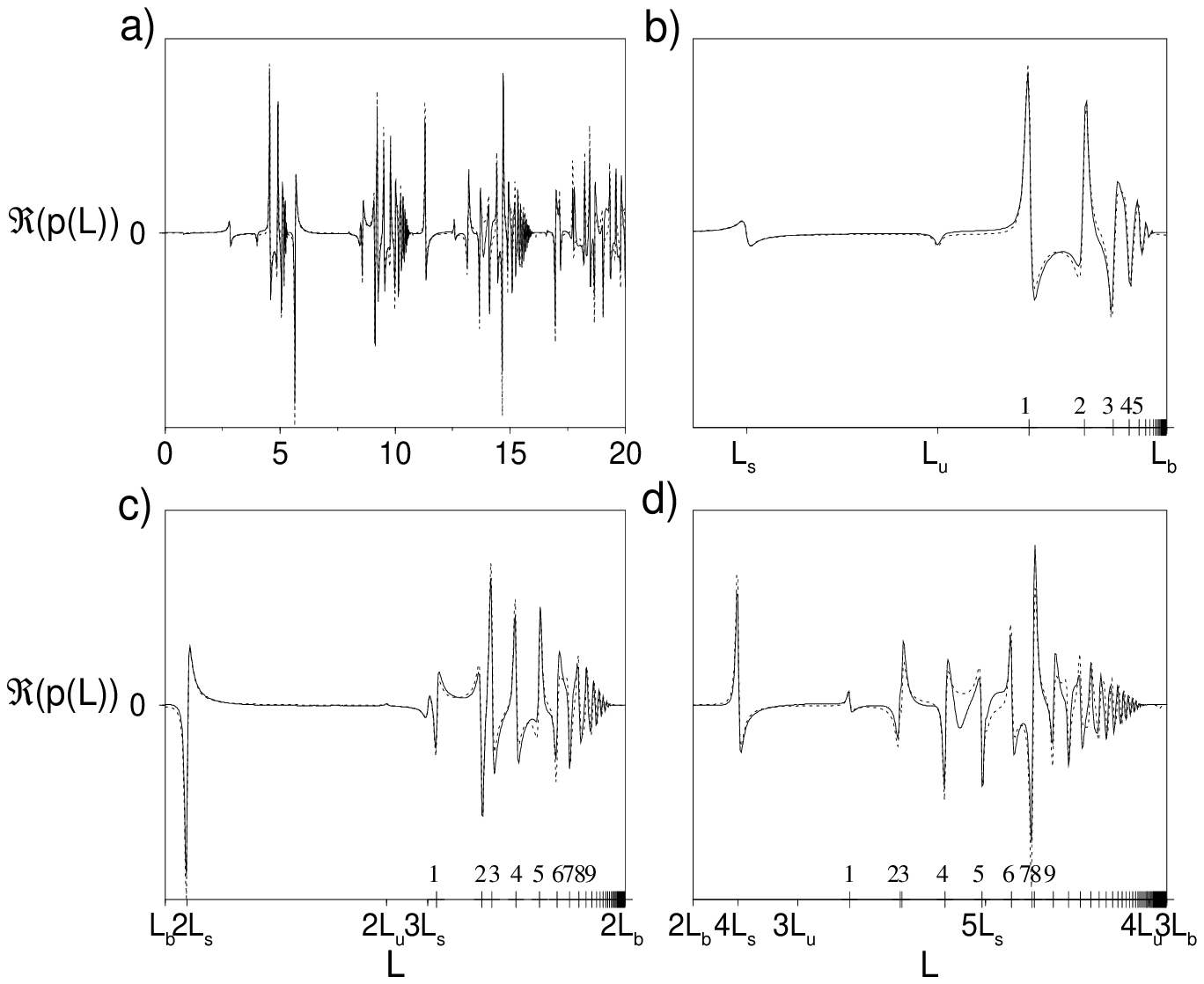,width=12.0cm}}}
\FIGo{fig:fouriertransformed}{\figfouriertransformed}{\FIGfouriertransformed}
\end{Filesave}
\rem{
\begin{Filesave}{bilder}
\begin{figure}
  \centerline{\psfig{figure=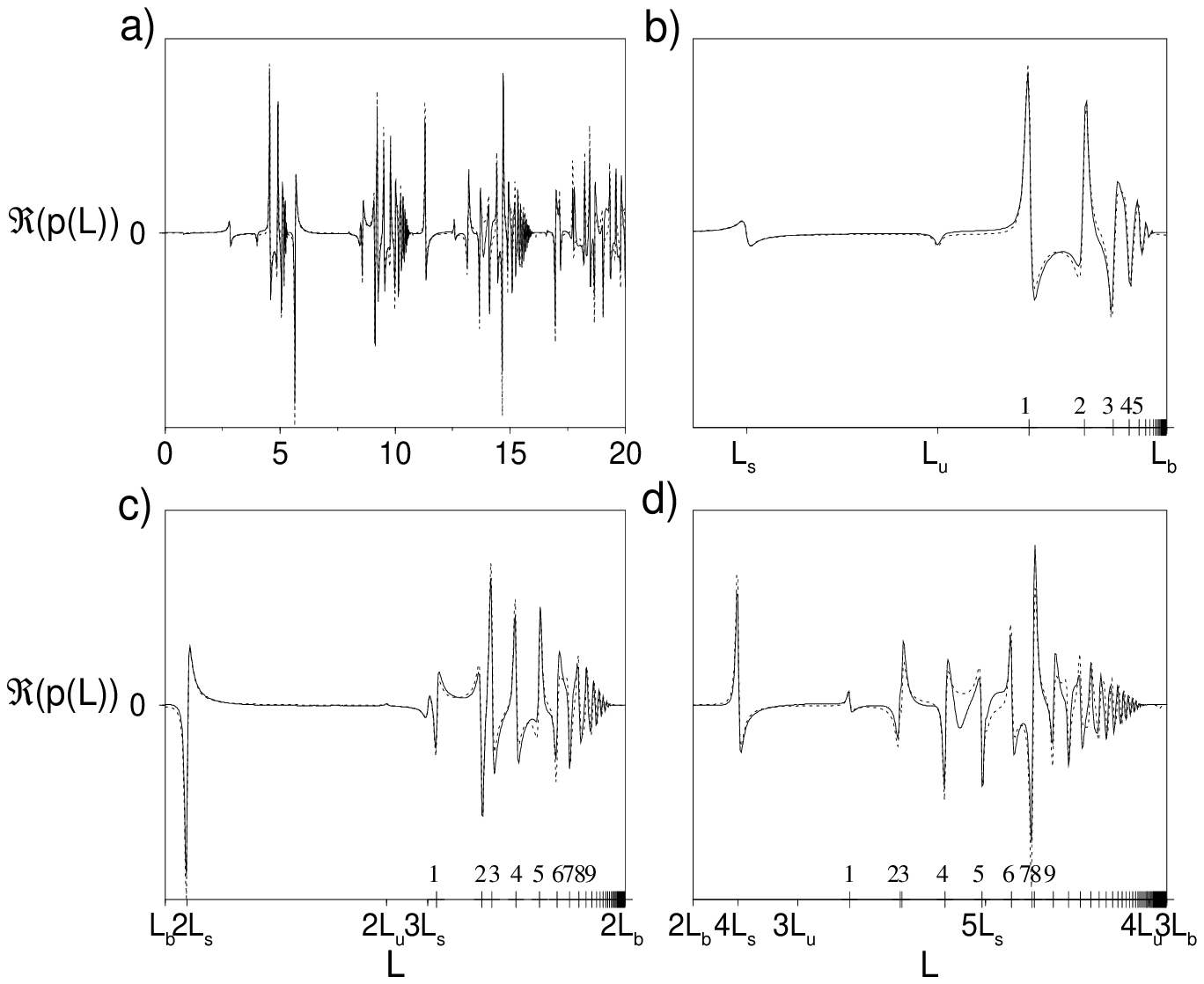,width=12.0cm}}
  \caption[]{\label{fouriertransformed} \capsty Comparison of
the exact to the semiclassical result for $\Re(p(L))$ ($a=1/\sqrt{2}$). 
The solid line is the cosine transform of the exact $n_{\text{osc}}(E)$, the
dotted line marks the semiclassical curve. (b),(c) and (d) give
magnifications of  the
length  ranges $[2.5,5.4]$, $[5.4,10.8]$ and $[10.8,16.2]$, respectively.}
\end{figure}
\end{Filesave}
}
The results obtained from these formulas are shown in
\fig~\ref{fig:fouriertransformed}. Here resonant tori with winding number $w=\rtopo_{\ang}/\rtopo_{\rad}$
and $\rtopo_{\ang},\rtopo_{\rad}\in\{1,...,50\}$ are included as far
as they are realized in phase space. 
Figure (a) again shows the clustering of
the contributions any time a multiple of the sliding orbit length
$\lengthb$ is met. In Figures (b), (c) and (d) we show
magnifications of the ranges between the $(q-1)$th and $q$th  multiple
of $\lengthb$ for $q=1,2,3$. The small ticks are labeled by the length
of the multiple traversals of $\ob$, $\os$ and $\ou$ that can be found
in that range. The large ticks labeled by numbers above belong to a
selection of resonant tori 
listed in \tab~\ref{tableorbitdata}. 
\begin{table}[!h]
\tabstart
\centering
\small
\begin{tabular}{|c|c|c|c|c||c|c|c|c|c||c|c|c|c|c|}
\hline
 (b)    & type  & $w$    & $q$& $L$       & 
 (c)    & type  & $w$    & $q$& $L$       &
 (d)    & type  & $w$    & $q$& $L$      \\ \hline
 $1$    & $\tR$ & $1/3 $ &$ 1$&  $ 4.56$  &
 $1$    & $\tR$ & $2/5 $ &$ 1$&  $ 8.58$  &
 $1$    & $\tR$ & $3/7 $ &$ 1$&  $12.59$ \\
 $2$    & $\tR$ & $1/4 $ &$ 1$&  $ 4.90$  &
 $2$    & $\tR$ & $1/3 $ &$ 2$&  $ 9.12$  &
 $2$    & $\tR$ & $3/8 $ &$ 1$&  $13.16$ \\ 
 $3$    & $\tR$ & $1/5 $ &$ 1$&  $ 5.07$  &
 $3$    & $\tO$ & $2/3 $ &$ 1$&  $ 9.23$  &
 $3$    & $\tO$ & $3/4 $ &$ 1$&  $13.18$ \\ 
 $4$    & $\tR$ & $1/6 $ &$ 1$&  $ 5.17$  &
 $4$    & $\tR$ & $2/7 $ &$ 1$&  $ 9.51$  &
 $4$    & $\tR$ & $1/3 $ &$ 3$&  $13.68$ \\
 $5$    & $\tR$ & $1/7 $ &$ 1$&  $ 5.23$  &
 $5$    & $\tR$ & $1/4 $ &$ 2$&  $ 9.80$  &
 $5$    & $\tR$ & $3/10$ &$ 1$&  $14.10$ \\ 
        &       &        &    &           &
 $6$    & $\tR$ & $2/9 $ &$ 1$&  $10.00$  &
 $6$    & $\tR$ & $3/11$ &$ 1$&  $14.43$ \\
        &       &        &    &           &
 $7$    & $\tR$ & $1/5 $ &$ 2$&  $10.15$  &
 $7$    & $\tO$ & $3/5 $ &$ 1$&  $14.67$ \\
        &       &        &    &           &
 $8$    & $\tR$ & $2/11$ &$ 1$&  $10.26$  &
 $8$    & $\tR$ & $1/4 $ &$ 3$&  $14.70$ \\ 
        &       &        &    &           &
 $9$    & $\tR$ & $1/6 $ &$ 2$&  $10.34$  &
 $9$    & $\tR$ & $3/13$ &$ 1$&  $14.91$ \\
\hline
\end{tabular}
\tabend
\caption[]{\capsty Data for the periodic orbit representatives of
resonant tori marked in \fig~\ref{fig:fouriertransformed}. $w$ is the
winding ratio, $L$ the length and $q$ the number of traversals.}
\label{tableorbitdata} 
\end{table}
The agreement between the exact and semiclassical curve is remarkably
good. It gets a little worse when $L$ becomes larger and the 
density of peaks grows. Then the 
sum gives an enormous mixture where the
distinction of the individual contributions becomes more or less
impossible.
In \fig~\ref{fig:fourierspecials} we show magnifications 
around some individual periodic objects. The third row shows how the
amplitude of the contribution of the traversal of $\os$ alternates in
magnitude. For any even number of traversals the contribution is that
of a thin torus and the amplitude is of the same order in magnitude as
the torus contributions in the first two rows. 
%
The remaining traversals contribute as ordinary stable 
isolated periodic orbits. 
The fourth row shows
the fast decrease of amplitudes of the unstable periodic orbit $\ou$
with a growing number of traversals. 
This feature is familiar from
hyperbolic systems where all orbits are of this kind and the
exponential decay of amplitudes with orbit length justify,
e.g., the cycle expansion of quantum mechanical as well as
classical dynamical Zeta functions (see, e.g., \cite{Eckhardt93}).
\begin{Filesave}{bilder}
\def\figfourierspecials{Comparison of the
exact (solid) and semiclassical (dotted) result for $\Re(p(L))$ 
in the neighborhood of the
following periodic objects (a)(i)-(iii) traversals of the type $\tR$
torus with $w=1/3$, (b)(i) and (ii) traversals of type $\tO$ torus
with $w=2/3$, (b)(iii) type $\tO$ torus with $w=3/5$, (c) traversals
of the stable periodic orbit $\os$ and (c)(i)-(iii) traversals of the
unstable periodic orbit $\ou$. 
The width for all pictures is 0.4, the heights are 140 or 20.}
\def\FIGfourierspecials{\centerline{\psfig{figure=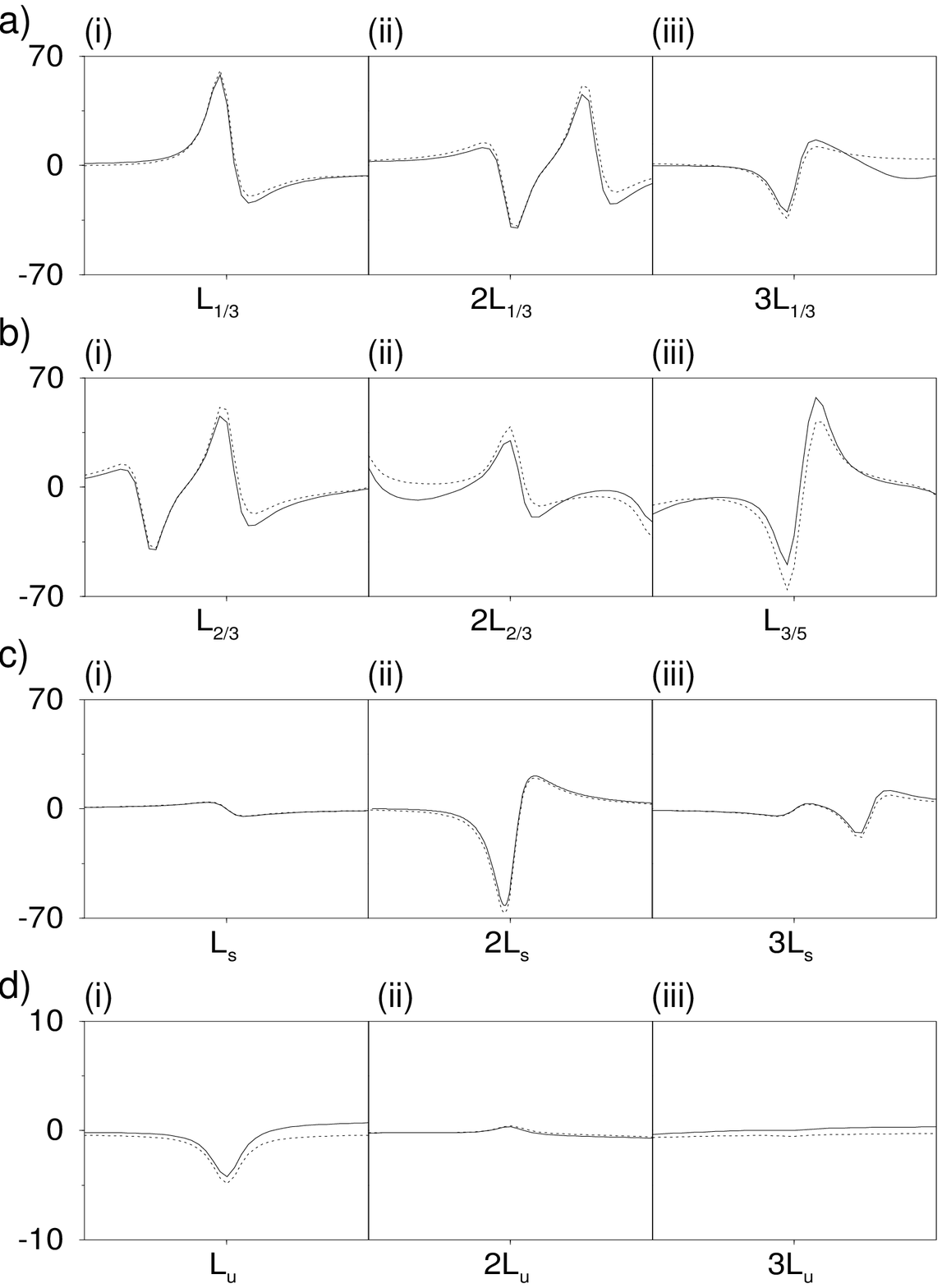,width=8.0cm}}}
\FIGo{fig:fourierspecials}{\figfourierspecials}{\FIGfourierspecials}
\end{Filesave}
\rem{
\begin{Filesave}{bilder}
\begin{figure}
  \centerline{\psfig{figure=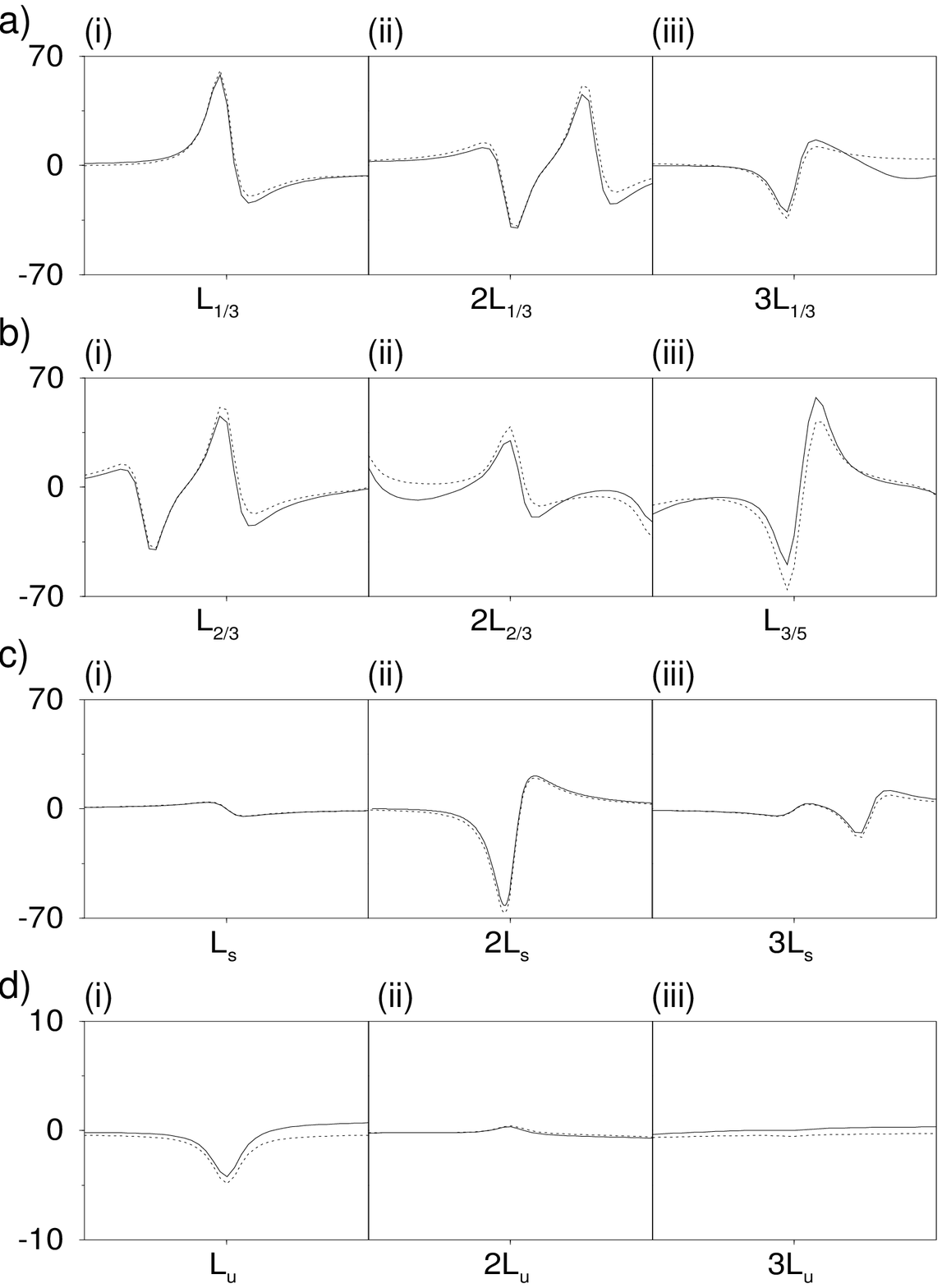,width=8.0cm}}
  \caption[]{\label{fourierspecials} \capsty Comparison of the
exact (solid) and semiclassical (dotted) result for $\Re(p(L))$ 
in the neighborhood of the
following periodic objects (a)(i)-(iii) traversals of the type $\tR$
torus with $w=1/3$, (b)(i) and (ii) traversals of type $\tO$ torus
with $w=2/3$, (b)(iii) type $\tO$ torus with $w=3/5$, (c) traversals
of the stable periodic orbit $\os$ and (c)(i)-(iii) traversals of the
unstable periodic orbit $\ou$. 
The width for all pictures is 0.4, the heights are 140 or 20.}
\end{figure}
\end{Filesave}
}
We performed the same calculations for $a=0.7$, where $\os$ is close to
resonant and also for $a=\cos(\pi(\sqrt{5}-1)/4)$ where $w$ becomes
equal to the golden mean, i.e. $\os$ becomes as far away from
resonant as possible. The results not represented here indicate that
any time a $q$-fold traversal of $\os$ has neighboring trajectories
that are almost closed in phase space it is better to replace the
stable orbit contribution by a thin torus contribution. 
\NEU
For $a=0.7$ this means we have a strong contribution for the first 
even traversal of the isolated stable periodic orbit which can be viewed
as a result of the real complex orbit corresponding to $w=1/2$. 
For $a=\cos(\pi(\sqrt{5}-1)/4)$ this is the case for $q=3,5,8,...$, 
the denominators of the continued fraction
approximations $2/3,3/5,5/8,...$ for the golden mean. 

The version of  Berry-Tabor formula in \equ~(\ref{eq:btuniform})  gives a
uniform expression that is valid for all winding numbers of stable
isolated periodic orbits.
In the non-rational case the contributions to the inverse spectrum can
no longer be calculated analytically. 
In \cite{Sieber96} the inverse spectrum was calculated numerically.
We restricted ourselves to a resonant case in order to be able to 
obtain analytical results.

\section{Conclusions and outlook}
\label{sec:conclusions}

From the classical point of view the billiard in the ellipse
is a typical integrable system: 
the frequencies change from one torus to another
and there exist both, stable and unstable isolated periodic orbits, 
the latter leading to a separatrix on the energy surface.
Extending the classical mechanics to the complex plane, we introduced
three kinds of complex orbits: (i) tunneling orbits and (ii) scattering
orbits, both with imaginary action (``imaginary complex orbit''), 
and (iii) orbits with complex turning points, complex momentum and 
real action (``real complex orbits'').
The classification of these orbits in terms of the position of the turning 
points and their connection by a Stokes or anti-Stokes
line applies to all separable systems.
The action of these orbits appears naturally in the semiclassical treatment 
of these systems. 

The separation of the Schr\"odinger equation leads to special 
cases of the spheroidal wave equation corresponding 
to ``spin'' $\pm1/2$.
With a simple transformation the two coupled boundary value problems can 
be turned into a form suitable for the application of a standard shooting 
method in order to efficiently calculate the eigenvalues.
The two discrete spatial symmetries of the ellipse lead to the distinction 
of four parities, while the time reversal symmetry leads to an
asymptotic degeneracy of tori involving a rotational degree of freedom.
As expected from the \WKB\ approximation the wave functions are 
concentrated on the projection of the classical torus onto 
configuration space. 
The isolated unstable periodic orbit induces a less pronounced 
``scar'' for states close to the separatrix.
Additionally these wave functions strongly localize at the focus points
of the ellipse. An explanation of this behavior is under investigation.

The eigenvalue spectrum can best be understood when transformed
to the classical action space, where eigenstates are located either on 
simple \EBK\ lattices far away from the separatrix or in a more
complicated transition regime close to the separatrix.
In order to semiclassically describe the transition between the
\EBK\ lattices in the vicinity of the separatrix a uniform
quantization scheme was employed, which smoothes the discontinuity of the
Maslov indices at the separatrix. The imaginary complex orbits
are used in this context to define 
a smooth Maslov phase, replacing the discontinuous Maslov index.
The semiclassical states are obtained as the intersections
of a set of lines in classical action space. 
Numerically these states agree quite well with the exact results, 
even for low quantum numbers.
The relative error shows regular patterns, except for separatrix states,
and it tends to be large for states with at least one small
quantum number and for states close to the separatrix.
A state with given quantum number is located inside a 
``semiclassical quantum cell'', and specifying its parity narrows its
position down to a ``parity box'' of width $\hbar/2$.
This interpretation of the uniform quantization allows for
an illuminating picture of the situation for states close to the separatrix.
Without uniformization, parity boxes intersected by the separatrix may 
contain a wrong number of \EBK\ states.

The Berry-Tabor trace formula rests on \EBK\ quantization. 
In the original uniform version it contains contributions from 
real complex orbits, but imaginary complex orbits are not incorporated.
A treatment ignoring the separatrix produces faulty peaks in the
semiclassical density of states. They occur for states that
belong to parity boxes with the wrong number of \EBK\ states.
Incorporating the imaginary complex orbit by a smoothly varying 
Maslov phase leads to a modified trace formula.
In the \WKBBT\ version these corrections are ignored in the
stationary phase approximation, resulting in a sum over 
resonant tori as in the original version. 
This, however, cannot correct all the peaks from states belonging to 
parity boxes with the wrong number of \EBK\ states. 
In order to obtain satisfactory peaks for these states
we had to fully take into account
the non-constant Maslov phase.
This leads to a \uniWKBBT\ trace formula quite similar to the original one, 
with a changed effective energy surface now depending on the energy.
The effective resonant tori of this effective energy surface do not 
correspond to periodic orbits of the billiard. 
With this sum over (in general) non-periodic orbits we were 
able to produce the correct peaks for all states.
The numerical effort increases considerably because the stationary points
have to be found anew for every energy.
Even though the \uniWKBBT\ trace formula might not be a useful 
tool for the practical calculation of eigenvalues, it can give
some hints on how to incorporate complex orbits into 
formulas of this type.
It remains an open question how a uniformization should come about
in terms of a Green's function derivation of the Berry-Tabor trace
formula \cite{BerryTabor77}. It is by no means obvious how the variable Maslov
phase should be emulated by, say, a summation over complex orbits. 

In the study of the length spectrum of our integrable system
we focused on the case where
the stable orbit has a rational winding number. In this case the
divergent Gutzwiller term has to be replaced by a thin torus
contribution obtained by Richens. The four types of dominant 
contributions in the trace formula can be analytically Fourier
transformed, such that the inverse spectrum can be 
explicitly written as a sum over periodic objects.
We found a remarkably good agreement, even though in this
approach only the real complex tori are partially incorporated in 
the Richens term. 
It is an open question why the inverse spectrum
is so much less sensitive to the presence of the separatrix
in our case.

Extensions of our work can be done in two directions. 
On the one hand one can take the elliptic billiard as a
starting point for the penetration into the nonintegrable regime by a
deformation of the billiard boundary. This is the content of \cite{Sieber96}. 
On the other hand one can consider the three dimensional billiards 
in the ellipsoid
\cite{RW95,WR96}, starting with the cases of prolate and oblate 
spheroids.
A first step in this direction can be found in \cite{AyantArvieu86a}. 
The Berry-Tabor trace formula for a system with three degrees of freedom 
is much more involved, because the resonant tori can no longer be 
labeled by a single rational winding number. 
Instead one has to find a complicated set of resonances on the 
energy surface \cite{WR96}.
For these 3D billiards, the exact quantum mechanical spectrum, 
its semiclassical quantum cells, and and its length spectrum 
are currently under investigation.

\section*{Acknowledgements}

We thank P.H.~Richter and M.~Sieber for illuminating discussions and M.~Sieber for
communicating his results before publication. This work was supported
by the Deutsche Forschungsgemeinschaft.

\bibliographystyle{unsrt}

\CLOBIL

\Capts{ 
\section*{Figure captions}
\setlength{\parindent}{0cm}
\printfigcap{fig:effpot}{\figeffpot}
\printfigcap{fig:esurface}{\figesurface}
\printfigcap{fig:elliorb}{\figelliorb}
\printfigcap{fig:waves}{\figwaves}
\printfigcap{fig:xietagraph}{\figxietagraph}
\printfigcap{fig:statesX}{\figstatesX}
\printfigcap{fig:statesI}{\figstatesI}
\printfigcap{fig:Mwaves}{\figMwaves}
\printfigcap{fig:qcell}{\figqcell}
\printfigcap{fig:errors}{\figerrors}
\printfigcap{fig:statescmp}{\figstatescmp}
\printfigcap{fig:EBKBT}{\figEBKBT}
\printfigcap{fig:WKBBT}{\figWKBBT}
\printfigcap{fig:effesurf}{\figeffesurf}
\printfigcap{fig:uniWKBBT}{\figuniWKBBT}
\printfigcap{fig:powerspectrum}{\figpowerspectrum}
\printfigcap{fig:fouriertransformed}{\figfouriertransformed}
\printfigcap{fig:fourierspecials}{\figfourierspecials}
\rem{
\newpage
\showfig{\FIGeffpot}
\showfig{\FIGesurface}
\showfig{\FIGelliorb}
\showfig{\FIGwaves}
\showfig{\FIGxietagraph}
\showfig{\FIGstatesX}
\showfig{\FIGstatesI}
\showfig{\FIGMwaves}
\showfig{\FIGqcell}
\showfig{\FIGerrors}
\showfig{\FIGstatescmp}
\showfig{\FIGEBKBT}
\showfig{\FIGWKBBT}
\showfig{\FIGeffesurf}
\showfig{\FIGuniWKBBt}
\showfig{\FIGpowerspectrum}
\showfig{\FIGfouriertransformed}
\showfig{\FIGfourierspecials}
}
}

\end{document}